\documentclass[a4paper,12pt, leqno]{article}
\usepackage{amsmath}
\usepackage{amssymb, latexsym}
\usepackage{amsfonts}
\usepackage{makeidx}
\usepackage{amssymb}
\usepackage{color}     

\date{}

\newtheorem{lem}{Lemma}[section]
\newtheorem{thm}{Theorem}[section]
\newtheorem{prop}{Proposition}[section]

\newtheorem{definition}{Definition}[section]

\numberwithin{equation}{section}
\newcommand{\dbar}{d\!\!\!{\lower-0.6ex\hbox{$-$}}\!}
\newcommand{\dslash}{d\!\!\!{\lower-0.6ex\hbox{$-$}}}
\newcommand{\e}{\varepsilon}
\newcommand{\h}{\hbar}
\newcommand{\ott}{\lower-0.4ex\hbox{${\scriptscriptstyle{\otimes}}$}}
\newcommand{\btt}{\lower-0.2ex\hbox{${\scriptscriptstyle{\bullet}}$}}
\newcommand{\ctt}{\lower-0.2ex\hbox{${\scriptscriptstyle{\circ}}$}}
\newcommand{\dtt}{\lower-0.2ex\hbox{${\scriptscriptstyle{\diamond}}$}}
\newcommand{\odt}{\lower-0.4ex\hbox{${\scriptscriptstyle{\odot}}$}}
\newcommand{\bsG}{\boldsymbol\varGamma}
\newcommand{\Cal}{\mathcal}
\begin{document}

\pagestyle{plain}

\title{Deformation Expression for Elements of Algebras (VIII)\\
-- $SU(2)$-vacuum and the regular representation space--}


\author{
     Hideki Omori\thanks{ Department of Mathematics,
             Faculty of Sciences and Technology,
        Tokyo University of Science, 2641, Noda, Chiba, 278-8510, Japan,
         email: omori@ma.noda.tus.ac.jp}
        \\Tokyo University of Science
\and  Yoshiaki Maeda\thanks{Department of Mathematics,
                Faculty of Science and Technology,
                Keio University, 3-14-1, Hiyoshi, Yokohama,223-8522, Japan,
                email: maeda@math.keio.ac.jp}
          \\Keio University
\and  Naoya Miyazaki\thanks{ Department of Mathematics, Faculty of
Economics, Keio University,  4-1-1, Hiyoshi, Yokohama, 223-8521, Japan,
        email: miyazaki@hc.cc.keio.ac.jp}
        \\Keio University
\and  Akira Yoshioka \thanks{ Department of Mathematics,
          Faculty of Science, Tokyo University of Science,
         1-3, Kagurazaka, Tokyo, 102-8601, Japan,
         email: yoshioka@rs.kagu.tus.ac.jp}
           \\Tokyo University of Science
     }

\maketitle



\setcounter{equation}{0}

\tableofcontents


\section{Introduction}

Rotations in the 3-dimensional space is expressed in terms of
vector analysis as  a linear combination of rotations  w.r.t. each axis
$$
R_t= 
(\tilde{e}_2{\times}\tilde{e}_3)p_1{+}(\tilde{e}_3{\times}\tilde{e}_1)p_2
{+}(\tilde{e}_2{\times}\tilde{e}_3)p_3,
\quad p_1^2{+}p_2^2{+}p_3^2{=}\mu^2.
$$
 
Consider the problem ``Give the equation of the conceptional rotations in ${\mathbb R}^3$
without using the parameter expressing individual rotations'', just as
the conceptional motion of constant velocity along straight lines
(Galiley motions) is expressed as $\frac{d^2}{dt^2}{\pmb r}{=}0$. 
The best answer may by given as follows:

Let $1, \tilde{e}_1, \tilde{e}_2, \tilde{e}_3$ be the natural basis of
quaternion field $({\mathbb Q}, *)$. The answer is the quaternion
valued one component equation  
\begin{equation}\label{bestans}
\partial_t\phi(t, x_1, x_2, x_3){=}
\frac{1}{2}[\tilde{e}_1\partial_{x_1}+\tilde{e}_2\partial_{x_2}{+}\tilde{e}_3\partial_{x_3},
\phi(t, x_1, x_2, x_3)]
\end{equation}
where $[\,\,,\,\,]$ is the commutator bracket in the quaternion
field. \eqref{bestans} splits into the 3-component equation by setting 
$\phi(t,{\pmb x}){=}
\phi_1(t,{\pmb x})\tilde{e}_1
{+}\phi_2(t,{\pmb x})\tilde{e}_2
{+}\phi_3(t,{\pmb x})\tilde{e}_3$ 
$$
\partial_{t}
\begin{bmatrix}
\phi_1\\\phi_2\\\phi_3
\end{bmatrix}{=}
\begin{bmatrix}
0&{-}\partial_{x_3}&\partial_{x_2}\\
\partial_{x_3}&0& -\partial_{x_1}\\
-\partial_{x_2}&\partial_{x_1}&0\\
\end{bmatrix}
\begin{bmatrix}
\phi_1\\\phi_2\\\phi_3
\end{bmatrix}
$$
Behind this equation, we see the partial differential  equation 
$$
i\partial_t\phi(t,x_1,x_2,x_3){=}
(\tilde{e}_1\partial_{x_1}+\tilde{e}_2\partial_{x_2}{+}\tilde{e}_3\partial_{x_3}){*}\phi(t, x_1, x_2, x_3).
$$
This is called the {\bf Weyl equation}.  

Setting 
$\phi(t,{\pmb x}){=}\phi_0(t,{\pmb x}){+}
\phi_1(t,{\pmb x})\tilde{e}_1
{+}\phi_2(t,{\pmb x})\tilde{e}_2
{+}\phi_3(t,{\pmb x})\tilde{e}_3$, we have 
$$
\begin{aligned}
i\partial_ti\phi_0(t,\pmb x)&{=}
i\partial_{x_1}\phi_1{+}i\partial_{x_2}\phi_2{+}i\partial_{x_3}\phi_3\\
i\partial_t\phi_1(t,\pmb x)&{=}\partial_{x_2}\phi_3{-}\partial_{x_3}\phi_2{+}\partial_{x_1}i\phi_0\\
i\partial_t\phi_2(t,\pmb x)&{=}\partial_{x_3}\phi_1{-}\partial_{x_1}\phi_3{+}\partial_{x_2}i\phi_0\\
i\partial_t\phi_3(t,\pmb x)&{=}\partial_{x_1}\phi_2{-}\partial_{x_2}\phi_1{+}\partial_{x_3}i\phi_0\\
\end{aligned}
$$
This is viewed as the equation of the conceptional $3$-dim rotations
taking the time parameter in mind.
Replacing $t'{=}-it$, the original equation changes into:
$$
\partial_{t'}
\begin{bmatrix}
i\phi_0\\\phi_1\\\phi_2\\\phi_3
\end{bmatrix}{=}
\begin{bmatrix}
0&-\partial_{x_1}&-\partial_{x_2}&-\partial_{x_3}\\
\partial_{x_1}&0&{-}\partial_{x_3}&\partial_{x_2}\\
\partial_{x_2}&\partial_{x_3}&0& -\partial_{x_1}\\
\partial_{x_3}&-\partial_{x_2}&\partial_{x_1}&0\\
\end{bmatrix}
\begin{bmatrix}
i\phi_0\\\phi_1\\\phi_2\\\phi_3
\end{bmatrix},
$$
these are $O(1,3)$-invariant w.r.t. $4$-components
 $(\phi_0,\phi_1,\phi_2,\phi_3)$. 

 At a first glance, 
there is no parameter indicating individual 
rotations.  
However, Fourier-transform   
$\hat{\phi}(t,{\pmb p}){=}\int_{{\mathbb R}^3}e^{i\sum  p_ix_i}\phi(t,{\pmb x})\dbar{\pmb x}$
changes the original equation into 
$$
\partial_{t}\hat{\phi}(t,{\pmb p}){=}
({\tilde{e}_1}p_1+{\tilde{e}_2}p_2{+}{\tilde{e}_3}p_3){*}\hat{\phi}(t,{\pmb p}),
$$
and individual rotations are involved in $p_1, p_2, p_3$. 

Namely, via Fourier transformation space-variables are transformed
into momentum-variables. As a result individual rotations appear 
instead of Lorentz covariance. 

\bigskip

The square of the Weyl equation is the (mass-less) Klein-Gordon equation:
$$
\partial^2_t\phi{=}(\partial^2_{x_1}{+}\partial^2_{x_2}{+}\partial^2_{x_3})\phi.
$$
Weyl equation is viewed as the ``field equation'' of relativistic
mass-less particles. The massive Klein-Gordon equation is 
$$
\partial^2_t\phi{=}(\partial^2_{x_1}{+}\partial^2_{x_2}{+}\partial^2_{x_3}{+}m^2)\phi.
$$
The reason why $m$ is the mass term 
is based on Einstein's relation $E^2{=}{\pmb p}^2{+}m^2c^2$. 

\medskip
On the other hand in the theory of Dirac, the equation of 
massive ($m{>}0$) relativistic particles is given by 
\begin{equation}\label{SinhCosh00}
i\partial_{\xi_0}
{\footnotesize
{\begin{bmatrix}
\Phi\\\Psi
\end{bmatrix}}}
{=}
{\footnotesize{
\begin{bmatrix}
\mu I_2&D\\
D&-\mu I_2
\end{bmatrix}}}
{\footnotesize
{\begin{bmatrix}
\Phi\\\Psi
\end{bmatrix}}},\quad 
{D}{=}\frac{1}{\h}(\rho_1\partial_{\xi_1}{+}
      \rho_2\partial_{\xi_2}{+}\rho_3\partial_{\xi_3})
\end{equation}
where 
$\rho_i{=}
{\footnotesize
{\begin{bmatrix}
\sigma_i&0\\
0&\sigma_i
\end{bmatrix}}}
$ 
with Pauli matrices $\sigma_i$ and $\mu$ is called the mass-term. 
As this is considered in the {\bf Clifford algebra} of 4 generators,  
this is viewed as an equation for ``Fermions''. 

However, in \cite{OMMY4} we found Clifford algebras are contained in 
the {\bf transcendentally extended} Weyl algebra. In contrast, it is known that 
the Clifford algebra of {\bf infinite generators} contains Weyl algebra. 

On the other hand in physics, equations written in Weyl algebra (not transcendentally extended)
is often called ``Bosonic equations''. Note that the Bose-Fermi
symmetry is widely believed in physics. Thus, it is mathematically 
interesting to write the ``Bosonic equation''  
corresponding to the (massive) Dirac equation. 

In this note, we propose an equation which is covariant (not invariant) 
under Lorentz transformations. The fundamental solution will be 
constructed in the form of product integrals. But unfortunately,  
it is still difficult to negotiate with causality conditions in 
the theory of relativity, as we do not have definite idea of treating 
the time parameter.

\bigskip
Now, we have to note that the term ``vacuums'' used often in this note may
not be the same that physicists uses. 
A $G$-vacuum is only a group $G$-invariant 
idempotent element in some extended Weyl algebra. It is still 
difficult to obtain $SU(2)$-vacuum. This is because we have to 
restrict the expression parameters to a very special class which is 
very easy to be broken. Much work seem to be done in simulating the 
$SU(2)$-vacuum, with the aim of learning more about the vacuum
structure of QCD. But our $SU(2)$-vacuum may have nothing to do with 
these. 

Readers will find many obstructions in mathematical detail, 
but we think these suggest many interesting subjects. 
In particular, throughout this series we are concerned 
with parameters of expressing elements of algebras. In a very 
beginning these parameters are thought inessential for physics. 
However, we found there are many phenomena that suggest 
such parameters must have some significance in physics. 
This note is only a trial to understand mathematically the
significance of expression parameters.

\section{Weyl algebra with hermitian structures and $*$-exponentials}

First of all, we define the Weyl algebra $(W_2, *)$ as the associative
algebra of 2 generators $u, v$ with the commutation relation 
$[u,v]_*{=}-i\h$. $W_2$ extends transcendentally under suitable topology 
so that the exponential functions of quadratic forms are treated.

An  hermitian structure is defined by setting 
$u^*=u, v^*{=}v, i^*{=}-i, [u,v]^*{=}[v,u]$ on generators $u, v$ as an  
involutive anti-automorphism.  Here $u, v$ are treated as
hermite elements. Hence $2u{\ctt}v{=}u{*}v{+}v{*}u$, $u^2{+}v^2$,
$u^2{-}v^2$ are hermite elements.

There is another hermite structure defined by $u^*{=}iv, v^*{=}iu$, 
and $i^*{=}-i$. Hence $u{\ctt}v$, $i(u^2{+}v^2)$, $(u^2{-}v^2)$ are hermite elements.

Although it is involutive only on even elements, if we define an 
anti-automorphism by $u^*{=}v$, $v^*{=}-u$, $i^*{=}-i$, then
$iu{\ctt}v$, $i(u^2{-}v^2)$, $u^2{+}v^2$ are hermite elements.

\subsection{Product formulas and linear change of generators}

To make a transcendental extension, we use a concrete product formula
by setting $(u,v){=}(u_1,u_2)$ 
\begin{equation}\label{prodform0}
f{*_{_K}}g{=}fe^{\frac{i\h}{2}
\sum_{ij}\overleftarrow{\partial}_i{\Lambda}_{ij}\overrightarrow{\partial}_j}g,
\quad \Lambda{=}K{+}J,\quad 
J={\footnotesize{
\begin{bmatrix}
0&{-}1\\
1&0
\end{bmatrix}}}
\end{equation}
where $K{=}(K_{ij})$ is any $2{\times}2$ complex symmetric matrix.
The algebraic structure does not depend on $K$, but the expression of
elements depend on $K$. $K$ is called an expression parameter. 
For instance,  $u_*^2$, $u{*}v$ are defined in $W_2$, 
but $u_{*_{_K}}^2{=}u^2{+}{2i\h}K_{11}$, $u{*_{_K}}v{=}uv{+}{2i\h}(K_{12}{-}1)$.
We denote these by 
$$
{:}u_*^2{:}_{_K}{=}u^2{+}{2i\h}K_{11},\quad 
{:}u{*}v{:}_{_K}{=}uv{+}{2i\h}(K_{12}{-}1).
$$
Such a manner to express elements is redundant  
while we treat $*$-polynomials or 
$*$-exponential functions of linear forms. However, expression
parameters play important roles in Jacobi's $\theta$-functions (cf.\cite{OMMY3}) 
and these play essential  roles when $*$-exponential functions of quadratic forms are
treated. Indeed it is a main target of this series of notes to
understand the significance of expression parameters.

If $K{=}0$, \eqref{prodform0} is called the 
{\bf Moyal product formula}. By restricting $(u,v)$ in ${\mathbb R}^2$, this formula  
may be given by the integral form 
\begin{equation}\label{Moualinteg}
f{*_0}g(u,v){=}
os{\text{-}}\!\!\iint_{\mathbb R^2}
f(u{+}\frac{\h}{2}s, v{+}\frac{\h}{2}t)
g(u{+}s',v{+}t')e^{i(ts'{-}st')}\dbar s\dbar t\dbar s'\dbar t'.
\end{equation}
If $K{=}K_0{=}
{\footnotesize{
\begin{bmatrix}
0&1\\
1&0
\end{bmatrix}}}$,
then \eqref{prodform0} is called the product formula of
pseudo-differential operators ({\bf{$\Psi$DO-product formula}} in
short). Its integral form is 
$$
f{*_{_{K_0}}}g(u,v){=}
os{\text{-}}\!\!\iint_{\mathbb R^2}
f(u, v{+}\h s)
g(u{+}t,v)e^{-ist}\dbar s\dbar t.
$$
These integral formulas are convenient to treat ``lower order terms'',
while \eqref{prodform0} is used to compute ``positive order terms''.
In \eqref{prodform0}, $(u,v)$ are viewed as complex variables.

\medskip
Next, we consider the effect of a linear change of generators  
$$
{u'}_i=\sum u_kS_i^k, \quad S\in S\!L(2,{\mathbb C}),\quad 
 ({\pmb u}'={\pmb u}S).
$$
By the help that 
$$
\partial_{u_i}=\sum S_i^k\partial_{u'_k},
$$
the product formula is rewritten 
by using new generators as 
\begin{equation} \label{eq:KK2}
 f*_{_{K}}g=fe^{\frac{i\h}{2}
(\sum\overleftarrow{\partial_{u'_i}}
({}^t\!S{\Lambda}S)^{ij}\overrightarrow{\partial_{u'_j}})}g.
\end{equation}
As ${}^t\!SJS=J$, the algebraic structure of 
$({\mathbb C}[\pmb u], *_{_K})$ does not change.
Thus the notation $*_{_K}$ is better to be replaced by 
$*_{_{K'}}$ where $K'{=}{}^t\!S{K}S$. 

For every $K, K'$, the intertwiner
is defined by  
\begin{equation}
\label{intertwiner}
I_{_K}^{^{K'}}(f)=
\exp\Big(\frac{i\h}{4}\sum_{i,j}(K'_{ij}{-}K_{ij})
\partial_{u_i}\partial_{u_j}\Big)f \,\,
(=I_{0}^{^{K'}}(I_{0}^{^{K}})^{-1}(f)). 
\end{equation}
This gives an isomorphism 
$I_{_K}^{^{K'}}:({\mathbb C}[{\pmb u}]; *_{_{K+J}})\rightarrow 
({\mathbb C}[{\pmb u}]; *_{_{K'+J}})$.
Namely, the following identity holds for any 
$f,g \in {\mathbb C}[{\pmb u}]:$ 
\begin{equation}\label{intertwiner2}
I_{_K}^{^{K'}}(f*_{_K}g)=
I_{_K}^{^{K'}}(f)*_{_{K'}}I_{_K}^{^{K'}}(g).
\end{equation}
Intertwiners do not change the algebraic structure $*$, 
but do change the expression of elements by the ordinary 
commutative structure.  

Thus, a symplectic change of generators is recovered by    
the intertwiner $I_{_K}^{^{{}^t\!SKS}}$.  
Change of generators are viewed often as coordinate transformations, 
but note here that $I_{_K}^{^{{}^t\!SKS}}$ 
is something like the ``square root'' of 
symplectic  coordinate transformations, and these behaves 
as $2$-to-$2$ mappings on the space of $*$-exponential functions of 
quadratic forms. (Cf. \cite{OMMY3}, \cite{OMMY4}.)

\subsection{Star-exponential functions of quadratic forms
and several properties}

The $K$-ordered expression of the $*$-exponential function 
$e_*^{t\frac{1}{2i\h}\langle{\pmb u}g, {\pmb u}g\rangle_*}$ for 
$g{\in}S\!L(2,{\mathbb C})$ is given by  
\begin{equation}\label{tildeKKK}
\begin{aligned}
{:}e_*^{\frac{t}{i\h}
\langle{\pmb u}g, {\pmb u}g\rangle_*}{:}_{_{K}}
&{=}
\frac{1}{\sqrt{\det(\cos t I{-}(\sin t){}^t\!gKg)}}
e^{\frac{1}{i\h}\langle{\pmb u}g 
\frac{\sin t}{\cos tI-\sin t \,{}^t\!g Kg},\,\,{\pmb u}g\rangle}.
\end{aligned}
\end{equation}
where $\pmb u{=}(u,v)$, $[u,v]{=}-i\h$, and $K$ is any complex symmetric matrix.
Note that the phase part and the inside of $\sqrt{\,\,\,}$ is
$\pi$-periodic (not $2\pi$-periodic).  It is easy to rewrite
\eqref{tildeKKK} as 
\begin{equation*}
\begin{aligned}
{:}e_*^{\frac{t}{i\h}
\langle{\pmb u}g, {\pmb u}g\rangle_*}{:}_{_{K}}
&{=}
\frac{1}{\sqrt{\det(\cos t I{-}(\sin t)g\,{}^t\!gK)}}
e^{\frac{1}{i\h}\langle{\pmb u}
\frac{\sin t}{\cos tI-\sin t \,g\,{}^t\!g K}g\,{}^t\!g,\,\,{\pmb u}\rangle}.
\end{aligned}
\end{equation*}

\bigskip 

Now setting $t{=}\pi$ in \eqref{tildeKKK}, we have 
${:}e_*^{\frac{\pi}{i\h}
\langle{\pmb u}g, {\pmb u}g\rangle_*}{:}_{_{K}}{=}1/{\sqrt{1}}$.
For a while we assume $\det K{=}1$. To manage the sign of 
$\sqrt{\,\,\,\,}$, we first fix the value
at $t=0$ as ${:}e_*^{\frac{0}{i\h}
\langle{\pmb u}g, {\pmb u}g\rangle_*}{:}_{_{K}}{=}1$. 

Now let $\mu, \nu$ be the eigenvalues of ${}^t\!g Kg$. Then, 
$\mu\nu{=}\det{}^t\!g Kg{=}1$ and $\mu{+}\nu{=}{\rm{tr}.}({}^t\!g Kg)$
$$
\begin{aligned}
\sqrt{\det(\cos t I{-}(\sin t){}^t\!gKg)}&{=}
\sqrt{(\cos t{-}\mu\sin t)(\cos t{-}\nu\sin t)}\\
&{=}\frac{1}{2}\sqrt{(1{+}i\mu)(1{+}i\nu)}e^{-it}
\sqrt{e^{2it}{-}\frac{i{-}(-\mu)}{-i{-}(-\mu)}}
\sqrt{e^{2it}{-}\frac{i{-}(-\nu)}{-i{-}(-\nu)}}.
\end{aligned}
$$
Note that $\big|\frac{i{-}(-\mu)}{-i{-}(-\mu)}\big|$ is the ratio 
of the distance of $-\mu$ from $i$ and $-i$.  Hence if $(-\mu)$ is 
in the upper half-plane, then
$\big|\frac{i{-}(-\mu)}{-i{-}(-\mu)}\big|<1$ and therefore 
$\sqrt{e^{2it}{-}\frac{i{-}(-\mu)}{-i{-}(-\mu)}}$ changes the sign 
when $t$ moves from $0$ to $\pi$. (Note that the inside of
$\sqrt{\,\,}$ is $\pi$-periodic.) 
Denote for simplicity 
$$
R_{\mu}{=}\frac{i{-}(-\mu)}{-i{-}(-\mu)},\quad R_{\nu}{=}\frac{i{-}(-\nu)}{-i{-}(-\nu)} 
$$
Furthermore considering ${re^{2it}{-}R_{\mu}}$
for some $0{<}r{<}1$, we see singular points appears $\pi$-periodically 
on the line in lower half plane parallel to the real axis.

As $\det({}^t\!gKg){=}\det K{=}\mu\nu=1,$ 
if $\mu, \nu \not\in {\mathbb R}$, then 
${:}e_*^{\frac{t}{i\h}
\langle{\pmb u}g, {\pmb u}g\rangle_*}{:}_{_{K}}$ 
is not singular on $t{\in}{\mathbb R}$ and 
only one of
$\sqrt{e^{2it}{-}\frac{i{-}(-\mu)}{-i{-}(-\mu)}}$, 
$\sqrt{e^{2it}{-}\frac{i{-}(-\nu)}{-i{-}(-\nu)}}$ changes the sign
when $t$ moves from $0$ to $\pi$. As $e^{-it}$ changes sign when $t$ moves 
from $0$ to $\pi$,  $\sqrt{\det(\cos t I{-}(\sin t){}^t\!gKg)}$ does not
change sign. It follows  
${:}e_*^{\frac{\pi}{i\h}
\langle{\pmb u}g, {\pmb u}g\rangle_*}{:}_{_{K}}{=}1$. 

\begin{lem}\label{key}
If ${\rm{tr}.}({}^t\!gKg)$ is not a real number, or if 
${\rm{tr}.}({}^t\!gKg)$ is a real number with $({\rm{tr}.}({}^t\!gKg))^2<4$,
then $\mu, \nu$ are not real number and 
$\sqrt{\det(\cos t I{-}(\sin t){}^t\!gKg)}$ does not
change sign.
\end{lem}

\bigskip
Setting $t{=}\pi/2$ in \eqref{tildeKKK}, we have 
$$
{:}e_*^{\frac{\pi}{2i\h}
\langle{\pmb u}g,{\pmb u}g\rangle_*}{:}_{_{K}}{=}
\frac{1}{\sqrt{\det K}}
e^{-\frac{1}{i\h}\langle{\pmb u}\frac{1}{K},{\pmb u}\rangle}
$$
which looks independent of $g{\in}S\!L(2{\mathbb C})$. 
But recall here that the formula of changing generators give 
$$
{:}e_*^{\frac{t}{i\h}
\langle{\pmb u}g, {\pmb u}g\rangle_*}{:}_{_{K}}{=}
{:}e_*^{\frac{t}{i\h}
\langle{\pmb u}, {\pmb u}\rangle_*}{:}_{_{{}^t\!g\,Kg}}.
$$
Change of expression parameters sometimes give a change of generators.
The above observation relating to Lemma \ref{key} shows this must
depend on $g$ discontinuously.

\medskip
Denoting 
${\e}_{00}{=}e_*^{\frac{\pi}{2i\h}\langle{\pmb u}g,{\pmb u}g\rangle_*}$,
we call this the {\bf polar element}.  
The exponential law gives in general  
$$
{:}{\e}_{00}^2{:}_{_{K}}{=}{\pm}1
$$
depending on $K$.  

\bigskip

For the use in the later section, we give the formula of
$*$-exponential function of degenerate quadratic from 
\begin{equation}
 \label{eq:expquad}
{:}e_*^{t\frac{1}{i\h}(\xi_1u{+}\xi_2v)_*^2}{:}_{_K}{=}
\frac{1}{\sqrt{1{-}\tau t}}\,
e^{\frac{t}{1{-}\tau t}(\xi_1u{+}\xi_2v)_*^2}, \quad \tau{=}\langle{\xi K,\xi\rangle}
\end{equation}
which is decreasing in $t$ of the order $\sqrt{|t|}^{-1/2}$.

\subsection{Notes from linear algebra}\label{linear}

Now setting 
$g={\footnotesize{
\begin{bmatrix}
a&b\\
c&d 
\end{bmatrix}}}\in S\!L(2,{\mathbb C})$, we have 
$$
\langle{\pmb u}g,{\pmb u}g\rangle_*{=}
({u},{v})
\begin{bmatrix}
a^2{+}b^2& ac{+}bd\\
ac{+}bd& c^2{+}d^2
\end{bmatrix}
\begin{bmatrix}
{u}\\{v}
\end{bmatrix}{=}
(a^2{+}b^2){u}^2{+}(c^2{+}d^2){v}^2{+}2(ac{+}bd){u}{\ctt}{v}.
$$

The space 
\begin{equation}\label{XYZ}
\begin{bmatrix}
a^2{+}b^2& ac{+}bd\\
ac{+}bd& c^2{+}d^2
\end{bmatrix}{=}
\begin{bmatrix}
X& Z\\
Z& Y
\end{bmatrix}
\end{equation}
is the space $D_{-1}{=}\{XY{-}Z^2{=}1\}$ in ${\mathbb C}^3$ (The space
of all quadratic forms of discriminant $-1$). Replacing 
$X{=}x{+}y$, $Y{=}x{-}y$, $Z{=}z$, we have $x^2{-}y^2{-}z^2{=}1$. 
$D_{-1}$ may be understood as the complexified 2-sphere $x^2{+}(iy)^2{+}(iz)^2{=}1$. 

To get the real sphere, we set as follows together with $\alpha^2{+}\beta^2{+}\gamma^2{=}1$ 
\begin{equation}\label{x,y,rho}
\begin{bmatrix}
a^2{+}b^2& ac{+}bd\\
ac{+}bd& c^2{+}d^2
\end{bmatrix}{=}
\begin{bmatrix}
\alpha{+}i\beta& i\gamma\\
i\gamma & \alpha{-}i\beta
\end{bmatrix}, \quad \alpha, \beta, \gamma \in {\mathbb R},\quad
\alpha{+}i\beta{=}\sqrt{1{-}\gamma^2}\,e^{i\theta}, \quad |\gamma|\leq 1.
\end{equation}
We denote this set by $\tilde{S}^2$. This is the space of all
symmetric elements in $SU(2)$. It is easy to see that 
$\tilde{S}^2{=}SU(2)/SO(2)$.
We denote $\mathcal{S}'{=}\{g; \,\,g\,{}^t\!g{\in}
  {\tilde{S}}^2\}$, e.g. 
$$
\begin{bmatrix}
a& b\\
c& d
\end{bmatrix}{=}
\begin{bmatrix}
\sqrt[4]{1{-}\gamma^2}\,e^{i\theta/2}\cosh\xi&\sqrt[4]{1{-}\gamma^2}\,ie^{i\theta/2}\sinh\xi \\
\sqrt[4]{1{-}\gamma^2}\,ie^{-i\theta/2}\sinh\eta&\sqrt[4]{1{-}\gamma^2}\,e^{-i\theta/2}\cosh\eta\\ 
\end{bmatrix}, 
$$
where $\xi,\,\eta,\,\theta {\in} {\mathbb R}$,
If $\gamma\,{=}\pm 1$, we set 
$
\begin{bmatrix}
a& b\\
c& d
\end{bmatrix}{=}
\frac{1}{\sqrt{2}}
\begin{bmatrix}
1& i\\
i& 1
\end{bmatrix}$.

\begin{lem}\label{key2}
The eigenvalue of $T\in SU(2)$ is real if and only if $T{=}\pm I$.  
If $K{=}I$, and ${}^t\!gg\in {\tilde{S}^2}$, then except the case ${}^t\!gg=\pm I$,  
$\sqrt{\det(\cos t I{-}(\sin t){}^t\!gKg)}$ does not
change sign.
\end{lem}

\noindent
{\bf Proof}\,\, The first one is wellknown. Next one is the case 
${\rm{tr}}.({}^t\!gKg))$ is real and 
$2\alpha$ in \eqref{x,y,rho}. As $\alpha^2{+}\beta^2{+}\gamma^2{=}1$, 
${\rm{tr}}.({}^t\!gKg))^2\leq 4$, and the equality takes place only 
by $\alpha{=}1$, $\beta{=}\gamma{=}0$.
\hfill $\Box$

\medskip
Suppose $g\,{}^t\!g$ be a complex matrix   
$g\,{}^t\!g{=}
{\footnotesize{
\begin{bmatrix}
\xi{-}i\eta&-i\zeta\\
-i\zeta&\xi{+}i\eta
\end{bmatrix}}}$, $\xi^2{+}\eta^2{+}\zeta^2{=}1$,  
$\xi, \eta,\zeta\in {\mathbb C}$  
and suppose 
$\begin{bmatrix}
\alpha{+}i\beta&i\gamma\\
i\gamma&\alpha{-}i\beta
\end{bmatrix}$ be in $SU(2)/SO(2)$.
We take $K{=}
R\begin{bmatrix}
\alpha{+}i\beta&i\gamma\\
i\gamma&\alpha{-}i\beta
\end{bmatrix}$ as an expression parameter. 
The multiplier $R{=}re^{i\theta}$ will be
used for singular points to avoid real line.

Then, a direct computation gives 
by setting ${\pmb \xi}{=}(\xi, \eta, \zeta)$,
$\pmb\alpha{=}(\alpha,\beta,\gamma)$ that 
$$
\begin{aligned}
&{\rm{tr}.}
{\footnotesize{
\begin{bmatrix}
\xi{-}i\eta&-i\zeta\\
-i\zeta&\xi{+}i\eta
\end{bmatrix}}}
R
{\footnotesize{
\begin{bmatrix}
\alpha{+}i\beta&i\gamma\\
i\gamma&\alpha{-}i\beta
\end{bmatrix}}}= 2R\langle\pmb\alpha,\pmb\xi\rangle,\\
&\det\left(\cos tI{-}(\sin t)
R{\footnotesize{
\begin{bmatrix}
\xi{-}i\eta&-i\zeta\\
-i\zeta&\xi{+}i\eta
\end{bmatrix}}}
{\footnotesize{
\begin{bmatrix}
\alpha{+}i\beta&i\gamma\\
i\gamma&\alpha{-}i\beta
\end{bmatrix}}}\right){=}
(\cos t{-}R\langle\pmb\alpha,\pmb\xi\rangle\sin t)^2{+}
\sin^2t R^2\langle\pmb\alpha\times\pmb\xi,\pmb\alpha\times\pmb\xi\rangle 
\end{aligned}
$$ 
where $\langle\pmb\alpha{\times}\pmb\xi, \pmb\alpha{\times}\pmb\xi\rangle{=}
(\xi\beta{-}\eta\alpha)^2{+}(\xi\gamma{-}\zeta\alpha)^2{+}(\eta\gamma{-}\zeta\beta)^2.$
Split $\pmb\xi{=}\pmb x{+}i\pmb x'$ by the real and the imaginary
parts. As $\langle{\pmb\xi},{\pmb\xi}\rangle{=}1$ by 
the condition $\det g\,{}^t\!g=1$, we see 
$\langle\pmb x, \pmb x'\rangle{=}0$, $\|\pmb x\|^2{-}\|\pmb x'\|^2{=}1$ and
$$
\langle\pmb\alpha,\pmb\xi\rangle{=}\langle\pmb\alpha,\pmb x\rangle{+}i\langle\pmb\alpha,\pmb x'\rangle,
\quad 
\pmb\alpha{\times}\pmb\xi{=}\pmb\alpha{\times}\pmb x+i\pmb\alpha{\times}\pmb x'.
$$
By setting $t{=}\frac{\pi}{2}$, the determinant gives 
$R^2\langle\pmb\alpha,\pmb\xi\rangle^2{+}R^2\langle\pmb\alpha\times\pmb\xi,\pmb\alpha\times\pmb\xi\rangle
{=}R^2$.

It follows 
$$
\langle\pmb\alpha,\pmb x\rangle\langle\pmb\alpha,\pmb x'\rangle{+}
\langle\pmb\alpha{\times}\pmb x, \pmb\alpha{\times}\pmb
x'\rangle=0,\quad
\langle\pmb\alpha,\pmb x\rangle^2{-} \langle\pmb\alpha,\pmb x'\rangle^2
+\langle\pmb\alpha{\times}\pmb x, \pmb\alpha{\times}\pmb x\rangle-
\langle\pmb\alpha{\times}\pmb x',\pmb\alpha{\times}\pmb
x'\rangle{=}1
$$
Hence if $\pmb x'{=}0$, all terms are nonnegative and  
$(\cos t{-}\langle\pmb\alpha,\pmb x\rangle\sin t)^2{+}
\sin^2t\langle\pmb\alpha\times\pmb x, \pmb\alpha\times\pmb x\rangle$
has a double multiple root only at  
$$
\cos t{-}\sin t\langle\pmb\alpha,\pmb x\rangle{=}0,\quad 
{\pmb x}=\pm\pmb\alpha, \quad t{=}\frac{\pi}{4},\,\,\frac{3\pi}{4}.
$$
But these are in fact the same point as 
$\cos t{-}\sin t\langle\pmb\alpha,\pmb x\rangle$ is $\pi$-periodic. 
As this does not depend on $R$, it is impossible to move the singular
point away from the real line by selecting suitable $R$.

\medskip 

If $\pmb x{\not=}\pm\pmb\alpha$ and $\pmb x'{=}0$, then 
$$
\sqrt{(\cos t{-}\langle\pmb\alpha,\pmb x\rangle\sin t)^2{+}
\sin^2t(\pmb\alpha\times\pmb x)^2}
$$
does not change sign when $t$ moves from $0$ to $\pi$. (Note this is $\pi$-periodic.)

\bigskip
If $\pmb x'{\not=}0$, then choosing $\pmb\alpha{=}\pmb x$, we see 
$\langle\pmb\alpha,\pmb x'\rangle{=}0$ and 
$$
\langle\pmb\alpha,\pmb x\rangle^2{=}1{+}\|\pmb\alpha\times\pmb x'\|^2.
$$
It follows 
$(\cos t{-}\langle\pmb\alpha,\pmb x\rangle\sin t)^2{+}
\sin^2t(\pmb\alpha\times\pmb x)^2=0$
has two different simple real roots.

Thus, taking  
$
K{=}
r{\footnotesize{
\begin{bmatrix}
\alpha{+}i\beta&i\gamma\\
i\gamma&\alpha{-}i\beta
\end{bmatrix}}},\quad r>1,$ or $r<1$,  
as an expression parameter, we see there is 
${\footnotesize{
\begin{bmatrix}
\alpha{+}i\beta&i\gamma\\
i\gamma&\alpha{-}i\beta
\end{bmatrix}}} \in SU(2)$ such that 
$$
\sqrt{\det\left(\cos tI{-}(\sin t)
{\footnotesize{
\begin{bmatrix}
\xi{-}i\eta&-i\zeta\\
-i\zeta&\xi{+}i\eta
\end{bmatrix}}}
r{\footnotesize{
\begin{bmatrix}
\alpha{+}i\beta&i\gamma\\
i\gamma&\alpha{-}i\beta
\end{bmatrix}}}\right)}
$$
changes sign when $t$ moves from $0$ to $\pi$. That is, this is
alternating $\pi$-periodic. Furthermore,  
taking a suitable $e^{i\theta}$ one can make $\mu, \nu$ pure imaginary
numbers. Hence, 
$$
\sqrt{\det\left(\cos tI{-}(\sin t)
{\footnotesize{
\begin{bmatrix}
\xi{-}i\eta&-i\zeta\\
-i\zeta&\xi{+}i\eta
\end{bmatrix}}}
e^{i\theta}
{\footnotesize{
\begin{bmatrix}
\alpha{+}i\beta&i\gamma\\
i\gamma&\alpha{-}i\beta
\end{bmatrix}}}\right)}
$$
does not change sign 
when $t$ moves from $0$ to $\pi$. That is, this is $\pi$-periodic.

\noindent
{\bf Remark 1}.\,\,By switching $g\,{}^t\!g$ and $K$, we regard 
$\begin{bmatrix}
\xi{-}i\eta& -i\zeta\\
-i\zeta& \xi{+}i\eta
\end{bmatrix}$ as an expression parameter and we regard 
$\begin{bmatrix}
\alpha{+}i\beta&i\gamma\\
i\gamma&\alpha{-}i\beta
\end{bmatrix}$ as $g\,{}^t\!g$. 
Fix 
$\begin{bmatrix}
\xi{-}i\eta& -i\zeta\\
-i\zeta& \xi{+}i\eta
\end{bmatrix}$ so that $\pmb x'\not=0$. Then 
$\langle\pmb x,\pmb x\rangle{-}\langle\pmb x',\pmb x'\rangle{=}1$ 
gives $\langle\pmb x,\pmb x\rangle>1$ and there is $\alpha$ such 
that $\langle\pmb\alpha,\pmb\xi\rangle{=}\langle\pmb\alpha,\pmb x\rangle{>}1$.  
This implies if we multiply $r{\not=}0$ to the expression parameter 
$K{=}r
\begin{bmatrix}
\xi{-}i\eta& -i\zeta\\
-i\zeta& \xi{+}i\eta
\end{bmatrix}$, then there is $g\in \mathcal S'$ such that
$\sqrt{\det(\cos t I{-}(\sin t){}^t\!gKg)}$ changes sign. Note that
expression parameters are not necessarily $\det K{=}1$.

\bigskip
Now suppose $g\,{}^t\!g$ satisfies 
$\langle\pmb x',\pmb\alpha\rangle=0$. This is the case where 
${\rm{tr}.}({}^t\!gKg)$ is a real number, but 
in what follows we restrict $g\in\mathcal S'$ i.e. $g{}^t\!g\in {\tilde S}^2$,  
and investigate where eigenvalues of ${}^t\!gKg$ are real numbers. 
In such a restricted case, we see that $\pmb\xi{=}\pmb x$ and 
$\langle\pmb x,\pmb x\rangle{=}1$. Thus, 
$$
{\rm{tr}.}(
\begin{bmatrix}
\xi{-}i\eta& -i\zeta\\
-i\zeta& \xi{+}i\eta
\end{bmatrix}
K){=}2\langle\pmb\xi,\pmb\alpha\rangle{\leq}2
$$
The equality holds only if $\pmb\xi{=}\pm\pmb\alpha$ and 
$g\,{}^t\!gK{=}\pm I$.

Thus we have 
\begin{thm}
For an arbitrarily fixed $K$ in ${\tilde S}^2$, 
$\sqrt{\det(\cos t I{-}(\sin t){}^t\!gKg)}$ does not
change sign for any $g{\in}\mathcal S'$ except the case 
$g{}^t\!gK{=}\pm I$.
\end{thm}

This theorem shows that excluding only one point $g_{_{K}}$ 
in $\mathcal S'$ such that 
$g{}^t\!g{=}\pm K^{-1}$, the $*$-exponential function 
${:}e_*^{t\frac{1}{2i\h}\langle{\pmb u}g, {\pmb u}g\rangle}{:}_{_K}$ 
is $2\pi$-periodic and has no singular point on ${\mathbb R}$. 

At a first glance it looks we have two exceptional points. 
In fact these are the same point, because   
the same singular point is expressed by opposite parameter $-t$  
on the interval $(0,2\pi)$. 

\bigskip
\noindent
{\bf Note}\,\,Proposition 3.2 in the previous note \cite{ommy9} is not
correct where the singular point $g_{_{K}}$ is not cared. 
However we can apply the result in \cite{OMMY4}. 

\bigskip
Thus, we have 
\begin{prop}\label{111}
For every $g{\in}\mathcal S'$, $g{\not=}g_{_{K}}$  
${:}e_*^{t\frac{1}{2i\h}\langle{\pmb u}g,{\pmb u}g\rangle_*}{:}_{_{K}}$ is 
$2\pi$-periodic, and the singular points distributed
$2\pi$-periodically on two lines sitting in both upper and lower half planes. 
The real line is between these. At the point $g{=}g_{_{K}}$, 
${:}e_*^{t\frac{1}{2i\h}\langle{\pmb u}g,{\pmb u}g\rangle_*}{:}_{_{K}}$
has a not branching singular point at $t{=}\pi/2$, but alternating $2\pi$-periodic.  
\end{prop}

\begin{definition}
An expression parameter $K$ is called a {\bf nice expression parameter} 
if all $g\in \mathcal S'$ except only one 
${:}e_*^{t\frac{1}{2i\h}\langle{\pmb u}g,{\pmb u}g\rangle_*}{:}_{_{K}}$ is 
$2\pi$-periodic, and the two lines of singular 
points are sitting in both upper and lower half planes.
\end{definition}

Proposition\,\ref{111} shows that any $K\in \tilde{S}^2$ is a nice
expression parameter, but  the above Remark 1 shows that any other 
expression parameters are not nice expression parameters. 
Note that $\tilde{S}^2$ is only 2-dimensional, while complex symmetric 
matrices of determinant 1 is 4-dimensional. This is the reason why 
we said $SU(2)$-vacuum is easy to be broken.

\bigskip
Recall now $\tilde{S}^2{=}SU(2)/SO(2)$. Hence 
$SU(2)$ may be viewed as the collection of one parameter subgroups 
$S_{\pmb\alpha}$ parameterized by $\pmb\alpha\in\tilde{S}^2$. 
(This is not the Hopf fibration by itself.)

For any expression parameter $K$ chosen in $\tilde{S}^2$, the set 
\begin{equation}\label{setset}
\{{:}e_*^{t\frac{1}{2i\h}\langle{\pmb u}g, {\pmb u}g\rangle}{:}_{_K};
\,\, t\in [0,\pi],\,\, g{\in}\mathcal S'\}
\end{equation}
covers the group $SU(2)$ except only one $1$-parameter subgroup  
$S_{g_{_K}}{=}
{:}e_*^{t\frac{1}{2i\h}\langle{\pmb u}g_{_{_K}},{\pmb u}g_{_{K}}\rangle}{:}_{_K};
 t{\in}[0,2\pi]$.
Denote by $\infty$ the singular point 
$$
{:}e_*^{\frac{\pi}{4i\h}
\langle{\pmb u}g_{_K},{\pmb u}g_{_K}\rangle}{:}_{_K},\,\,
g_{_K}{}^t\!g_{_K}{\in}{\tilde{S}^2}.
$$ 
We identify ${\tilde{S}^2}{\setminus}\{\infty\}$ with the complex
plane $\mathbb C$ by stereographic projection. 
$SU(2){\setminus}S_{g_{_K}}$ is identified 
$S^1{\times}{\mathbb C}$, which is naturally embedded in 
${\mathbb C}^2$. 
By these observation, we see  
\begin{prop}\label{holoholo}
In any $K$-ordered expression such that $K{\in}\tilde{S}^2$, 
${:}e_*^{t\frac{1}{i\h}\langle{\pmb u}g, {\pmb u}g\rangle}{:}_{_K}$
is holomorphic w.r.t. $g$ in a suitable complex domain involving 
$g\,{}^t\!g{\in}\tilde{S}^2{\setminus}\{\infty\}$.
\end{prop}
 
Note that $\gamma{=}(\cos t, \sin t (x, y, z))$ corresponds to the
element 
$$
\gamma_*{=}
e_*^{it(\frac{1}{\h}x(\tilde{u}^2{+}\tilde{v}^2)
{+}iy\frac{1}{i\h}(\tilde{u}^2{-}\tilde{v}^2){+}2i\rho\frac{1}{i\h}\tilde{u}{\ctt}\tilde{v})}
\,\, {\in} \,\,SU(2)
$$

On the other hand, by viewing ${\tilde{S}^2}$ as the Riemann sphere
the group $S\!L(2,{\mathbb C})/\{\pm 1\}$ acts
transitively on ${\tilde{S}^2}$ as M{\"o}bius transformations. 
This fact supports also  Proposition \ref{holoholo}.  

\subsubsection{Quaternion group} 

Fix arbitrarily $K\in \tilde{S}^2$ such that 
$$
{:}e_*^{\frac{ti}{i\h}{u}{\ctt}{v}}{:}_{_K}, \quad 
{:}e_*^{\frac{t}{2i\h}({u}^2+{v}^2)}{:}_{_K}, \quad 
{:}e_*^{\frac{ti}{2i\h}({u}^2-{v}^2)}{:}_{_K}
$$
are welldefined. Then, at $t{=}\pi$ these give the same element ${\e}_{00}$,
called  the polar element. We have seen in \cite{OMMY5} that square roots of 
polar element ${\e}_{00}$:
$$
{e}_1{=}ie_*^{\frac{\pi i}{2i\h}{u}{\ctt}{v}},\,\, 
{e}_2{=}ie_*^{\frac{\pi}{4i\h}({u}^2+{v}^2)},\,\,
{e}_3{=}ie_*^{\frac{\pi i}{4i\h}({u}^2-{v}^2)}
$$
form something like a double cover of the 
quaternion group. That is, in a nice expression parameter $K$, it holds
$$
{e}_i{*}{e}_j{=}{\e}_{00}{*}{e}_j{*}{e}_i.
$$
Note that $\frac{1}{2}(1{\pm}{\e}_{00})$ are idempotent elements and 
$$
1{=}\frac{1}{2}(1{-}{\e}_{00}){+}\frac{1}{2}(1{+}{\e}_{00}),\quad 
\frac{1}{2}(1{-}{\e}_{00}){*}\frac{1}{2}(1{+}{\e}_{00}){=}0.
$$
Hence using the projection $\pi{=}\frac{1}{2}(1{-}{\e}_{00}){*}$,
where $\pi({\e}_{00})$ is treated as $-1$, we have the equation same
to \eqref{bestans} 
$$
\partial_t\pi\phi(t,\pmb x){=}
\frac{1}{2}[\pi{e}_1\partial_{x_1}+\pi{e}_2\partial_{x_2}{+}\pi{e}_3\partial_{x_3},
\pi\phi(t, \pmb x)].
$$
On the other hand, quadratic forms on the phase part 
$$
{le}_1{=}\frac{1}{i\h}{u}{\ctt}{v},\,\,   
{le}_2{=}\frac{1}{2\h}({u}^2{+}{v}^2),\,\,
{le}_3{=}\frac{1}{2i\h}({u}^2{-}{v}^2)     
$$
satisfies in any ordered expression the same commutation 
relations as 
$\pi(e_1), \pi(e_2), \pi(e_3)$. That is 
$$
[{le}_1,{le}_2]_*{=}2i{le}_3, \quad
[{le}_2,{le}_3]_*{=}2i{le}_1, \quad 
[{le}_3,{le}_1]_*{=}2i{le}_2.
$$

Thus, the equation  
$i\partial_t{\phi}_t{=}
\frac{1}{2\h}\Big[(le_1)\partial_{\xi_1}{+}
(le_2)\partial_{\xi_2}{+}(le_3)\partial_{\xi_3},\,\,
{\phi}_t\Big]$
express the conceptional 3-dimensional rotations mentioned in the  
introduction. 
Now under the thought of the ``second quantization'' 
we have to treat the equation behind this: 
$$
i\partial_t{\phi}_t{=}
\frac{1}{\h}\Big((le_1)\partial_{\xi_1}{+}
(le_2)\partial_{\xi_2}{+}(le_3)\partial_{\xi_3}\Big){*}{\phi}_t.
$$
This may be viewed as the counter part of Weyl equation. Taking the
Fourier transform, this is rewritten as 
\begin{equation}\label{counterWW}
\partial_t{\phi}_t{=}
\frac{1}{i\h}\Big((le_1)\alpha{+}
(le_2)\beta{+}(le_3)\gamma\Big){*}{\phi}_t.
\end{equation}
In the next section it will be shown this is covariant under Lorentz transformations.

\bigskip
On the other hand, although  its square does not become 
Klein-Gordon equation without taking 
$2{\times}2$-matrix representation, we regard    
\begin{equation}\label{Kconter}
\partial^2_t{\phi}_t{=}
\frac{1}{\h^2}\Big(\big((le_1)p_1{+}
(le_2)p_2{+}(le_3)p_3\big)^2{+}m^2\Big){*}{\phi_t}
\end{equation}
as the counter part of Klein-Gordon equation, where 
$p_1, p_2, p_3$ are regarded as parameters. 
In the later section, it will be shown that one can take its square
root as a differential equation of infinite components of order one 
without using Clifford algebra.

\section{Action of Lorentz group to the space of quadratic forms}\label{actLorenz}

By taking $X,Y$ in \eqref{XYZ} independent pure imaginary numbers, 
$$
\begin{bmatrix}
i\beta& \rho{+}i\gamma\\
{-}\rho{+}i\gamma & i\beta'
\end{bmatrix}, \quad \beta, \beta', \gamma, \rho\in {\mathbb R}
$$
is the space ${\mathfrak{sh}}(2)$ of all skew-hermitian
matrices. Note that 
$$
i{\mathfrak{h}}(2){=}{\mathfrak{sh}}(2),\quad 
{\mathfrak{sl}}(2,{\mathbb C}){=}{\mathfrak{h}}_0(2){+}{\mathfrak{su}}(2)
$$
where ${\mathfrak{h}}_0(2)$ is the space of all traceless hermitian
matrices.  

\bigskip
For $x_0, x_1, x_2, x_3\in {\mathbb R}$, the space of all hermite matrices
$$
{\mathfrak{h}}(2){=}
\left\{{\footnotesize{
\begin{bmatrix}
x_0{+}x_3& x_1{+}ix_2\\
x_1{-}ix_2& x_0{-}x_3
\end{bmatrix}; 
x_i{\in}{\mathbb R}
}}
\right\}\quad {\text{with}}\,\,,\det{X}{=}x_0^2{-}(x_1^2{+}x_2^2{+}x_3^2)
$$
is the Minkowski space with  Minkowski-metrix $\det X$, on which 
$S\!L(2,{\mathbb C})$ acts as 
$X\to AXA^*$, $A{\in}S\!L(2,{\mathbb C})$. Strictly speaking, the 
Minkowski space is the 4-dimensional affine space with metric tensor 
of $(1,3)$-type. The expression above is one of coordinate expressions.

Note that 
Lorentz group $SO(1,3)$ is isomorphic to $S\!L(2,{\mathbb C})/\{\pm 1\}$. 
and also that ${\mathfrak{sl}}(2,{\mathbb C})J{=}Q(2)$ space of quadratic forms.
Hence we see 
$$
{\mathfrak{su}}_1(2)J{=}SU(2)/SO(2){=}S^2, \quad
{\mathfrak{su}}(2)J{=}{\mathbb R}S^2{=}{\mathbb R}^3,
$$
where ${\mathfrak{su}}_1(2)$ is the space of all skew-hermitian
matrices with determinant $1$. 
Although $iI$ is not traceless, consider the space
$$
{\mathfrak{sh}}(2)J{=}({\mathbb R}iI\oplus{\mathfrak{su}}(2))J 
$$
and we view this as 
$$
(u,v)({\mathbb R}iI\oplus{\mathfrak{su}}(2))J
\begin{bmatrix}
u\\v
\end{bmatrix} 
$$
where $(u,v)iJ
\begin{bmatrix}
u\\v
\end{bmatrix}
$
is $i(-u{*}v{+}v{*}u){=}-\h$. By this way the term of non-vanishing trace may be
changed into a term $\rho I$, and hence 
Lorentz group acts on 
the space $({\mathbb R}iI\oplus{\mathfrak{su}}(2))J$ by 
{\color{red}$X{\to}AXJ\,\bar{A}^{-1}{=}AXA^{*}J$}.

\bigskip
Now, note that Lorentz group $SO(1,3)$ is generated by $SU(2)$ and
$diag(\lambda, \lambda^{-1}), \lambda{\not=}0$. 
Note that 
$$
\begin{bmatrix}
\lambda&0\\
0&\lambda^-1
\end{bmatrix}
\begin{bmatrix}
a&b\\
b&c
\end{bmatrix}
\begin{bmatrix}
\lambda^{-1}&0\\
0&\lambda
\end{bmatrix}
{=}
\begin{bmatrix}
a&\lambda^2b\\
\lambda^{-2}b&c
\end{bmatrix}.
$$
Using these, one can write down the Lorentz action concretely. 

\bigskip
Note first that 
$i{\mathfrak{h}}(2){=}{\mathbb R}iI\oplus{\mathfrak{su}}(2)$. We have 
$$
A({\mathbb R}iI\oplus{\mathfrak{su}}(2))J{\bar{A}}^{-1}{=}
{\color{red}{\mathbb R}iAA^*J\oplus A{\mathfrak{su}}(2)A^*J}.
$$
If $A\in SU(2)$ then 
$A({\mathbb R}iI\oplus{\mathfrak{su}}(2))J{\bar{A}}^{-1}{=}{\mathbb R}iI\oplus{\mathfrak{su}}(2))J$.

If $A{=}diag(\lambda,\lambda^{-1})$, then decomposing 
$$
iAA^*J{=}{\color{red}\frac{1}{2}}{tr.}{AA^*}iJ{+}(iAA^*J{-}{\color{red}\frac{1}{2}}{tr.}{AA^*}iJ), \quad 
$$
$$
A{\mathfrak{su}}(2))A^*{=}
{\color{red}\frac{1}{2}}tr.(A{\mathfrak{su}}(2))A^*)J{+}
A{\mathfrak{su}}(2))A^*{-}{\color{red}\frac{1}{2}}tr.(A{\mathfrak{su}}(2))A^*)
$$
and involving $(iAA^*J{-}{\color{red}\frac{1}{2}}{tr.}{AA^*}iJ)$ to the traceless term 
and involving 
$diag(A{\mathfrak{su}}(2))A^*){-}{\color{red}\frac{1}{2}}tr.(A{\mathfrak{su}}(2))A^*$ term  
to the constant term by using the relation $u{*}v{-}v{*}u{=}-i\h$, 
these are rewritten in the same form. 
As Lorentz group is generated by 
$SU(2)$, $diag(a,a^{-1})$, $a>0$, this gives an action of Lorentz group. 

\subsection{Lorentz covariance of the counterpart of Wey's equation }

Now joining a new variable $\rho$ corresponding to the diagonal part,
we set 
$$h(t,\alpha,\beta,\gamma, \rho){=}
e_*^{t\frac{1}{i\h}(\langle{\pmb u}g,{\pmb u}g\rangle_*{+}\rho)}
=e_*^{t\frac{1}{i\h}\langle{\pmb u}g,{\pmb u}g\rangle_*}
e^{t\frac{1}{i\h}\rho}.
$$ 
This satisfies the equation 
\begin{equation}
\begin{aligned}\label{Dirac00}
\partial_th(t,\alpha,\beta,\gamma,\rho)&{=}
\frac{1}{i\h}(\langle{\pmb u}g,{\pmb u}g\rangle{+}\rho)
{*}h(t,\alpha,\beta,\gamma,\rho)\\
&{=}
\frac{1}{\h}(\rho{+}\alpha(le_2){+}\beta(le_3){+}\gamma(le_1)){*}
h(t,\alpha,\beta,\gamma,\rho).
\end{aligned}
\end{equation}
This means that
$\tilde{h}(t,\alpha,\beta,\gamma,\rho)
{=}{\rm{Ad}}(h(t,\alpha,\beta,\gamma,\rho))(h(0,\alpha,\beta,\gamma,\rho))$ is 
a solution of the equation 
$$
\partial_t\tilde{h}(t,\alpha,\beta,\gamma,\rho){=}
[\frac{1}{i\h}(\rho{+}\alpha(le_2){+}\beta(le_3){+}\gamma(le_1)), \tilde{h}(t,\alpha,\beta,\gamma,\rho)]. 
$$

Note that if we regard $t\rho$ as an independent variable, then as  
$\partial_{t\rho}e^{t\frac{1}{i\h}\rho}{=}\frac{1}{i\h}e^{t\frac{1}{i\h}\rho}$, 
the equation \eqref{Dirac00} is written as 
$$
i(\partial_t{-}\partial_{t\rho})h(t,\alpha,\beta,\gamma,\rho){=}
\frac{1}{\h}(\alpha(le_2){+}\beta(le_3){+}\gamma(le_1)){*}
h(t,\alpha,\beta,\gamma, \rho).
$$
In precise, viewing $(t,\rho)$ as coordinate variables we take a singular coordinate change 
$(\xi_0, z){=}(t, t\rho)$. (i.e. $t{=}\xi_0$, $\rho{=}z/\xi_0$.)
Then 
$$
\left\{
\begin{matrix}
\medskip
\partial_{\xi_0}&{=}\,\,\,
\partial_t{-}\frac{\rho}{\xi_0}\partial_{\rho}&=\partial_t{-}\partial_{t\rho}\\
\partial_{z}&\!\!\!{=}\rho\partial_{\rho}&{}
\end{matrix}
\right.
$$
Constant term which appear by an adjoint transformation is 
involved in the (universal) time parameter $\xi_0$. 
Hence we have 
\begin{prop}
The counterpart of the Weyl equation 
\begin{equation}\label{counterWW}
\partial_{\xi_0}{\phi}{=}
\frac{1}{i\h}\Big((le_1)\alpha{+}
(le_2)\beta{+}(le_3)\gamma\Big){*}{\phi}.
\end{equation}
is covariant under $S\!L(2,{\mathbb C})/\{\pm 1\}$. 
\end{prop}

\medskip
\noindent
{\bf Remark 2}\,\,In the later section, we have to use singular coordinate 
transformations such as
$$
\partial_{tb}e^{t\frac{1}{i\h}b(u,v,\pmb\alpha)}
{=}\frac{1}{i\h}e^{t\frac{1}{i\h}b(u,v,\pmb\alpha))}
$$ 
for every fixed $(u,v)\in {\mathbb R}^2$ to obtain the Lorentz covariance of 
the counterpart of the massive Dirac's equation. The problem is that 
this procedure changes the Minkowski structure.

\subsection{Joining  massive terms }

As Weyl equation may be viewed as a 2-component equation, the massive
Dirac equation is written as a 4-component equation. The reason is 
one has to use the identity 
$
{\footnotesize
{\begin{bmatrix}
\mu&D\\
D&-\mu
\end{bmatrix}}}^2{=}
{\footnotesize{
\begin{bmatrix}
D^2{+}\mu^2&0\\
0&D^2{+}\mu^2
\end{bmatrix}}}
$
where $D$ may be an arbitrary operator.   
Hence, if 2-component equations are permitted, then 
one can put a mass term to \eqref{counterWW}.
Namely setting 
$Q{=}\frac{1}{i\h}\Big((le_1)\alpha{+}(le_2)\beta{+}(le_3)\gamma\Big)$, we
have an equation 
\begin{equation}
\partial_{\xi_0}
{\footnotesize{
\begin{bmatrix}
\Phi\\\Psi
\end{bmatrix}}}{=}
{\footnotesize
{\begin{bmatrix}
\mu&Q\\
Q&-\mu
\end{bmatrix}}}
{\footnotesize{
\begin{bmatrix}
\Phi\\\Psi
\end{bmatrix}}},
\end{equation}
which gives 
\begin{equation}\label{2-comp}
\partial_{\xi_0}^2
{\footnotesize{
\begin{bmatrix}
\Phi\\\Psi
\end{bmatrix}}}{=}
{\footnotesize
{\begin{bmatrix}
Q^2{+}\mu^2&0\\
0&Q^2+\mu^2
\end{bmatrix}}}
{\footnotesize{
\begin{bmatrix}
\Phi\\\Psi
\end{bmatrix}}}.
\end{equation}

This may be regarded as the equation corresponding to \eqref{Kconter}. 
However when we treat algebra valued equations, this equation 
is written formally as one component equation (cf. the next section). 

\subsubsection{Making matrix algebra $M(2)$}\label{MakingMat}

Recall the polar element ${\e}_{00}$ is defined by setting ${=}\pi/2$ 
in \eqref{tildeKKK}. As a nice expression parameter is used here 
we see ${:}{\e}_{00}^2{:}_{_{K}}{=}1$. In addition to the fact that 
$\frac{1}{2}(1{\pm}\tilde{\e}_{00})$ are idempotent elements
such that 
$$
1{=}\frac{1}{2}(1{+}{\e}_{00}){+}\frac{1}{2}(1{-}{\e}_{00}),\quad 
\frac{1}{2}(1{+}{\e}_{00}){*}\frac{1}{2}(1{-}{\e}_{00})=0
$$  
the bumping identity used often in previous notes (cf. \cite{OMMY4}
for instance) gives 
\begin{prop}
In generic ordered expression, 
${\e}_{00}$ anti-commutes with generators $u, v$.
Hence ${\e}_{00}$ commutes with every even element.
\end{prop}

In what follows, we use half-inverses (cf. \cite{ommy6}) defined as follows: 
$$
{u}^{\btt}{=}{v}{*}({u}{*}{v})_{*-}^{-1},\quad
 ({u}{*}{v})_{*-}^{-1}
{=}-\frac{1}{i\h}\int_0^{\infty}e_*^{s\frac{1}{i\h}{u}{*}{v}}ds
$$
These have the properties  
$$
{u}{*}{u}^{\btt}{=}1,\,\, 
{u}^{\btt}{*}{u}{=}1{-}\overline{\varpi}_{00},\,\, 
\overline{\varpi}_{00}{*}\overline{\varpi}_{00}{=}\overline{\varpi}_{00},\,\,
{u}{*}\overline{\varpi}_{00}{=}0{=}\overline{\varpi}_{00}{*}{u}^{\btt},
$$
where $\overline{\varpi}_{00}=\lim_{s\to\infty}\frac{1}{i\h}e_*^{s\frac{1}{i\h}{v}{*}{u}}$.
We see that 
$({u}^{\btt})^k{*}\overline{\varpi}_{00}{*}{u}^{\ell}$ are  $(k,\ell)$-matrix element.

Moreover as $({u}{*}{v})_{*-}^{-1}$ is an even element, we see 
$$
{\e}_{00}{*}u{=}-u{*}{\e}_{00},\quad 
{\e}_{00}{*}u^{\btt}{=}-u^{\btt}{*}{\e}_{00}. 
$$

\bigskip 
We now define  
\begin{equation}\label{phidagger}
\phi{=}(\frac{1}{2}(1{+}{\e}_{00}){*}u){=}
 (u\frac{1}{2}(1{-}{\e}_{00})), \quad 
\psi{=}(\frac{1}{2}(1{-}{\e}_{00}){*}{u}^{\btt})
{=}u^{\btt}{*}\frac{1}{2}(1{+}{\e}_{00})).
\end{equation}
 
\medskip
Next formulas are easy to see
\begin{equation}\label{phidagger2}
{\phi}^2{=}0{=}{\psi}^2,\quad \phi{*}\psi{+}\psi{*}\phi
{=}1.
\end{equation}

As $\phi, \psi$ commute with every {\bf even} elements, 
one can rewrite \eqref{2-comp} as a one component equation 
$$
i\partial_tf_t{=}
((\phi{*}\psi{-}\psi{*}\phi){*}\mu{+}(\phi{+}\psi){*}
\frac{1}{i\h}\Big((le_1)\alpha{+}(le_2)\beta{+}(le_3)\gamma\Big){*}f_t. 
$$
The solution is obtained by computing 
$$
e_*^{it((\phi{*}\psi{-}\psi{*}\phi){*}\mu{+}(\phi{+}\psi){*}Q)}.
$$
Taking Fourier transform, we have 
$$
\partial_t\hat{f}_t{=}
((\phi{*}\psi{-}\psi{*}\phi){*}\mu{+}(\phi{+}\psi){*}
\frac{1}{i\h}\Big((le_1)\partial_{\xi_1}{+}(le_2)\partial_{\xi_2}{+}(le_3)\partial_{\xi_3}\Big){*}\hat{f}_t.
$$
This may be viewed as the counter part of the Dirac equation.

\subsubsection{Symmetry by Fourier transform}

Note that \eqref{SinhCosh00} may be written as 
$$
{\footnotesize
{\begin{bmatrix}
i\partial_t {-}\mu&0\\
0&i\partial_t{+}\mu
\end{bmatrix}}}
{\footnotesize
{\begin{bmatrix}
\Phi_t(\alpha,\beta,\gamma)\\\Psi_t(\alpha,\beta,\gamma)
\end{bmatrix}}}
{=}
{\footnotesize{
\begin{bmatrix}
0&Q\\
Q&0
\end{bmatrix}}}
{\footnotesize
{\begin{bmatrix}
\Phi_t(\alpha,\beta,\gamma)\\\Psi_t(\alpha,\beta,\gamma)
\end{bmatrix}}}.
$$
Dirac equation is given by taking its Fourier transform
w.r.t. $(\alpha,\beta,\gamma)$. Now, consider the Fourier transform
w.r.t. $(t,\mu)$. Take the standard Fourier transform
$
f(t,\mu){=}\frac{1}{2\pi}\int_{\mathbb R^2}
\hat{f}(m,\tau)e^{-i(mt{+}\mu\tau)}dmd\tau$.
Then we have 
$$
i\partial_tf(t,\mu){=}\int_{\mathbb R^2}
(-m)\hat{f}(m,\tau)e^{-\frac{1}{i\h}(mt{+}\mu\tau)}\dbar m\dbar\tau, \quad 
\mu f(t,\mu){=}\int_{\mathbb R^2}
i\partial_{\tau}\hat{f}(m,\tau)e^{-\frac{1}{i\h}(mt{+}\mu\tau)}\dbar m\dbar\tau.
$$
Hence the above equation becomes 
$$
i\partial_t
{\footnotesize
{\begin{bmatrix}
\hat\phi(\tau, m, \alpha,\beta,\gamma)\\\hat\psi(\tau,m,\alpha,\beta,\gamma)
\end{bmatrix}}}
{=}
{\footnotesize{
\begin{bmatrix}
m&-Q\\
Q&-m
\end{bmatrix}}}
{\footnotesize
{\begin{bmatrix}
\hat\phi(\tau, m, \alpha,\beta,\gamma)\\\hat\psi(\tau,m,\alpha,\beta,\gamma)
\end{bmatrix}}}.
$$
This is acceptable, recalling that the time and the energy are
canonical conjugate variables in general mechanics.

\section{The square root of the counterpart of Klein-Gordon equation }

Recall the counter part of the Klein-Gordon equation  \eqref{Kconter} is written in the
form 
$$
-\partial^2_t\phi_t{=}\frac{1}{\h^2}(Q^2_{\pmb\alpha}(u,v){+}m^2){*}\phi_t,\quad 
Q_{\pmb\alpha}(u,v){=}\langle{\pmb u}g,{\pmb u}g\rangle_*, \quad 
g\,{}^t\!g{=}{\pmb\alpha}\in\tilde{S}^2. 
$$
As $Q_{\pmb\alpha}(u,v)$ is given by $g\,{}^t\!g \in \tilde{S}^2$, this is
viewed as a mapping from $\tilde{S}^2$ into the Weyl algebra $W_2$. 

We want to consider the square root of the equation 
$$
i\partial_t\phi_t{=}\pm\frac{1}{\h}\sqrt[*]{(Q_{\pmb\alpha}(u,v))^2{+}m^2}{*}\phi_t, \quad
\sqrt[*]{Q_{\pmb\alpha}^2{+}m^2}{=}\pm\sqrt[*]{1{+}m^2Q_{\pmb\alpha}^{-2}}\,\,{*}Q_{\pmb\alpha}.
$$
We want to change this into an infinite component differential
equation of order one written in the form 
$$
i\partial_t\Phi_t(u,v){=}M(u,v){*}
(le_1\partial_{\xi_1}{+}le_2\partial_{\xi_2}{+}le_3\partial_{\xi_3}){*}\Phi_t(u,v). 
$$
The point of this trick is that we regard 
${:}\sqrt[*]{1{+}m^2Q_{\pmb\alpha}^{-2}}{:}_{_K}$ as a function of  
$\pmb\alpha$ on the space $\tilde{S}^2$ and we use expansions of 
spherical functions.

\bigskip
To this aim, we recall first the formula \eqref{tildeKKK} in a
nice expression parameter $K$.   
\begin{equation*}
\begin{aligned}
{:}e_*^{\frac{t}{i\h}
\langle{\pmb u}g, {\pmb u}g\rangle_*}{:}_{_{K}}
&{=}
\frac{1}{\sqrt{\det(\cos t I{-}(\sin t){}^t\!gKg)}}
e^{\frac{1}{i\h}\langle{\pmb u}g 
\frac{\sin t}{\cos tI-\sin t \,{}^t\!g Kg},\,\,{\pmb u}g\rangle}
\end{aligned}
\end{equation*}
and note 
$$
g\frac{\sin t}{\cos tI-\sin t \,{}^t\!g Kg}\,{}^t\!g
{=}\frac{1}{\cos t{-}(\sin t)g\,{}^tgK}g\,{}^tg.
$$ 
Recall in a nice expression parameter $K$ 
singular points w.r.t. $t$ appear in very restricted way:  

\noindent
$(1)$: If $t$ is not real, then singular points are double branched
simple singularity distributed $\pi$-periodically. This will be used 
in the next section for the case $t$ is the pure imaginary number.

\noindent
$(2)$: If $t$ is real, such a singular point appears only at
$t{=}\pi/2$. Hence, this case does not appear if $t$ is restricted 
in the pure imaginary number.

\subsection{Computation of $\sqrt[*]{1{+}m^2Q_{\pmb\alpha}^{-2}}$}

First we note the integral 
\begin{equation}\label{invinv}
\int_{-\infty}^0e_*^{it\frac{1}{i\h}{\langle{\pmb u}g,{\pmb u}g\rangle}_*}dt
\end{equation}
in generic $K$-ordered expression converges to give a $*$-inverse 
$(\frac{1}{\h}{\langle{\pmb u}g,{\pmb u}g\rangle})^{-1}_{*+}$. If
there is a singular point on $(-\infty, 0]$, then make a detour by
small half circle to avoid the singularity. As singular points are 
branching singularity in generic ordered expression, the secondary 
residue does not appear (cf.\cite{ommy6}) and the integral 
does not depend on the detour. 

In the case of nice expression parameter, as in the remark mentioned in the
last part of the previous section, singularities appeared in 
$(-\infty, 0]$ are double branched simple singularity. 
Hence the integral \eqref{invinv} is welldefined for every $g\in S'$ 
to give  a smooth function of $\pmb\alpha{=}g\,{}^tg\in\tilde{S}^2$.

By using the exponential law with $e^{tz}$, we see if $|{\rm{Re}}\,z|<1/2$, 
then the integral gives an inverse  
$(z{+}Q_{\pmb\alpha})_{*+}^{-1}{=}\int_{-\infty}^0e_*^{t(z{+}Q_{\pmb\alpha})}dt$
and its derivative 
$\frac{d^k}{dz^k}(z{+}Q(u,v))_{*+}^{-1}$ for every $k$: 
\begin{equation}\label{fine-order}
\frac{d^k}{dz^k}\frac{1}{\langle{\pmb u}g,{\pmb u}g\rangle_*{+}z\rho}
{=}(-1)^k\frac{k!\rho^{k}}{(\langle{\pmb u}g,{\pmb u}g\rangle_*{+}z\rho)^{k{+}1}}
{=}\frac{1}{\h}\int_{-\infty}^0(t^k\rho^{k})
e_*^{it\frac{1}{i\h}(\langle{\pmb u}g,{\pmb u}g\rangle_*{+}z\rho)}dt
\end{equation}
Hence roughly speaking 
$\sqrt[*]{1{+}m^2Q(u,v)_{\pmb\alpha}^{-2}}$ 
 may be viewed as 
\begin{equation}\label{fine-order22}
\sqrt[*]{1{+}m^2Q(u,v)_{\pmb\alpha}^{-2}}{=}
1{+}\frac{1}{2}m^2Q(u,v)_{\pmb\alpha}^{-2}{-}
\frac{1}{4}m^4Q(u,v)_{\pmb\alpha}^{-4}{+}\cdots.
\end{equation}
In precise, we take the Taylor expansion for $|z|<1$,  
$\sqrt{1{+}z}=\sum_{k\geq 0}c_{k}z^{k}$. We see then 
$$
\sqrt[*]{1{+}m^2Q(u,v)_{\pmb\alpha}^{-2}}{=}
1{-}\frac{1}{\h}\int_{-\infty}^0\sum_{k\geq 1}
c_k\frac{m^{2k}t^{2k{-}1}}{(2k{-}1)!}
e_*^{t\frac{1}{\h}(\langle{\pmb u}g,{\pmb u}g\rangle)}dt.
$$
As $\sum_{k\geq 1}c_k\frac{m^{2k}t^{2k{-}1}}{(2k{-}1)!}$
is an entire function, we get
${:}\sqrt[*]{1{+}m^2Q_{\pmb\alpha}^{-2}}{:}_{_K}$ as a smooth function 
of $\pmb\alpha$ and $(u,v)$ where $(u,v)$ is regarded as coordinate
functions of ${\mathbb R}^2$. 
Denote this by $a(\pmb\alpha; u,v)$. 
By this concrete form $a(\pmb\alpha; u,v)$ belongs to the symbol class
$\tilde{\Sigma}^0(\mathbb R^2)$ defined in the next section. 

For every fixed $u,v$ we take the expansion by spherical functions by
setting $\pmb\alpha=(\theta,\phi)$ 
$$
a(\pmb\alpha; u,v){=}
\sum_{n=0}^{\infty}(A_{n,0}(u,v)P_n(\cos\phi){+}
\sum_{m=1}^{n}(A_{n,m}(u,v)\cos m\phi{+}B_{n,m}(u,v)\sin m\phi)P_n^m(\cos\theta). 
$$
Now, setting 
$$
\Phi(\theta,\phi){=}\sum_{n=0}^{\infty}(x_{n,0}(u,v)P_n(\cos\phi){+}
\sum_{m=1}^{n}(x_{n,m}(u,v)\cos m\phi{+}y_{n,m}(u,v)\sin m\phi)P_n^m(\cos\theta),
$$
we have an equation of infinite components
\begin{equation}\label{msDirac}
\partial_t\Phi(\theta,\phi){=}
\frac{1}{i\h}a(\pmb\alpha; u,v){*_{_K}}(le_1\alpha{+}le_2\beta{+}le_3\gamma){*_{_K}}\Phi(\theta,\phi).
\end{equation}
As $le_i$ are quadratic form of $u,v$, the first ${*_{_K}}$-products are
welldefined.  

By Fourier transform this is changed into a differential equation of
first order:
\begin{equation}\label{cpDirac}
\partial_t\hat{\Phi}(\theta,\phi){=}
\frac{1}{\h}a(\pmb\alpha; u,v){{*_{_K}}}
(le_1\partial_{\xi_1}{+}le_2\partial_{\xi_2}{+}le_3\partial_{\xi_3}){{*_{_K}}}\hat{\Phi}(\theta,\phi).
\end{equation} 
As it will be shown in the next section, 
this is an equation written in the algebra 
$(\tilde{\Sigma}^0(\mathbb R^2)_{*,_{_K}}, *_{_K})$. 
This algebra is the non-positive part of the extended Weyl algebra. 
Note that $(\tilde{\Sigma}^0(\mathbb R^2)_{*,_{_K}}, *_{_K})$ plays
the same role as the Clifford algebra in Dirac's equation.  

The fundamental solution of \eqref{msDirac} is given by 
$$
\Phi(\theta,\phi){=}e_{*_{_K}}^{it(a(\pmb\alpha; u,v){*_{_K}}(le_1\alpha{+}le_2\beta{+}le_3\gamma))}.
$$
As $a(\pmb\alpha; u,v){*_{_K}}(le_1\alpha{+}le_2\beta{+}le_3\gamma)$
is an element of 
$le_1\alpha{+}le_2\beta{+}le_3\gamma{+}b(u,v,\alpha),
b(u,v,\alpha)\in \widetilde{\Sigma}_{ev}^{0}(\mathbb R^2)$, this is obtained
as a one parameter subgroup of Fourier integral operators.

\bigskip
\noindent
{\bf Remark 3}\,\, $U(1)$-gauge principle may be applied 
to treat \eqref{cpDirac} under the effect of
electro-magnetic fields. That is to make a $U(1)$-connection 
and replace $\partial_{\xi_i}$ by $\nabla_{\xi_i}$. Another word, 
electro-magnetic fields are written by the terms involving
differentials by the space-time coordinates. Thus, a quantization of 
electro-magnetic fields under the thought of deformation quantization is 
to write differentials by the space-time coordinates as adjoint
operators of Weyl algebras. Such a procedure will be suggested in 
the later section \S\,\ref{univenv}.

\subsubsection{Weyl symbol class $\tilde{\Sigma}^0(\mathbb R^2)$ and 
the product formula}

In this section, $u,v$ are regarded as the coordinate function of
${\mathbb R}^2$ and let $\rho{=}\sqrt{1{+}u^2{+}v^2}$.    
$\tilde{\Sigma}^0(\mathbb R^2)$ is a class of $C^{\infty}$ functions
$f$ on ${\mathbb R}^2$ having asymptotic expansions 
$$
f{\sim} f_0(\theta){+}\rho^{-1}f_{-1}(\theta){+}\cdots{+}\rho^{-k}f_{-k}(\theta){+}\cdots 
$$
where $f_j(\theta)$ is a smooth function on $S^1$.  

Depending on the expression parameter $K$ we will make $*$-functions for
every $f \in \tilde{\Sigma}^0(\mathbb R^2)$ in what follows.

First of all, keeping the Fourier transform of $1$ in mind, we define 
$*$-{\it delta functions of full-variables} 
$\delta_*^{(\mathbb R^2)}({\pmb u}{-}{\pmb x})$ as in \cite{ommy9}  by 
$$
{:}\delta_*^{(\mathbb R^2)}({\pmb u}{-}{\pmb x}){:}_{_K}{=}\!
\int_{\mathbb R^2}
{:}e_*^{\frac{1}{i\h}
\langle\pmb\xi ,{\pmb u}{-}{\pmb x}\rangle}{:}_{_K}\dbar{\pmb\xi}
{=}\int_{\mathbb R^2}
{:}e_*^{\frac{1}{i\h}
 (\xi({u}{-}{x}){+}
{\eta}({v}{-}y))}
{:}_{_K}\dbar{\xi}\dbar{\eta}
$$

We note here that the exponential law gives 
$$ 
e_*^{\frac{1}{i\h}\langle{\pmb\xi},{\pmb u}{-}{\pmb x}\rangle}{=}
e_*^{\frac{1}{i\h}\langle{\pmb\xi},{\pmb u}\rangle}
e^{-\frac{1}{i\h}\langle{\pmb\xi},{\pmb x}\rangle},\quad
{\text{and}}\quad 
\int_{\mathbb R^2} e_*^{\frac{1}{i\h}\langle{\pmb\xi},{\pmb u}{-}{\pmb x}\rangle}
\dbar{\pmb x}
{=}\delta^{(\mathbb R^2)}(\pmb\xi)e_*^{\frac{1}{i\h}\langle{\pmb\xi},{\pmb u}\rangle}.
$$ 
It follows  
\begin{equation}
\int_{\mathbb R^2}\delta^{(\mathbb R^2)}_*({\pmb u}{-}{\pmb x})\dbar{\pmb x}{=}1. 
\end{equation}

\medskip
Let $f({\pmb x})$ be a tempered distribution on
${\mathbb R}^2$ and let 
$\check{f}^{({\mathbb R}^2)}(\pmb\xi){=}\int_{{\mathbb R}^2}f({\pmb x})
e^{-\frac{1}{i\h}\langle{\pmb\xi},\,{\pmb x}\rangle}\dbar{\pmb  x}$
be the inverse Fourier transform. 
Noting the wellknown reciprocity formula 
$f({\pmb x}){=}
\int_{{\mathbb R}^2}\check{f}^{(\mathbb R^2)}({\pmb\xi})
e^{\frac{1}{i\h}\langle{\pmb\xi},\,{\pmb x}\rangle}\dbar{\pmb\xi}$, 
we define $*$-function corresponding to $f(\pmb x)$ as 
\begin{equation}\label{StarFunc}
{:}f_*^{(\mathbb R^2)}({\pmb u}{-}{\pmb x}){:}_{_K}{=}
\int_{\mathbb R^2}\check{f}^{({\mathbb R}^2)}({\pmb\xi})
{:}e_*^{\frac{1}{i\h}\langle{\pmb\xi},\,{\pmb u}{-}{\pmb x}\rangle}{:}_{_K}
\dbar{\pmb x}
{=}
\int_{{\mathbb R}^2}\int_{{\mathbb R}^2}{f}({\pmb x}')
 e^{-\frac{1}{i\h}\langle{\pmb\xi},\,{\pmb x}'\rangle}
{:}e_*^{\frac{1}{i\h}\langle{\pmb\xi},\,{\pmb u}{-}{\pmb x}\rangle}{:}_{_K}
\dbar{\pmb x'}\dbar{\pmb\xi}.
\end{equation}
We denote by $\tilde{\Sigma}^0(\mathbb R^2)_{*_{_K}}$ the totality of 
obtained $*$-functions by this way.

\medskip

The Weyl ordered ($K{=}0$) expression of 
$\delta_*^{({\mathbb R}^2)}(\pmb u{-}\pmb x)$ is given by 
${:}\delta_*^{({\mathbb R}^2)}(\pmb u{-}\pmb x){:}_{0}
{=}\delta^{({\mathbb R}^2)}(\pmb u{-}\pmb x)$. 
Thus we see 
\begin{equation}\label{WeylWeyl}
{:}f^{({\mathbb R}^2)}_*(\pmb u){:}_0
{=}\int f(\pmb x)\delta^{({\mathbb R}^2)}(\pmb u{-}\pmb x)\dbar{\pmb x}{=}f(\pmb u).
\end{equation}
\begin{prop}\label{$*$-funcWeyl}
The inverse of the correspondence $f(\pmb x)\to f^{({\mathbb R}^2)}_*(\pmb u)$ is
given by its Weyl ordered expression and replacement of $\pmb u$ by
$\pmb x$. 
\end{prop}

Hence the $*$-product formula of these $*$-functions is given by the
Moyal product formula given in the integral form 
\eqref{Moualinteg}. Namely
$$
{:}f^{({\mathbb R}^2)}_*(\pmb u){*}g^{({\mathbb R}^2)}_*(\pmb u){:}_{_K}{=}
(f{*_0}g)_*^{({\mathbb R}^2)}.
$$
Hence we have 
\begin{prop}
$(\tilde{\Sigma}^0(\mathbb R^2)_{*,_{K}}; *_{_K})$ becomes an associative algebra.
\end{prop}

Now, by using  the formula of Laplace transform, we define   
$$
\rho_*^{-1}{=}\frac{1}{\sqrt{\pi}}\!
\int_0^{\infty}\frac{1}{\sqrt{t}}e_*^{-t\frac{1}{i\h}(u^2{+}v^2{+}1)}dt, 
\quad 
\rho_*{=}(u^2{+}v^2{+}1){*}
\frac{1}{\sqrt{\pi}}\!\int_0^{\infty}\frac{1}{\sqrt{t}}e_*^{-t\frac{1}{i\h}(u^2{+}v^2{+}1)}dt.
$$
In fact, $(\tilde{\Sigma}^0(\mathbb R^2)_{*,_{K}}; *_{_K})$ is a
$\rho_*^{-1}$-regulated algebra defined in \cite{ommy8}. Roughly speaking 
this is filtered by $\rho_*^{-1}$ satisfying 
$$
[\rho_*^{-1},\tilde{\Sigma}^0(\mathbb R^2)_{*_{_K}}]
\subset \rho_*^{-1}{*}\tilde{\Sigma}^0(\mathbb R^2)_{*_{_K}}{*}\rho_*^{-1}
$$
and 
$$
[\tilde{\Sigma}^0(\mathbb R^2)_{*_{_K}}, \tilde{\Sigma}^0(\mathbb R^2)_{*_{_K}}]_*
\subset \rho_*^{-1}{*}\tilde{\Sigma}^0(\mathbb R^2)_{*_{_K}},\quad
[\rho_*, \tilde{\Sigma}^0(\mathbb R^2)_{*_{_K}}]\subset 
\tilde{\Sigma}^0(\mathbb R^2)_{*_{_K}}.
$$
The factor space is given by 
$$
\tilde{\Sigma}^0(\mathbb R^2)_{*_{_K}}/\rho_*^{-1}{*}\tilde{\Sigma}^0(\mathbb R^2)_{*_{_K}}
\cong C^{\infty}(S^1).
$$

As for Clifford algebras it is not necessary  to consider transcendental
extension because these are finite dimensional. On the contrary, Weyl
algebras are infinite dimensional and elements are represented by 
``unbounded operators''. $\tilde{\Sigma}^0(\mathbb R^2)_{*_{_K}}$
is  the part of ``bounded operators'' in the extended Weyl algebra.  
This $\rho_*^{-1}$-regulated algebra will be used also in the next section.

\bigskip
In fact, as ${\rm{ad}}(\rho_*^2)$ leaves the space of quadratic forms 
invariant, it is sometimes convenient to restrict our system to 
a smaller class by using only ``even elements''. 
Now, let 
$\tilde{\Sigma}_{ev}^0(\mathbb R^2)$ is the class of $C^{\infty}$ functions
$f$ on ${\mathbb R}^2$ having asymptotic expansions 
$$
f{\sim} f_0(\theta){+}\rho^{-2}f_{-2}(\theta){+}\cdots{+}\rho^{-2k}f_{-2k}(\theta){+}\cdots 
$$
where $f_j(\theta)$ is a smooth function on $S^1$. 
As $[\rho_*^{2},\tilde{\Sigma}_{ev}^0(\mathbb R^2)_{*_{_K}}]
\subset\tilde{\Sigma}_{ev}^0(\mathbb R^2)_{*_{_K}}$
$\tilde{\Sigma}_{ev}^0(\mathbb R^2)$ is in fact a
$\rho_*^{-2}$-regulated algebra filtered by
$\rho_*^{-2k}{*}\tilde{\Sigma}_{ev}^0(\mathbb R^2)_{*_{_K}}$ such that 
$$
[\rho_*^{-2k}\tilde{\Sigma}_{ev}^0(\mathbb R^2)_*,
\rho_*^{-2\ell}\tilde{\Sigma}_{ev}^0(\mathbb R^2)_*]_{*_{_K}}
\subset
\rho_*^{-2(k{+}\ell+1)}\tilde{\Sigma}_{ev}^0(\mathbb R^2)_{*_{_K}}. 
$$

As for \eqref{msDirac}, 
$a(\pmb\alpha; u,v){*_{_K}}(le_1\alpha{+}le_2\beta{+}le_3\gamma)$ is
written in the form 
$$
(le_1\alpha{+}le_2\beta{+}le_3\gamma){+}{\rho_*^{-2}\tilde{\Sigma}_{ev}^0(\mathbb R^2)} 
$$
without the term of order $0$. Hence the fundamental
solution is obtained by the $*$-exponential function given in the form
of Fourier integral operators:
Recalling \eqref{fine-order22} we set 
$$
a(\pmb\alpha; u,v){*_{_K}}(le_1\alpha{+}le_2\beta{+}le_3\gamma){=}
\langle{\pmb u}g, {\pmb u}g\rangle{+}b(\pmb\alpha; u,v),\quad 
g\in {\mathcal S}',\quad 
b(\pmb\alpha; u,v)\in {\rho}_*^{-2}{*}{\tilde{\Sigma}}^{0}_{ev}({\mathbb R}^2).
$$
and write 
$$
e_{*_{_K}}^{t\frac{1}{i\h}(\langle{\pmb u}g, {\pmb u}g\rangle{+}b(\pmb\alpha; u,v))}.
=
{:}e_*^{t\frac{1}{i\h}(\langle{\pmb u}g,{\pmb u}g\rangle}{:}_{_K}{*_{_K}}f(t,u,v). 
$$
Then the equation we have to solve is 
$$
\partial_tf(t,u,v){=}
{:}{\rm{Ad}}
(e_{*}^{-t\frac{1}{i\h}\langle{\pmb u}g, {\pmb u}g\rangle})(b(\pmb\alpha; u,v)){:}_{_K}
{*_{_K}}f(t,u,v).
$$
As 
${:}{\rm{Ad}}(e_{*}^{-t\frac{1}{i\h}\langle{\pmb u}g, {\pmb u}g\rangle})(b(\pmb\alpha; u,v))$
is a smooth curve in ${\rho}_*^{-2}{*}{\tilde{\Sigma}}^{0}_{ev}({\mathbb R}^2)$ the
existence of the solution written in the form 
$e_*^{t\frac{1}{i\h}(1{+}c(t,u,v))}$, 
$c(t,u,v)\in {\rho}_*^{-2}{*}{\tilde{\Sigma}}^{0}_{ev}({\mathbb R}^2)$, 
is ensured by the product integrals, although the concrete form is hard to write down.
A precise treatment of these product integrals will be seen in the future note.

\bigskip
If we apply 
${\rm{Ad}}
(e_{*}^{t\frac{1}{i\h}\langle{\pmb u}g',{\pmb u}g'\rangle}))$, $g'{\in}S\!L(2,{\mathbb C})$, 
to both sides of \eqref{msDirac}, then a constant term (the term of
order $0$) appears in the phase. This term will be involved in the
time parameter by using a singular coordinate transformation as in
Remark 2.

\section{$SU(2)$-vacuum}

Note first that the proof of existence of $\Omega_*$ in \cite{ommy9} is not correct,
as the property of the singular point was not cared. First, we give a 
correction. 
\begin{prop}
Under a nice expression parameter $K$, the integral 
$\frac{1}{2\pi}\int_0^{2\pi}{:}e_*^{\frac{t}{2i\h}\langle{\pmb u}g, 
{\pmb u}g\rangle_*}{:}_{_{K}}$ which gives a pseudo-vacuum 
does not depend on $\pmb\alpha{=}g\,{}^t\!g$.
\end{prop}

\noindent
{\bf Proof}\,\, For a fixed $K$, one parameter subgroup 
$e_*^{\frac{t}{2i\h}\langle{\pmb u}g, {\pmb u}g\rangle_*}$
is welldefined except one $g_{_K}$. Non-vanishing 
of this integral is proved in \cite{ommy9}. 

Now, let  $g(s)$, $s\in [0,1]$ be a curve avoiding $g_{_K}$. We see 
$$
e_*^{\frac{0}{2i\h}\langle{\pmb u}g(s), {\pmb u}g(s)\rangle_*}{=}1,
\quad 
e_*^{\frac{\pi}{2i\h}\langle{\pmb u}g(s), {\pmb u}g(s)\rangle_*}{=}{\e}_{00}.
$$
Hence by minding Proposition \ref{holoholo} on the domain $(t,s)\in [0,\pi]{\times}[0,1]$,
Cauchy's integration theorem shows that the integral does not depend
on $s$. As a result, we have no need to care about the singular
point. ${}$\hfill $\Box$

Hence we have the nontrivial existence of the integral  
$$
\Omega_*{=}\int_{(t,g){\in}SU(2)}
{:}e_*^{\frac{t}{2i\h}\langle{\pmb u}g, 
{\pmb u}g\rangle_*}{:}_{_{K}}d\mu
{=}\frac{1}{8\pi^2}\int_{\tilde{S}^2}\int_0^{2\pi}
{:}e_*^{\frac{t}{2i\h}\langle{\pmb u}g, 
{\pmb u}g\rangle_*}dtd\pmb\alpha{:}_{_{K}}
$$
where $d\mu$ is the invariant volume form with total volume $1$. 
Let us call this the $SU(2)$-{\bf vacuum}. 

It is clear that 
$e_*^{t\frac{1}{2i\h}\langle{\pmb u}g,{\pmb u}g\rangle_*}{*}\Omega_*{=}\Omega_*$,\quad 
$({\mathfrak{su}}(2)J){*}\Omega_*{=}\{0\}$.
By viewing as quadratic forms, we see 
$$
({\mathfrak{sl}}(2,{\mathbb C})J){*}\Omega_*{=}{\mathfrak{h}}_0(2)J{*}\Omega_*.
$$
The next one is trivial
\begin{prop} 
${\mathfrak{h}}_0(2){=}i{\mathfrak{su}}(2)$ is a Lie algebra over
$\mathbb R$ under the bracket product $[|X,Y|]{=}i[iX,iY]$. Hence 
${\mathfrak{h}}(2){=}{\mathbb R}\oplus{\mathfrak{h}}_0(2)$ is 
a Lie algebra under this new bracket product. This is viewed as the 
Lie algebra of $U(2)$.
\end{prop}

\bigskip
\noindent
{\bf Remark 4}.\,\, Note that the complementary subspace of 
${\mathfrak{su}}(2)$ is not unique. 
Furthermore, as ${\mathfrak{su}}(2){*}\Omega_*{=}0$ we have to
restrict the coefficients to ${\mathbb R}$. 

\subsection{What are remainded under the vacuum ${*}\Omega_*$}
\label{univenv}

The universal enveloping algebra of $({\mathfrak{h}}(2); [|\,\,,\,\,|])$ is 
an infinite dimensional noncommutative algebra over ${\mathbb R}$ generated by 
Pauli-matrices $\sigma_1, \sigma_2, \sigma_3$ and $I$ with only the 
commutation relations
$$
\sigma_1{*}\sigma_2{-}\sigma_2{*}\sigma_1{=}2\sigma_3,\,\,
\sigma_2{*}\sigma_3{-}\sigma_3{*}\sigma_2{=}2\sigma_1,\,\,
\sigma_3{*}\sigma_1{-}\sigma_1{*}\sigma_3{=}2\sigma_2,\,\,
I{*}\sigma_i{=}\sigma_i{*}I
$$
We denote this by $Env({\mathfrak{h}}(2); [|\,\,,\,\,|])$.
Note that Pauli-matrices are the bases of $2{\times}2$-hermite
matrices
$$
I{=}
\begin{bmatrix}
1&0\\
0&1
\end{bmatrix},\quad 
\sigma_1{=}
\begin{bmatrix}
0&1\\
1&0
\end{bmatrix},\quad
\sigma_2{=}
\begin{bmatrix}
0&-i\\
i&0
\end{bmatrix},\quad
\sigma_3{=}
\begin{bmatrix}
1&0\\
0&-1
\end{bmatrix}.
$$
However, we do not have  $I^2{=}I$, $\sigma_i^2{=}I$
e.t.c. Furthermore, we do not set ${-}\sigma_k{=}(i\sigma_k)^2$ e.t.c..
We have only the commutation relations.  

Note also that 
$$
({\mathfrak{sl}}(2,{\mathbb C}); [\,\,,\,\,]_{mat})\cong (Q(u,v);
[\,\,,\,\,]_*)\quad Q(u,v){=}\{\text{quadratic forms}\}
$$
Hence a linear base of ${\mathfrak{h}}_0(2)J$ is  
$$
{he}_1{=}\frac{1}{i\h}{u}{\ctt}{v},\,\,   
{he}_2{=}\frac{1}{2\h}({u}^2{+}{v}^2),\,\,
{he}_3{=}\frac{1}{2i\h}({u}^2{-}{v}^2).   
$$
They satisfy in any expression the commutation relations  
$$
[|{he}_1,{he}_2|]_*{=}2{he}_3, \quad
[|{he}_2,{he}_3|]_*{=}2{he}_1, \quad 
[|{he}_3,{he}_1|]_*{=}2{he}_2.
$$
The Casimir element in the enveloping algebra vanishes: 
${\zeta}{=}{he}_1^2{+}he_2^2{+}he_3^2{=}0$.

Now for every $g\in S\!L(2,{\mathbb C})$, Lie algebra isomorphism ${\rm{Ad}}(g):
  {\mathfrak{sl}}(2,{\mathbb C})\to {\mathfrak{sl}}(2,{\mathbb C})$
  maps ${\mathfrak{h}}(2)$ to other space ${\mathfrak{h}}'(2)$. 
By the observation in \S\,\ref{actLorenz} one can make a projection  
 ${\mathfrak{h}}'(2)\to{\mathfrak{h}}(2)$.
\begin{prop}\label{AdjointLO}
$S\!L(2,{\mathbb C})/\{\pm 1\}$ acts on  
$({\mathfrak{h}}(2); [|\,\,,\,\,|])$
as Lie algebra isomorphisms. Hence the action extends on 
its universal enveloping algebra. 
\end{prop}

\noindent
{\bf Proof}\,\,It is enough to prove the first one. For every $X,
Y\in{\mathfrak h}_0(2)$, we set for every $g\in S\!L(2,{\mathbb C})$
${\rm{Ad}}(g)X{=}X'{+}c(X)I$, ${\rm{Ad}}(g)Y{=}Y'{+}c(Y)I$.
Then, $[|X,Y|]{\in }{\mathfrak h}_0(2)$ and 
$$
{\rm{Ad}}(g)[|X,Y|]{=}[|{\rm{Ad}}(g)X, {\rm{Ad}}(g)Y|]{=}[|X',Y'|].
$$  
It follows ${\rm{Ad}}(g)$ is a Lie homomorphism. \hfill  ${\Box}$

\noindent
{\bf Remark 5}. It is difficult to join linear terms to the Lie
algebra ${\mathfrak{h}}_0(2)$ so that the enveloping algebra
$Env({\mathfrak g})$ of the real Lie algebra 
$({\mathfrak g},[|\,\,,\,\,|]){=}
\{a{+}be^{i\xi}u{+}ce^{i\eta}v{+}({\mathfrak{h}}_0(2)J); a,b,c \in
{\mathbb R}\}$ over $\mathbb R$ do not crash out by the multiplication 
${*}\Omega$ from the r.h.s.. 

Here we giave a proof under the assumption that the constant term of 
the enveloping algebra forms a field over ${\mathbb R}$. 
If  
$e^{i\xi{+}i\eta}{=}\pm e^{\frac{\pi iq}{p}}$, then 
$[|e^{i\xi}u,e^{i\eta}v|]{=}\mp\h e^{\frac{\pi iq}{p}}$.
If $p{=}2^{m{+}1}$, then $(\mp\h e^{\frac{\pi iq}{p}})^{2^{m}}={\pm i}\h^{2^{m}}$, 
and 
$\pm i{\mathfrak{h}}_0(2){=}\pm {\mathfrak{su}}(2)$. 
In general, under the assumption, the  constant elements $e^{i(\xi{+}\eta)}$
is in the enveloping algebra. Multiplying its inverse to 
$e^{i\xi}u, e^{i\eta}v$, we see that $e^{i(\xi{-}\eta)}u, e^{i(\eta{-}\xi)}v$ 
are in $Env({\mathfrak g})$. Thus, 
$$
[|(e^{i(\xi{-}\eta)}u)^2,
(e^{i(\eta{-}\xi)}v)^2|]{=}-i[u^2,v^2]{=}4\h^2 i.
$$
Hence $Env({\mathfrak g}){\supset}i{\mathfrak{h}}_0(2){=}{\mathfrak{su}}(2)$. 

\subsubsection{Lie subalgebras contained  in a suitably extended 
universal enveloping algebra }

Although ${\mathfrak{su}}(2){*}\Omega_*{=}0$, the complementary
subspace is not unique. What we have to consider is the universal 
enveloping algebra of 
${\mathfrak{sl}}(2,{\mathbb C})/{\mathfrak{su}}(2)$. 
Although ${he}_3{\pm}i{he}_1{\not\in}{\mathfrak{h}}(2)$,
a certain nontrivial element may remain in   
$$
e_*^{\frac{t}{i\h}({he}_3{\pm}i{he}_1}{*}{\Omega_*}.
$$
Now, note that 
$$
2i\h({he}_3{+}i{he}_1){=}(u{+}iv)_*^2,\quad 
2i\h({he}_3{-}i{he}_1){=}(u{-}iv)_*^2. 
$$
In this section, we make 
$\sqrt[*]{2({he}_3{+}i{he}_1)}$, $\sqrt[*]{2({he}_3{-}i{he}_1)}$
in a suitably extended 
$\widetilde{Env}({\mathfrak{h}}(2); [|X,Y|])$ of
$Env({\mathfrak{h}}(2); [|X,Y|])$. 
To this end, we use the formula of Laplace transform 
$$
\frac{\sqrt{\pi}}{\sqrt{p}}{=}\int_0^{\infty}\frac{1}{\sqrt{t}}e^{-pt}dt
$$
First we recall \eqref{eq:expquad} that
\begin{equation*}
{:}e_*^{t\frac{1}{i\h}(u{\pm}iv)_*^2}{:}_{_K}{=}
\frac{1}{\sqrt{1{-}\tau t}}\,
e^{\frac{t}{1{-}\tau t}(u{\pm}iv)^2}, \quad \tau{=}{K}_{11}{-}K_{22}{+}2iK_{12}
\end{equation*}
In a case of nice expression parameter, we see
$\tau{=}2i(\beta+\gamma)$. But the argument below 
can be applied in the case $\tau\not=0$ in general.  Now we see the integral  
$$
\int_0^{\infty}{:}\frac{1}{\sqrt{t}}e_*^{-t\frac{1}{i\h}(u{\pm}iv)_*^2}{:}_{_K}dt
{=}\int_0^{\infty}\frac{1}{\sqrt{t}}
\frac{1}{\sqrt{1{+}\tau t}}\,
e^{-\frac{t}{1{+}\tau t}(u{\pm}iv)^2}dt
$$
converges. Define elements 
$$
{:}L(u{+}iv){:}_{_K}{=}:\frac{1}{i\h}(u{+}iv)_*^2{*}
\frac{1}{\sqrt{\pi}}\!\int_0^{\infty}\frac{1}{\sqrt{t}}e_*^{-t\frac{1}{i\h}(u{+}iv)_*^2}dt{:}_{_K}
$$ 
$$
{:}L(u{-}iv){:}_{_K}{=}:\frac{1}{i\h}(u{-}iv)_*^2{*}
\frac{1}{\sqrt{\pi}}\!\int_0^{\infty}\frac{1}{\sqrt{t}}e_*^{-t\frac{1}{i\h}(u{-}iv)_*^2}dt{:}_{_K}
$$ 
These satisfy 
$$
\big(L(u{+}iv){-}(u{+}iv)\big){*}\big(L(u{+}iv){+}(u{+}iv)\big){=}0{=}
\big(L(u{-}iv){-}(u{-}iv)\big){*}\big(L(u{-}iv){+}(u{-}iv)\big).
$$
These are elements of $\rho_*{*}\tilde{\Sigma}^0(\mathbb R^2)_{*_{_K}}$,
which may be written as 
$$
{:}L(u{+}iv){:}_{_K}{=}u{+}iv{+}\phi_*, \quad
{:}L(u{-}iv){:}_{_K}{=}u{-}iv{+}\psi_*, \quad \phi_*,\,\psi_*\in 
\tilde{\Sigma} ^0(\mathbb R^2)_{*_{_K}}. 
$$
Some care will be required, for $\phi_*$, $\psi_*$ are $0$-divisors,
but by suitable replacements (cf.\cite{om3},\S\,XIII.5), one can reduce the remainder
terms arbitrarily ``small'' $\phi_*,\,\psi_*\in \rho_*^{-N}{*}
\tilde{\Sigma}^0(\mathbb R^2)_{*,_{K}}$, but these do not vanish in
general. This is because $L(u{+}iv)$ commutes with the polar element ${\e}_{00}$ while 
 $u{+}iv$ anti-commutes with ${\e}_{00}$. 
  These remain as terms of ``smoothing operators''. 

Hence, we have 
\begin{prop}\label{envenv}
Suitably extended $\widetilde{Env}({\mathfrak{h}}(2))$ contains for instance a 
Lie algebra over ${\mathbb R}$: 
$$
\tilde{\mathfrak g}{=}\{x_0{+}x_1u{+}x_2iv{+}\sigma\frac{1}{2\h}(u^2{-}v^2)\}{+}
{\text{some elements in }}\,\,
\rho_*^{-N}{*}\tilde{\Sigma}^0(\mathbb R^2)_{*,_{K}}.
$$  
\end{prop} 

\medskip
Note that if we use all quadratic forms $Q(u,v)$, then 
${\rm{Ad}}(e_*^{\frac{1}{i\h}Q(u,v)})$ generates the group
$S\!L(2,{\mathbb C})$. 

By Proposition\,\ref{AdjointLO}, we can apply ${\rm{Ad}}(e_*^{\frac{1}{i\h}Q(u,v)})$
to the above $\widetilde{Env}({\mathfrak{h}}(2))$ and obtain a 
family of Lie algebras over ${\mathbb R}$
$$
\left\{{\rm{Ad}}(e_*^{\frac{1}{i\h}Q(u,v)})\tilde{\mathfrak g}; \,\,
Q(u,v)={\text{quadratic forms}}\right\}
$$
parameterized by  $S\!L(2,{\mathbb C})/\varGamma$ where $\varGamma$ is
the stabilizer of ${\mathbb C}{\otimes}\tilde{\mathfrak g}$.  

{\color{red} Although they are not equal to $({\mathfrak h}(2), [|\,\,,\,\,|])$ as Lie algebras,
their enveloping algebras equal to
$\widetilde{Env}({\mathfrak{h}}(2))$. Thus, one can make the
complexifications of these Lie algebras and 
one may use their enveloping algeba as if it were the regular representation space of $\Omega_*$. 
}

Several comments about the stabilizer will be worthwhile. In a strict sense,
$\varGamma$ is the identity. However what we are concerned is the ``space-time'' 
as the Minkowski space, or a certain class of specific coordinate expressions.
Here, we concern only on Minkowski space with a light-cone frame. 
In the next section, we see that the  Lie algebra ${\mathfrak g}$
gives a light-cone frame of the Minkowski space. Hence, we see 
the following:
\begin{prop}
The light-cone frame of the Minkowski space given by 
the Lie algebra $\tilde{\mathfrak g}$ is determined by the highest term 
$\frac{1}{2\h}(u^2{-}v^2)$. Hence the stabilizer $\varGamma$ 
is $e_*^{z\frac{1}{2\h}(u^2{-}v^2)}$, $z\in {\mathbb C}$ and 
the factor space  $S\!L(2,{\mathbb C})/\varGamma$ is a 4-dimensional space.
\end{prop}

\subsection{Noncommutative Minkowski spaces}

In this section we treat the Lie algebra  
$$
{\mathfrak g}
{=}\left\{\{x_0{+}x_1u{+}x_2iv{+}\sigma\frac{1}{2\h}(u^2{-}v^2)\},\quad
[\,\,,\,\,]_*, \quad {\color{red}x_i, \sigma\in {\mathbb R}}\right\}.
$$ 
This is linearly isomorphic  over ${\mathbb R}$ to the linear space 
$$
{\mathfrak h}(2){=}
\left\{
\frac{1}{2}
\begin{bmatrix}
x_0{+}\sigma& x_1{+}ix_2\\
x_1{-}ix_2& x_0{-}\sigma
\end{bmatrix},\,\,\,[|\,\,\,,\,\,\,|] \right\}
$$
by the natural correspondence. Namely, Proposition\,\ref{envenv} shows
there are a family of $*$-Lie algebras over $\mathbb R$ that give same 
universal enveloping algebra as  
 $({\mathfrak h}(2), [|\,\,,|])$. Note the latter is isomorphic to 
the Lie algebra of $U(2)$. 

\bigskip
In this section, we show the following
\begin{thm}\label{MinkMink}
The Lie group structure with the Lie algebra ${\mathfrak g}$ is 
constructed on the space ${\mathfrak g}$ itself and this group 
has an adjoint invariant Lorentz metric.
\end{thm}
The group structure is given as follows: 
First we take the central extension of ${\mathbb R}^2$ by the
skew-symmetric form $J$ on ${\mathbb R}^2$. We denote this by 
${\mathbb R}_J{\times}{\mathbb R}^2$. The group $\tilde{G}_0$ we want to make 
 is the semi-direct product
$$
({\mathbb R}_J{\times}{\mathbb R}^2){\rtimes}_{_{\rm{Ad}}}{\mathbb R}
$$
by the adjoint action of $e_*^{\sigma\frac{1}{2i\h}(u^2{-}v^2)}$.

\bigskip
To be precise, recall first the $*_{_K}$-product of 
$*_{_K}$-exponential functions is given by  
\begin{equation}\label{prodform22}
{:}e_{*}^{\frac{1}{i\h}\langle{\pmb{\xi}},\pmb{u}\rangle}{*}
e_{*}^{\frac{1}{i\h}\langle{\pmb{\eta}},\pmb{u}\rangle}{:}_{_K}
{=}
e^{\frac{1}{2i\h}\langle{\pmb{\xi}}J,\pmb{\eta}\rangle}
{:}e_{*}^{\frac{1}{i\h}\langle{\pmb{\xi}{+}\pmb{\eta}},\pmb{u}\rangle}{:}_{_K},\quad
\pmb{\xi}, \pmb{\eta}\in {\mathbb C}^3, \quad
\Lambda{=}K{+}J,\quad {\pmb u}{=}(u,v).
\end{equation}
Using this formula consider next the semi-direct product group 
${\mathbb C}_{\times}e_*^{\frac{1}{i\h}\langle{\pmb\xi},{\pmb u}\rangle}
{*}e_*^{t\frac{i}{2i\h}(u^2{-}v^2)}$. To fix the group structure, we
note first that 
$$
{\rm{ad}}(\frac{1}{2i\h}(u^2{-}v^2)
\begin{bmatrix}
u\\v
\end{bmatrix}{=}
{-}
\begin{bmatrix}
0&i\\
i&0
\end{bmatrix}
\begin{bmatrix}
u\\v
\end{bmatrix},\quad 
{\rm{Ad}}(e_*^{\sigma\frac{i}{2i\h}(u^2{-}v^2)})
\begin{bmatrix}
u\\v
\end{bmatrix}{=}
\Theta(\sigma) 
\begin{bmatrix}
u\\v
\end{bmatrix},\quad 
\Theta(\sigma){=}
 \begin{bmatrix}
\cos\sigma&-i\sin\sigma\\
-i\sin\sigma&\cos\sigma
\end{bmatrix}.
$$
The group structure is given by  
$$
\Big(ae_*^{\frac{1}{i\h}\langle{\pmb\xi},{\pmb u}\rangle}
{*}e_*^{\sigma\frac{i}{2i\h}(u^2{-}v^2)}\Big)
{*}
\Big(be_*^{\frac{1}{i\h}\langle{\pmb\eta},{\pmb u}\rangle}
{*}e_*^{\tau\frac{i}{2i\h}(u^2{-}v^2)}\Big)
{=}ab\,\,e_*^{\frac{1}{i\h}\langle{\pmb\xi},{\pmb u}\rangle}{*}
e_*^{\frac{1}{i\h}\langle{\pmb\eta}{\Theta(-\sigma)},{\pmb u}\rangle}
{*}e_*^{(\sigma{+}\tau)\frac{i}{2i\h}(u^2{-}v^2)}.
$$
\eqref{prodform22} is used to compute the first $*$-product of r.h.s..
Denote this group by $G_{J}$.  

\medskip
Using this in particular case, we see  
$$ 
e^{\frac{1}{\h}\tau}
e_*^{\frac{1}{i\h}(x_1 u{+}x_2 iv)}
{*}e_*^{\sigma\frac{i}{i\h}(u^2{-}v^2)}\quad 
\tau, x_1, x_2,\sigma \in{\mathbb R},
$$
forms a group $G_0$. We denote by ${\tilde{G}}_0$ its universal
covering group. It will be shown below that 
$\tilde{G}_0$ is given by defining the group structure on 
the tangent space ${\mathfrak g}_0{=}{\mathbb R}^4$ at the identity

Indeed, 
$e^{\frac{1}{\h}s}
e_*^{\frac{1}{i\h}(x_1u{+}x_2 iv)}$ is a 
central extension of the group 
$e^{\frac{1}{i\h}(x_1u{+}x_2 iv)}$ by the 2-cocycle 
$e^{\frac{1}{2\h}\langle{\pmb x}J, {\pmb y}\rangle}$. The universal covering 
group is the central extension of ${\mathbb R}^2$ by the
skew-symmetric form $J$ on ${\mathbb R}^2$. 

The tangent space at the identity $1$ forms a Lie algebra 
$$
{\mathfrak g}_{0}{=}
\{\frac{1}{\h}\tau{+}
\frac{1}{i\h}(x_1u{+}x_2 iv){+}\sigma\frac{i}{2i\h}(u^2{-}v^2)\}
$$
with bracket products
$$
[\frac{1}{i\h}(x_1u{+}x_2iv),\frac{1}{i\h}(y_1u{+}y_2iv)]
{=}\frac{1}{\h}(x_2y_1{-}x_1y_2),
\quad
[\frac{1}{i\h}(x_1u{+}x_2iv),
 \frac{1}{2\h}(u^2{-}v^2)]{=}
-\frac{1}{i\h}(x_2u{-}x_1iv).
$$

\medskip
We define a Lorentz metric on 
${\mathfrak g}_{0}{=}{\mathbb R}{\oplus}E^2{\oplus}{\mathbb R}$
by the bilinear form $Q$ as follows: 
\begin{equation}\label{minkow}
Q{=}\langle(\dot\tau,\dot{x}_1,\dot{x}_2,\dot\sigma), \,
(\dot\tau',\dot{y}_1,\dot{y}_2,\dot\sigma')\rangle
{=}\frac{1}{2}(\dot\tau\dot\sigma'{+}\dot\sigma\dot\tau')
{-}\langle{\dot{\pmb x}}, {\dot{\pmb y}}\rangle,
\end{equation}
where $\langle{\dot{\pmb x}}, {\dot{\pmb y}}\rangle$ is the 
Euclidean inner product on ${\mathbb R}^2$.

It is not hard to see that the Lorentz metric \eqref{minkow} is adjoint invariant.  
Thus, this extends to an invariant bilinear metric on 
$\tilde{G}_{0}$ by left-translations. 

\section{Pseudo-vacuum representations}\label{Pseudo}

We start with giving some comments about various vacuums. 
For every $K$, the $K$-ordered expression of  
$e_*^{(s{+}it)\frac{1}{2i\h}\langle{\pmb u}g,{\pmb u}g\rangle}$
has a remarkable periodicity property in $t$. 
There is an interval $[a,b]$, called the {\bf exchanging interval}, such that 
$$
{:}e_*^{(s{+}it)\frac{1}{2i\h}\langle{\pmb u}g,{\pmb u}g\rangle}{:}_{_K}{=}
\left\{
\begin{matrix}
{\text{alternating $2\pi$-periodic}} & s<b\\
{\text{$2\pi$-periodic}}  & a<s<b \\
{\text{alternating $2\pi$-periodic}}& b<s
\end{matrix}
\right.
$$
In Weyl ordered expression, $K{=}0$, the exchanging interval of
$e_*^{(s{+}it)\frac{1}{i\h}u{\ctt}v}$ is $a{=}b{=}\frac{\pi}{2}$, and 
in fact 
${:}e_*^{(s{+}it)\frac{1}{i\h}u{\ctt}v}{:}_0$ is singular at $t{=}\frac{\pi}{2}$. 
In the normal ordered expression, the exchanging interval of
$e_*^{(s{+}it){1}{i\h}u{\ctt}v}$ looks to be $a{=}-\infty,
b{=}\infty$ as this is an entire element, but it should be regarded 
that there is no exchanging interval because ${:}e_*^{(s{+}it){1}{i\h}u{\ctt}v}{:}_{_{K_0}}$
is alternating $2\pi$-periodic. (Cf.\cite{OMMY4}.)

We see $a<b$ in generic $K$. Pseudo-vacuums are defined 
whenever $a<0<b$ by 
$$
{:}\varpi_*(0){:}_{_K}{=}\frac{1}{2\pi}\int_0^{2\pi}
{:}e_*^{(s{+}it)\frac{1}{2i\h}\langle{\pmb u}g,{\pmb u}g\rangle}{:}_{_K}dt.
$$
as an idempotent element. 
This does not depend on $s$ whenever $a<s<b$ by Cauchy's integration
theorem. By the periodicity mentioned above, we have 
\begin{equation}
\frac{1}{4\pi}
\int_{-2\pi}^{2\pi}{:}e_*^{(s{+}it)\frac{1}{2i\h}\langle{\pmb u}g,{\pmb u}g\rangle}{:}_{_K}{=}
\left\{
\begin{matrix}
0 & s<b\\
{:}\varpi_*(0){:}_{_K}& a<s<b \\
0 & b<s
\end{matrix}
\right.
\end{equation}

\bigskip 

In what follows of  this section  we fix a nice expression parameter $K$. 
For every $g\in \mathcal{S}'$ we see that 
$e_*^{t\frac{1}{i\h}\langle{\pmb u}g,{\pmb u}g\rangle_*}$ is
$\pi$-periodic and singular points are distributed $\pi$-periodically 
along two lines sitting upper and lower half-plane. 
In what follows, we treat the case 
$\langle{\pmb u}g,{\pmb u}g\rangle{=}
2u{\ctt}v$ as a representative of general case by setting 
$g{=}\frac{1}{2}
\begin{bmatrix}
1&i\\
i&1
\end{bmatrix}$. All others are obtained by taking adjoint
transformations given by 
$e_*^{t\frac{1}{i\h}\langle{\pmb u}g,{\pmb u}g\rangle_*}$, $g\in
{\mathcal S}'$, which gives the adjoint action of $SU(2)$. 

Furthermore, argument below can be applied by taking adjoint 
transformations 
${\rm{Ad}}(e_*^{t\frac{1}{i\h}\langle{\pmb u}g,{\pmb u}g\rangle_*})$ 
by using any $g{\in} S\!L(2,{\mathbb C})$. Noting that 
$\langle{\pmb u}g,{\pmb u}g\rangle_*$ covers all quadratic forms 
with discriminant $-1$,  $2u{\ctt}v$ is a representative 
of all quadratic forms with discriminant $-1$. 
Note that these are Lorentz transformations bigger than the above.  

\bigskip
In \cite{ommy9}, we see the integral 
$$
{:}\varpi_{*}(0){:}_{_K}{=}\frac{1}{2\pi}\int_0^{2\pi}{:}e_*^{it\frac{1}{i\h}(u{\ctt}v)}{:}_{_K}dt
$$
gives an idempotent element, and we called this the pseudo-vacuum. 

Different from other vacuums used in \cite{OMMY4}, \cite{OMMY5} e.t.c. the regular
representation space does not form an algebra by the $*$-product, but 
$$
(u{\ctt}v){*}\varpi_{*}(0){=}0{=}\varpi_{*}(0){*}(u{\ctt}v),\quad
{\mathbb C}[u,v]{*}\varpi_{*}(0){=}({\mathbb C}[u]{+}{\mathbb C}[v]){*}\varpi_{*}(0).
$$｡｡
Regular representations on this space 
$({\mathbb C}[u]{+}{\mathbb  C}[v]){*}\varpi_{*}(0)$ is already
discussed in \cite{ommy6} by using Laurent expansions. As a result, we
have bilateral matrix elements $\{D_{k,l}; \,\,k, l\in {\mathbb Z}\}$
such that $\sum_{k\in {\mathbb Z}}D_{k,k}$ converges to $1$. 
(See comments given in this note \S\,\ref{comments}.) 

\medskip
In what follows, we give another treatment by using 
half-inverses (cf. \cite{ommy6}) defined in the previous section \S\,\ref{MakingMat}: 
$$
{u}^{\btt}{=}{v}{*}({u}{*}{v})_{*-}^{-1},\quad 
 ({u}{*}{v})_{*-}^{-1}
{=}-\frac{1}{i\h}\int_0^{\infty}e_*^{s\frac{1}{i\h}{u}{*}{v}}ds\,\,
$$
$$
{u}{*}{u}^{\btt}{=}1,\,\, 
{u}^{\btt}{*}{u}{=}1{-}\overline{\varpi}_{00},\,\, 
\overline{\varpi}_{00}{*}\overline{\varpi}_{00}{=}\overline{\varpi}_{00},\,\,
{u}{*}\overline{\varpi}_{00}{=}0{=}\overline{\varpi}_{00}{*}{u}^{\btt},
$$
where $\overline{\varpi}_{00}=\lim_{s\to\infty}\frac{1}{i\h}e_*^{s\frac{1}{i\h}{v}{*}{u}}$.
We see that 
$({u}^{\btt})^k{*}\overline{\varpi}_{00}{*}{u}^{\ell}$ are  $(k,\ell)$-matrix element.

\noindent
{\bf Note}\,\,The algebra above is isomorphic to the algebra of
``calculus''by setting 
$$
u^{\btt}{=}\int_0^xdx, \quad u{=}\frac{d}{dx},\quad
\overline{\varpi}_{00}{=}\delta_0, \quad {\text{where }}
\quad \delta_0(f(x)){=}f(0).
$$

However note that the double integral 
$$
\int_0^{\infty}\int_0^{2\pi}
e_*^{s\frac{1}{i\h}{u}{*}{v}}{*}e_*^{it\frac{1}{i\h}{u}{\ctt}{v}}
dsdt
$$
does not converge suffered by a singular point in the domain. 
In spite this, the repeated integrals give certain values.
Note that 
$$
e_*^{s\frac{1}{i\h}{u}{*}{v}}{*}e_*^{it\frac{1}{i\h}{u}{\ctt}{v}}{=}
e_*^{(s{+}it)\frac{1}{i\h}{u}{\ctt}{v}}e^{{-}\frac{1}{2}s}{=}
e_*^{(s{+}it)\frac{1}{i\h}{u}{*}{v}}e^{{+}\frac{i}{2}t}
$$
$$
\frac{1}{4\pi}\int_{-2\pi}^{2\pi}{:}e_*^{(s{+}it)\frac{1}{i\h}{u}{\ctt}{v}}{:}_{_K}dt{=}
\left\{
\begin{matrix}
{:}\varpi_*(0){:}_{_K}& a<s<b \\
0& b<s
\end{matrix}
\right..
$$
Hence 
$$
\int_0^{\infty}\frac{1}{4\pi}
\Big(\int_{-2\pi}^{2\pi}{:}e_*^{(s{+}it)\frac{1}{i\h}{u}{\ctt}{v}}{:}_{_K}dt e^{{-}\frac{1}{2}s}\Big)ds
{=}
\int_0^{b}e^{{-}\frac{1}{2}s}ds{=}\frac{1}{1/2}(1{-}e^{{-}\frac{1}{2}b}){:}\varpi_*(0){:}_{_K}.
$$
It follows ${:}u^{\btt}{*}\varpi_*(0){:}_{_K}{=}2(1{-}e^{{-}\frac{1}{2}b})\,v{*}\varpi_*(0)$. Next, we
compute  
$(u{*}v)_{*-}^{-1}{*}{v}^{l}{*}\varpi_*(0)$ as follows:
$$
\int_{0}^{\infty}
\!\!\frac{1}{4\pi}\int_{-2\pi}^{2\pi}\!\!
{:}v^l{*}e^{s(\frac{1}{i\h}u{\ctt}v{-}(l{+}\frac{1}{2})}{*}e^{it\frac{1}{i\h}u{\ctt}v}{:}_{_K}dt ds
{=}
{:}v^{l}{*}\int_0^be^{-s(l{+}\frac{1}{2})}ds\varpi_*(0){:}_{_K}{=}
\frac{1}{l{+}\frac{1}{2}}(1{-}e^{-b(l{+}\frac{1}{2})}){:}v_*^{\ell}{*}\varpi_*(0){:}_{_K}.
$$
Hence we see 
\begin{equation}
{:}(u^{\btt})^n{*}\varpi_*(0){:}_{_K}{=}
\frac{i^n}{\h^n(1/2)_n}
\Big(\prod_{l{=}1}^{n{-}1}(1{-}e^{-b(l{+}\frac{1}{2})})\Big){:}v^n{*}\varpi_*(0){:}_{_K}.
\end{equation}

Note that the associativity is broken:
$$
\begin{aligned}
{:}\varpi_*(0){:}_{_K}={:}(u^n{*}(u^{\btt})^n){*}\varpi_*(0){:}_{_K}{\not=}
{:}u^n{*}\big(u^{\btt n}{*}\varpi_*(0)\big){:}_{_K}&{=}
{:}u^n{*}
\frac{i^n}{\h^n(1/2)_n}
\prod_{l{=}1}^{n{-}1}(1{-}e^{-b(l{+}\frac{1}{2})})v^{n}{*}\varpi_*(0){:}_{_K}\\
&{=}
\prod_{l{=}1}^{n{-}1}(1{-}e^{-b(l{+}\frac{1}{2})}){:}\varpi_*(0){:}_{_K}
\end{aligned}
$$
It is remarkable that the result is very 
sensitive on expression parameters.
It follows ${:}u^{\btt}{*}\varpi_*(0){:}_{_K}{=}2(1{-}e^{{-}\frac{1}{2}b})\,v{*}\varpi_*(0)$. Next, we
compute  
$(u{*}v)_{*-}^{-1}{*}{v}^{l}{*}\varpi_*(0)$ as follows:
$$
\int_{0}^{\infty}
\!\!\frac{1}{4\pi}\int_{-2\pi}^{2\pi}\!\!
{:}v^l{*}e^{s(\frac{1}{i\h}u{\ctt}v{-}(l{+}\frac{1}{2})}{*}e^{it\frac{1}{i\h}u{\ctt}v}{:}_{_K}dt ds
{=}
{:}v^{l}{*}\int_0^be^{-s(l{+}\frac{1}{2})}ds\varpi_*(0){:}_{_K}{=}
\frac{1}{l{+}\frac{1}{2}}(1{-}e^{-b(l{+}\frac{1}{2})}){:}v_*^{\ell}{*}\varpi_*(0){:}_{_K}.
$$
Hence we see 
\begin{equation}
{:}(u^{\btt})^n{*}\varpi_*(0){:}_{_K}{=}
\frac{i^n}{\h^n(1/2)_n}
\Big(\prod_{l{=}1}^{n{-}1}(1{-}e^{-b(l{+}\frac{1}{2})})\Big){:}v^n{*}\varpi_*(0){:}_{_K}.
\end{equation}

Note that the associativity is broken:
$$
\begin{aligned}
{:}\varpi_*(0){:}_{_K}={:}(u^n{*}(u^{\btt})^n){*}\varpi_*(0){:}_{_K}{\not=}
{:}u^n{*}\big(u^{\btt n}{*}\varpi_*(0)\big){:}_{_K}&{=}
{:}u^n{*}
\frac{i^n}{\h^n(1/2)_n}
\prod_{l{=}1}^{n{-}1}(1{-}e^{-b(l{+}\frac{1}{2})})v^{n}{*}\varpi_*(0){:}_{_K}\\
&{=}
\prod_{l{=}1}^{n{-}1}(1{-}e^{-b(l{+}\frac{1}{2})}){:}\varpi_*(0){:}_{_K}
\end{aligned}
$$
It is remarkable that the result is very 
sensitive on expression parameters.
The equality holds only in the case $b{=}\infty$. At a first glance,
this looks the case of normal ordered expression, but the exchanging
interval and the pseudo-vacuum $\varpi_*(0)$ do not appear in the 
normal ordered expression. 

\bigskip
Anyhow, loosing associativity does not suffer to the argument.
The regular representation space 
$({\mathbb C}[u]{+}{\mathbb  C}[v]){*}\varpi_{*}(0)$  is linearly isomorphic to the
space ${\mathbb C}\{u^{\btt},u]{*}\varpi_{*}(0)$, where 
${\mathbb C}\{u^{\btt},u]$ is viewed simply as a linear space
$$
\cdots{+}
a_n(u^{\btt})^n{+}\cdots{+}a_1u^{\btt}{+}a_0{+}b_1u{+}\cdots
{+}b_m u^m{+}\cdots\,\,\,{\text{(finite sum)}}. 
$$

\subsection{Generated algebra}

On the other hand, let ${\mathcal A}$ be the algebra generated by $u, u^{\btt}$:
$$
{\mathcal A}{=}{\mathbb C}\{u^{\btt},u]\oplus {\mathcal M}, \quad 
{\mathcal M}{=}\{({u}^{\btt})^k{*}\overline{\varpi}_{00}{*}{u}^{\ell};
k,\ell\in {\mathbb N}\}.
$$
Beside the regular representation space mentioned above, the algebra 
${\mathcal A}$ is represented on the linear space spanned by 
$\{({u}^{\btt})^k{*}\overline{\varpi}_{00}; k\in {\mathbb N}\}$ as
infinite matrices 
$$
\sum_{|k{-}l|=finite}A_{k,l}({u}^{\btt})^k{*}\overline{\varpi}_{00}{*}u^l
$$
e.g.
\begin{equation}\label{matrep}
u^{\ell}{=}\sum_{k{\geq}0}(u^{\btt})^k\overline{\varpi}_{00}{*}{u}^{k{+}\ell},\quad 
(u^{\btt})^{\ell}{=}\sum_{k{\geq}0}(u^{\btt})^{k{+}\ell}\overline{\varpi}_{00}{*}{u}^k.
\end{equation}
As the factor algebra ${\mathcal A}/{\mathcal M}$ is naturally
isomorphic to the algebra ${\mathbb C}[z,z^{-1}]$ of all Laurent
polynomials, 
$$
0\to {\mathcal M}\to {\mathcal A}\to {\mathbb C}[z^{-1},z]\to 0.
$$  
${\mathcal A}$ may be regarded as a nontrivial extension of 
${\mathbb C}[z^{-1},z]$ by ${\mathcal M}$ such that 
$$
z{*}z^{-1}{=}1,\quad z^{-1}{*}z{=}1-\overline{\varpi}_{00},\quad 
z^{-2}{*}z{=}z^{-1}{-}{u}^{\btt}{*}\overline{\varpi}_{00},\quad 
z^{-2}{*}z^2{=} 
1{-}\overline{\varpi}_{00}{-}{u}^{\btt}{*}\overline{\varpi}_{00}{*}{u},\,e.t.c.
$$ 
As ${\mathbb C}[z^{-1},z]$ is abelian,  we see 
$[{\mathcal M},{\mathcal M}]\subset[{\mathcal A},{\mathcal A}]\subset{\mathcal M}$.
As  ${\mathfrak{sl}}(n,{\mathbb C})$, $n>1$, are simple Lie algebras,   
we see ${\mathcal M}/[{\mathcal M},{\mathcal M}]\cong{\mathbb  C}$. 
All elements of $[{\mathcal M},{\mathcal M}]$ are traceless, but 
some elements of $[{\mathcal A},{\mathcal A}]$ have non-vanishing trace.
Hence, 
$\mathcal M{=}[{\mathcal A},{\mathcal A}]$,  $[{\mathcal M},{\mathcal M}]$ 
is a Lie ideal of ${\mathcal A}$. 
Thus we have a Lie algebra extension  
$$
0\to {\mathbb C}\to {\mathcal A}/[{\mathcal M},{\mathcal M}]\to{\mathbb C}[z^{-1},z]\to 0,
$$
where ${\mathbb C}[z^{-1},z]$ is viewed as abelian Lie algebra. The
projection ${\mathcal M}\to {\mathcal M}/[{\mathcal M},{\mathcal M}]$
is given by taking the trace.
Consider the complementary subspace 
${\mathcal A}_0$ of ${\mathbb C}$ in ${\mathcal A}/[{\mathcal M},{\mathcal M}]$, 
but we denote these as 
$$
\cdots{+}
a_n\hat{u}_{-n}{+}\cdots{+}a_1\hat{u}_{-1}{+}a_0{+}b_1\hat{u}_1{+}\cdots
{+}b_m \hat{u}_{m}{+}\cdots\,\,\,{\text{(finite sum)}}.
$$
${\mathcal A}_0$ generates a meta abelian Lie algebra ${\mathbb C}{\oplus}{\mathcal A}_0$
under the bracket product 
$$
[|X,Y|]{=}{\rm{Tr}}[X,Y], \quad [|{\rm{Tr}}[X,Y], Z|]{=}0,\quad [|\hat{u}_m,\hat{u}_n|]{=}m\delta_{m+n,0} 
$$
\begin{prop}\label{freeBoson}
The regular representation space ${\mathbb C}\{u^{\btt},u]$  
generates a Lie algebra ${\mathbb C}{\oplus}{\mathcal A}_0$ such that 
$[|\hat{u}_m, \hat{u}_{n}|]{=}m\delta_{m+n,0}$.
\end{prop}
$$
0\to {\mathbb C}\to {\mathbb C}{\oplus}{\mathcal A}_{0}\to {\mathbb C}[z^{-1},z]\to 0.
$$
Hence ${\mathbb C}{\oplus}{\mathcal A}_0$ is a central extension of the (abelian) Lie
algebra ${\mathbb C}[z^{-1},z]$.

Its enveloping algebra $\widetilde{\mathcal A}_0$ is isomorphic to 
the Weyl algebra with infinitely many generators. This is called often the
free Boson algebra. Elements can be expressed univalently as linear combinations of  
$$
{\pmb u}^{\alpha,\beta}{=}\hat{u}_{-m}^{\alpha_m}{\btt}\cdots{\btt}\hat{u}_{-1}^{\alpha_1}
{\btt}\hat{u}_0{\btt}\hat{u}_1^{\beta_1}{\btt}\cdots{\btt}\hat{u}_n^{\beta_n},\quad
({\text{called normal ordering}})
$$
by using the commutation relations given in Proposition\,\ref{freeBoson}.

There is a natural homomorphism $\pi$ of $\widetilde{\mathcal A}_0$ onto 
${\mathbb C}[z^{-1},z]$ defined by 
$$
\pi({\pmb u}^{\alpha,\beta}){=}z^{\beta_n{+}\cdots{+}\beta_1{-}\alpha_1{-}\cdots{-}\alpha_m}.
$$
such that ${\rm{Ker}}\,\pi$ is the ideal ${\mathcal I}$  
generated by $[|\widetilde{\mathcal A}_0,\widetilde{\mathcal A}_0|]$ i.e.
$$
0\to {\mathcal I}\to \widetilde{\mathcal A}_0\to {\mathbb C}[z^{-1},z]\to 0. 
$$
\medskip
The Fock representation of $\widetilde{\mathcal A}_0$ is given by
setting the vacuum $|0\rangle$ by $\hat{u}_n|0\rangle{=}0$ for 
$n\geq 0$. As this is equivalent with $u^n{*}\overline{\varpi}_{00}{=}0$, the 
representation is equivalent with \eqref{matrep}.

\bigskip
Once we have a vacuum, the regular reperesentation space is defined as 
the remainder space by the operation to the vacuum. Evolution
equations represented in the regular reperesentation space is refered 
often as a Schr{\"o}dinger equations. Now as in the procedure of the 
second quantization, we make a representation of the regular 
represenrtation space. But this is simply to make an embedding 
into the matrices \eqref{matrep}. 

\medskip
As any automorphism 
$\psi:\widetilde{\mathcal A}_0\to\widetilde{\mathcal A}_0$ 
leaves ${\mathcal I}$ invariant, $\psi$ must yield an isomorphism of 
${\mathbb C}[z,z^{-1}]$, that is, a holomorphic mapping of 
${\mathbb C}\setminus\{0\}$ onto itself leaving 
${\mathbb  C}[z^{-1},z]$ invariant. This is not 
a M{\"o}bius transformation. 
Similarly, any derivation 
$D:\widetilde{\mathcal A}_0\to\widetilde{\mathcal A}_0$
satisfies 
$$
D[|X,Y|]{=}[|DX,Y|]{+}[|X, DY|]\quad {\text{and hence}}\quad   
D{\mathcal I}\subset{\mathcal I}.
$$
Hence $D$ yields a derivation/complex
vector field $\tilde{D}$ on ${\mathbb C}[z^{-1},z]$. 

On the other hand, for every quadratic form 
$Q(\pmb u)\in \widetilde{\mathcal A}_0$ (finite summation), 
${\rm{ad}}(Q(\pmb u)){=}[|Q(\pmb u), {\,\,}|]$ is a derivation of 
$\widetilde{\mathcal A}_0$. However, as $[|Q(\pmb u), f|]{=}0$ for 
almost all $f\in \widetilde{\mathcal A}_0$,  and for every $z^n$ 
there is an element $f$ in its representative such that 
$[|Q(\pmb u), f|]{=}0$. Hence, this yields the 
trivial derivation $0$ on  ${\mathbb C}[z^{-1},z]$.

Similarly, by using the $*$-exponential function $e_*^{Q(\pmb u)}$,  
${\rm{Ad}}(e_*^{Q(\pmb u)})$ is an automorphism of 
$\widetilde{\mathcal A}_0$, but this yields the identity 
on  ${\mathbb C}[z^{-1},z]$ by the same reason.  

Note that $z^{n{+}1}\partial_z$ 
is a derivation of ${\mathbb C}[z^{-1},z]$.  
Its integral curve $\psi_t(z)$ starting at $z$ is given by the differential equation  
$$
\frac{d}{dt}\psi_t(z){=}(\psi_t(z))^{n{+}1},\quad \psi_0(z){=}z. 
$$
But the solution is 
$$
\psi_t(z)=\frac{z}{\sqrt[n]{1{-}ntz^n}},\,\,(n\geq 1),\quad 
{=}ze^t, \,\,(n{=}0),\quad 
{=}\sqrt[|n|]{z^{|n|}{-}{n^{-1}}t}, \,\,(n\leq{-1}). 
$$   
These are not in the group ${\rm{Aut}}({\mathbb C}{\setminus}\{0\})$,
except the case $n=0$. 
Exponential functions have multivalued nature in general just 
like $*$-exponential functions of quadratic forms. It is natural to expect that 
the Lie algebra of all derivations on $\widetilde{\mathcal A}_0$
generate a {\it blurred covering group} of 
${\rm{Aut}}({\mathbb C}{\setminus}\{0\}){=} 
{\color{red}\{e^z\to ce^{\pm z}; c{>}0\}}$.

\bigskip
Now note that ${\rm{ad}}({\tilde Q}(\pmb u)$ by a quadratic 
forms of infinite summation gives also a derivation of  
$\widetilde{\mathcal A}_0$. For instance, let 
$L_0{=}-\frac{1}{2}\sum_{k{\in}{\mathbb Z}}\hat{u}_{-k}{\ctt}\hat{u}_k$ 
where $X{\ctt}Y{=}\frac{1}{2}(X{\btt}Y{+}Y{\btt}X)$. 
Then  
$
[|L_0, \hat{u}_{m}|]_{\btt}{=}m\,\hat{u}_{m}$.  
Hence $L_0\,\, :\,\,\widetilde{\mathcal  A}_0\to \widetilde{\mathcal A}_0$\,\,
is a derivation corresponding to $z\partial_z$. Similarly, for every $n\in{\mathbb Z}$, let 
\begin{equation}\label{basevirasolo}
\begin{aligned}
L_n{=}-\frac{1}{2}\sum_{k{\in}{\mathbb Z}}\hat{u}_{n-k}{\ctt}\hat{u}_{k},  
\end{aligned}
\end{equation}
Then, we see 
$$
[|L_n, \hat{u}_m|]{=}m\hat{u}_{m{+}n}, \quad 
[|L_{-n}, \hat{u}_m|]{=}m\hat{u}_{m{-}n}, \quad e.t.c.
$$
Hence $L_n,\,n{\in}{\mathbb Z}$, are derivations on $\widetilde{\mathcal A}_0$
corresponding naturally to $z^{n{+}1}\partial_z$.  
It is easy to see that 
$$
[|L_k,[|L_{\ell}, \hat{u}_m|]|]{=}[|L_{\ell},[|L_{k}, \hat{u}_m|]|]
{=}[|(k{-}\ell)L_{k+\ell}, \hat{u}_m|]. 
$$
Hence $\{{\rm{ad}}(L_n); n{\in}{\mathbb Z}\}$ is a representation 
of the Lie algebra ${\mathbb C}[z,z^{-1}]\partial_z$ of all Laurent polynomial vector fields.

However, note that 
$$
\begin{aligned}
L_n\btt L_{-n}{-}L_{-n}\btt L_{n}&=
\frac{1}{4}\sum_{k,\ell}(\hat{u}_{-k}{\ctt}\hat{u}_{k+n})
\btt(\hat{u}_{-(\ell{+}n)}{\ctt}\hat{u}_{\ell})
-\frac{1}{4}\sum_{k,\ell}(\hat{u}_{-(\ell{+}n)}{\ctt}\hat{u}_{\ell})
\btt(\hat{u}_{-k}{\ctt}\hat{u}_{k+n})\\
&=2nL_0{-}\frac{1}{2}\sum_{k=1}^{n-1}(n{-}k)\hat{u}_{-k}\btt\hat{u}_k
{=}
2nL_0{-}\sum_{k=1}^{n-1}\frac{1}{2}k(n-k).
\end{aligned}
$$

Thus, 
\begin{equation}\label{Viras}
[|L_n, L_{-n}|]_{\btt}{=}
2nL_0{+}\frac{1}{12}(n{-}1)n(n{+}1).
\end{equation} 
It is easy to see that if $n{+}m{\not=}0$, then  
$[|L_n, L_{m}|]_{\btt}{=}(n-m)L_{n{+}m}$.

This implies that $L_n,\, n\in{\mathbb Z}$, generates a Lie algebra   
which is nontrivial central extension of ${\mathbb C}[z,z^{-1}]\partial_z$. 
${\mathbb C}[z,z^{-1}]\partial_z$ is called the Witt Lie algebra in 
the conformal field theory. 
Denoting ${\mathbb C}[z,z^{-1}]\partial_z$
by $\mathfrak g$ for simplicity, any central extension of $\mathfrak g$
is caused by a Chevalley 2-cocycle $\omega$, i.e. skew-symmetric
bilinar form $\omega: \mathfrak g{\times}\mathfrak g\to{\mathbb C}$ 
such that $d\omega{=}0$, that is, 
$$
\sum_{cyclic}\omega(X,[Y,Z]){=}0,\quad [|X,Y|]{=}[X,Y]{+}\omega(X,Y).
$$
The Chevalley 2-cohomology group of the Witt Lie algebra is known to
be 1 dimensional. The standard one is known as the Virasoro Lie
algebra given by 
\begin{equation}\label{Virs}
[L_n, L_m]{=}(n{-}m)L_{n{+}m}{+}\frac{c}{12}n(n^2{-}1)\delta_{n{+}m,0}.
\end{equation}
Denote this Lie algebra by $Vir(c)$. It will be mentioned a 
group theoretical treatment of this algebra from a view point of 
infinite dimensional Lie groups will be discussed in the next note.

Note here that there is no 
obstruction to restrict our system to the real coefficients. 

\subsection{Restriction to the real coefficients}

\bigskip
To make the situation clearer, we take the topological completion 
$C^{\infty}(S^1)$ (all smooth ${\mathbb C}$-valued functions on $S^1$)
of ${\mathbb C}[z,z^{-1}]$ by regarding these as polynomial functions  
on the unit circle $S^1$ and considering Fourier series. 
$C^{\infty}(S^1)$ is the space of the Fourier series 
$$
\{\sum a_ne^{in\theta}; \sum(1{+}n^2)^k|a_n|^2<\infty, \,\forall
k{\in}{\mathbb N}\,\}
$$

It is a little hard task to fix the topological completions  
$\overline{\mathcal A}$ and $\overline{\mathcal M}$ to obtain the
exact sequence 
$$
0\to \overline{\mathcal M}\to \overline{\mathcal A}\to C^{\infty}(S^1)\to 0.
$$
We have to leave the strict treatment to future notes, but it is
natural to think there is a central extension of the (abelian) Lie
algebra  
$$
0\to {\mathbb C}\to\overline{\mathcal A}_0\to C^{\infty}(S^1)\to 0. 
$$
$\overline{\mathcal A}_0$ is spanned by $\{u_n; n{\in}{\mathbb Z}\}$,
where $\hat{u}_n$ correspond to $e^{in\theta}$. 
We denote  by $\widetilde{\mathcal  A}_0$ the enveloping algebra of 
$\overline{\mathcal A}_0$. 

\medskip
Let $\bsG(T_{S^1})$ be the Lie algebra over ${\mathbb R}$ 
of the smooth diffeomorphism group $\Cal D(S^1)$ i.e. the space of
$C^{\infty}$ vector fields on $S^1$. $\bsG(T_{S^1})$ is linearly isomorphic 
to $C_{\mathbb R}^\infty(S^1)$. The Lie 
bracket of $\bsG(T_{S^1})$ is expressed in $C_{\mathbb R}^\infty(S^1)$ as 
$[f, g] = fg' - gf'$. 
We now restrict our attention to the derivations of   
$\widetilde{\mathcal A}_0$ which induce $C^{\infty}$-vector fields on
$S^1$. One may use the same deriavtions as in \eqref{basevirasolo} by
taking the real part. But it is known in \cite{GF} that a nontrivial 
2-cocycle $\alpha$ is given by 
$$ 
\alpha(f,h)=\int\limits_{S^1}(f'h{''} - f{''}h')dt
\quad{\text{(essentially same to \eqref{Virs})}}
$$
and any other 2-cocycles are cohomologus to $a\alpha$ for some
constant $a$. Thus, there is an  $\mathbb R$ central extension 
$\bsG(T_{S^1})\!\ltimes_{\alpha}\!\mathbb R$ of $\bsG(T_{S^1})$.  

Now, we want to lift this cocycle to the Hochschild cocycle on 
the group $\Cal D(S^1)$. Let $\Cal D_0(S^1)$ be the identity 
component of $\Cal D(S^1)$.
Recall that the homotopy type of $\Cal D_0(S^1)$ is  $S^1$. 
\par
Therefore the de Rham cohomology group $H^2(\tilde{\Cal D}_0(S^1))$ of 
the universal covering group $\tilde{\Cal D}_0(S^1)$ of $\Cal D_0(S^1)$ 
vanishes.  
\par
By a technique similar to Koszur construction, we can construct 
a Hochschild 2-cocycle $\omega$ from $\alpha$ (cf.\cite{om3}\,\S X.6).  
Thus, there exists a regular Fr{\'e}chet Lie group $\tilde G$ with the Lie algebra 
$\mathbb R\!\ltimes_{\alpha}\!\bsG(T_{S^1})$. $\tilde G$ is the  
$\mathbb R$ central extension of $\tilde{\Cal D}_0(S^1)$. Thus, the multivalued
nature appears when we consider on the space $S^1$.

\bigskip
\noindent
{\bf Future problems}

Recall here that $\frac{1}{i\h}{u}{\ctt}{v}$ was a representative 
of all quadratic forms $\frac{1}{i\h}\langle{\pmb u}g,{\pmb u}g\rangle$,
$g\in S\!L(2,{\mathbb C})$ that is a representative of 
all quadratic forms of discriminant $-1$. Hence, we have a
family of extensions: 
$$
0\to{\mathcal I}_{\pmb\gamma}\to\widetilde{\mathcal A}_{\pmb\gamma}
\to C^{\infty}_{\pmb\gamma}(S^1)\to 0
$$ 
parameterized by $\pmb\gamma\in\mathcal L{=}S\!L(2,{\mathbb C})/\varGamma$. 

As $S\!L(2,{\mathbb C})$ acts on $\mathcal L$, we have bridges
between two exact sequences:
$$
\begin{matrix}
0\to &{\mathcal I}_{\pmb\gamma}&\to &\widetilde{\mathcal  A}_{\pmb\gamma}&
\to &C^{\infty}_{\pmb\gamma}(S^1)&\to & 0\\
{  } & \downarrow         &{ } &  \downarrow &{  }        &\downarrow {} \\
0\to &{\mathcal I}_{\pmb\gamma'}&\to &\widetilde{\mathcal A}_{\pmb\gamma'}&\to
&C^{\infty}_{\pmb\gamma'}(S^1)&\to & 0\\
\end{matrix}
$$
On the other hand, infinitesimal adjoint action 
${\rm{ad}}(\frac{1}{i\h}Q(u,v)): 
\widetilde{\mathcal A}_{\pmb\gamma}\to
\widetilde{\mathcal  A}_{\pmb\gamma}  
$
is a derivation for every $\pmb\gamma\in \mathcal L$ 
and this yields a vector field on the space $\mathcal L{\times}S^1$.

Hence, what we realy have to consider is a central extension of the Lie
algebra of vector fields on $\mathcal L{\times}S^1$. This will be 
treated in forthcoming note.

\section{Spontaneous splitting of polar element}

$SU(2)$-vacuum is defined only by using a nice expression parameter,
which is by no means generic.  By Remark 1 in  
\S\,\ref{linear} we see that for a generic $K\in D_{-1}$, there is $g$
such that 
${:}e_*^{\pi\frac{1}{i\h}\langle{\pmb u}g,{\pmb u}g\rangle_*}{:}_{_{K}}=-1$. 

Now, we repeat the result in \cite{OMMY5} for the case $m=1$.
\begin{prop}\label{strange2}
Suppose there is $g\in Sp(1,{\mathbb C}){=}S\!L(2,\mathbb C)$ such that 
${:}e_*^{[0\to\pi]\frac{1}{i\h}
\langle{\pmb u}g,{\pmb u}g\rangle_*}{:}_{_{K}}=-1$. 
Then, there must exist 
$h\in Sp(1,{\mathbb C})$ such that 
${:}e_*^{[0\to\pi]\frac{1}{i\h}
\langle{\pmb u}h, {\pmb u}h\rangle_*}{:}_{_{K}}=1$, and 
$\hat{h}\in Sp(1,{\mathbb C})$ such that the path  
${:}e_*^{[0\to\pi]\frac{1}{2i\h}
\langle{\pmb u}\hat{h}, {\pmb u}\hat{h}\rangle_*}{:}_{_{K}}$
must hit a singular points.
\end{prop}

However, as singular points are double branched in genric ordered
expression, there must be a one parameter group going through 
 near the singular the singular point that cuts a slit and 
goes into another sheet and hence $1$ at $t{=}\pi$. Hence we have 

\begin{thm}
In generic ordered expression,  
${\mathcal S}'$splits into three open sets  
${S_0\cup S_{+}\cup S_-}$ such that 
$\overline{S_0\cup S_{+}\cup S_-}{=}{\mathcal S}'$ such that 
$$
\left\{
\begin{matrix}
g\in S_0 &e_*^{\pi\frac{1}{i\h}\langle{\pmb u}g,{\pmb u}g\rangle}{=}1:
&{\text{Real axis is between two lines of singular points}}\\
g\in S_{+} &e_*^{\pi\frac{1}{i\h}\langle{\pmb u}g,{\pmb
    u}g\rangle}{=}-1:&{\text{Singular points are in upper half-plane}}\\
g\in S_{-} &e_*^{\pi\frac{1}{i\h}\langle{\pmb u}g,{\pmb
    u}g\rangle}{=}-1:&{\text{Singular points are in lower half-plane}}\\
\end{matrix}
\right.
$$
\end{thm}
All of these are on one-parameter subgroups. To make the argument
simpler, we take one parameter groups 
$$
e_*^{t\frac{1}{2i\h}(u^2{+}v^2)},\quad 
e_*^{t\frac{1}{\h}u{\ctt}v}, \quad
e_*^{t\frac{1}{2\h}(u^2{-}v^2)}
$$
as representatives of $S_0$, $S_{+}$, $S_{-}$ respectively.
We denote their square
roots by  ${\e}_0$,  ${\e}^*$, ${\e}'$ respectively. These are the polar 
element under a nice expression parameters. 
We have also the 
square roots of these such that  
$$
e_1=e_*^{\frac{\pi i}{4i\h}(u^2{+}v^2)},\quad
e_2=e_*^{\frac{\pi i}{2i\h}u{\ctt}v}, \quad 
e_3=e_*^{-\frac{\pi}{4i\h}(u^2{-}v^2)}, 
$$
where we have $e_1^4=1$, $e_2^4=e_3^4={-}1$.

\subsection{Regular representation on each sector}\label{comments}

In spite of these difficulties, every one parameter subgroup makes 
vacuum or pseudo-vacuum and their own regular representation spaces. 
We discuss these in the case of each representative  
$$
e_*^{t\frac{1}{2i\h}(u^2{+}v^2)},\,\,e_*^{t\frac{1}{\h}u{\ctt}v},
\,\,e_*^{t\frac{1}{2\h}(u^2{-}v^2)}.
$$

\bigskip
\noindent
({\bf 1}) $e_*^{t\frac{1}{2i\h}(u^2{+}v^2)}$: By setting 
$u'{=}\sqrt{i}(u{+}iv)$, $v'{=}\sqrt{i}(u{-}iv)$ we have 
$u^2{+}v^2{=}u'{\ctt}v'$. 
This is the case discussed in \S\,\ref{Pseudo} where
we have a pseudo-vacuum $\varpi_*(0)$. 
If the expression parameter is restricted to nice expression
parameters, we have to treat only the case ({\bf  1}).

Thus, in what follows we treat $u', v'$ as $u, v$. 

Now, by the observation in \cite{ommy6},  if we use the convention    
\begin{equation}\label{pmconvention}
\zeta^k=
\left\{
\begin{matrix}
u^k,& k\geq 0\\
v^{|k|},& k<0,
\end{matrix}
\right.\qquad 
{\hat\zeta}^{\ell}=
\left\{
\begin{matrix}
v^\ell,& \ell\geq 0\\
u^{|\ell|},& \ell<0,
\end{matrix}
\right.
\end{equation}
then we have  
$$
D_{k,\ell}(K)= 
\frac{1}
{\sqrt{(\frac{1}{2})_{k}(\frac{1}{2})_{\ell}(i\h)^{|k|+|\ell|}}}
{\zeta}^k{*}{:}\varpi_*(0){:}_{_K}{*}{\hat\zeta}^{\ell}, \quad 
{:}\varpi_*(0){:}_{_K}{=}
\frac{1}{2\pi}\!\!\int_0^{2\pi}\!\!\!
{:}e_*^{it(\frac{1}{i\h}u{\ctt}v)}dt{:}_{_K}dt
$$
are matrix elements for every $k,\ell{\in}{\mathbb Z}$.

\subsubsection{Relation to matrix algebra $\mathcal A$}

To relate these to $\mathcal A$, we have to know more about the
property of $e_*^{(\sigma{+}it)\frac{1}{i\h}u{\ctt}v}$, $e_*^{(s{+}it)\frac{1}{i\h}v{*}u}$. 
First for ${:}e_*^{(s{+}it)\frac{1}{i\h}u{\ctt}v}{:}_{_K}$ in generic
ordered expression, there is an interval $[a,b]$ called the {\bf
  exchanging interval} (cf.\cite{ommy7}) such that 
$$\left\{
\begin{matrix}
s<a & {:}e_*^{(s{+}it)\frac{1}{i\h}u{\ctt}v}{:}_{_K} &
                 {\text{alternating }2\pi\text{-periodic}} \\
a<s<b&{:}e_*^{(s{+}it)\frac{1}{i\h}u{\ctt}v}{:}_{_K} &
                 {2\pi\text{-periodic}}\\
s<b& {:}e_*^{(s{+}it)\frac{1}{i\h}u{\ctt}v}{:}_{_K} &
                 {\text{alternating }2\pi\text{-periodic}}
\end{matrix}
\right.
$$
In a nice expression parameter $K$, the exchanging interval has the 
property $a<0<b$. Note that pseudo-vacuums  
requires only the property  $a<0<b$.

Furthermore, we see (cf.\cite{ommy6}, \cite{ommy7}) that $\varpi_*(0)$ is given also by  
$$
\varpi_*(0){=}\frac{1}{2\pi}\int_0^{2\pi}e_*^{(s{+}it)\frac{1}{i\h}{u}{\ctt}{v}}dt,\quad
a<\forall s<b
$$ 
Furthermore, we see easily  
$$
\overline{\varpi}_{00}{=}\frac{1}{2\pi}\int_0^{2\pi}e_*^{(s{+}it)\frac{1}{i\h}{v}{*}{u}}dt,\quad
b<\forall s.
$$ 
Now, noting $\frac{1}{i\h}v{*}u{=}\frac{1}{i\h}v{\ctt}u{+}\frac{1}{2}$
we compute as follows:
$$
\begin{aligned}
(2\pi)^2\overline{\varpi}_{00}{*}u^{l}{*}\varpi_*(0)&{=}
\iint e_*^{(s{+}it)\frac{1}{i\h}(v{\ctt}u{+}\frac{i\h}{2})}
e_*^{(\sigma{+}it')\frac{1}{i\h}(v{\ctt}u{+}{i\h}l)}dtdt'
{=}
\iint e_*^{s{+}\sigma{+}i(t{+}t')\frac{1}{i\h}v{\ctt}u}
e^{i(\frac{1}{2}t{+}lt')}e^{\frac{1}{2}s{+}l\sigma}\\
&=
\iint e_*^{(s{+}\sigma{+}i\tau)(\frac{1}{i\h}v{\ctt}u{-}il)}
e^{-it(l{-}\frac{1}{2})e^{l\sigma{+}\frac{1}{2}s}}dtd\tau.
\end{aligned}
$$
Now for a fixed $l$, we choose $s$ and $\sigma$ so that
$$
a<s{+}\sigma<b, \quad b<s,\quad l\sigma{+}\frac{1}{2}s{=}0.
$$
Then, we have 
$$
(2\pi)^2\overline{\varpi}_{00}{*}u^{l}{*}\varpi_*(0){=}
\frac{4}{2l{-}1}\varpi_*(0){*}u^l.
$$
Hence we have 
$$
(u^{\btt})^k{*}\overline{\varpi}_{00}{*}u^{l}{*}\varpi_*(0){=}
\frac{1}{(2l{-}1)\pi^2}(u^{\btt})^k{*}\varpi_{*}(0){*}u^{l},\quad k, l\in
\mathbb N.
$$
Hence by applying $\varpi_*(0)$ from r.h.s. the algebra 
${\mathcal A}$ in \S\,\ref{Pseudo} is translated to 
 uni-lateral matrices of $D_{k,l}$. 
It is interesting that this procedure mentioned above  
gives a method to change bilateral matrix-elements to  
uni-lateral matrices.

\bigskip
\noindent
({\bf 2}) $e_*^{it\frac{1}{i\h}u{\ctt}v}$: This gives the most typical
vacuum representations. The vacuum $\varpi_{00}$ is given by 
$$
\varpi_{00}{=}\frac{1}{2\pi}\int_0^{2\pi}e_*^{(s{+}it)\frac{1}{i\h}{v}{*}{u}}dt,\quad
\forall s<a.
$$
In generic ordered expressions,
$E_{p,q}=
\frac{1}{\sqrt{p!q!(i\h)^{p{+}q}}}u^p{*}\varpi_{00}{*}v^q$ 
is the $(p,q)$-matrix element, that is 
$E_{p,q}{*}E_{r,s}=\delta_{q,r}E_{p,s}$. The $K$-expression 
${:}E_{p,q}{:}_{_K}$ of $E_{p,q}$ will be denoted by 
$E_{p,q}(K)$. Note that $E_{0,0}(K){=}{:}\varpi_{00}{:}_{_K}$.
As the singular points are in the upper
half-plane, i.e. $0<a<b$ where $[a,b]$ is the exchanging interval,  the radius of convergence of 
$e_*^{\log w\frac{1}{i\h}u{*}v}$ is $>1$. 
Hence we have 
$$
\sum E_{n,n}{=}1.
$$

\bigskip
\noindent
({\bf 3}) $e_*^{it\frac{1}{2i\h}u^2{-}v^2}$: We set 
$u'{=}\frac{1}{\sqrt{2}}(u{+}v), v'{=}\frac{1}{\sqrt{2}}(u{-}v)$, and
treat $(u',v')$ as $(u,v)$. a similar calculations, we see 
that 
${\overline E}_{p,q}=
\frac{\sqrt{-1}^{p+q}}{\sqrt{p!q!(i\h)^{p{+}q}}}
v^p{*}{\overline{\varpi}}_{00}{*}u^q$ is the 
$(p,q)$-matrix element in generic ordered expressions.
 The $K$-expression of 
$\overline{E}_{p,q}$ will be denoted by $\overline{E}_{p,q}(K)$.
Note that $\overline{E}_{0,0}(K){=}{:}{\overline{\varpi}}_{00}{:}_{_K}$.

\bigskip
In general  the $*$-product ${\varpi}_{00}{*}{\overline{\varpi}}_{00}$
can not be defined univalently, as this depend on the manner of
calculation. 

The next identities are easy to see 
$$
\varpi_{00}{*}\frac{1}{i\h}u{\ctt}v=
\frac{1}{2}\varpi_{00},\quad
\frac{1}{i\h}u{\ctt}v{*}\overline{\varpi}_{00}=
{-}\frac{1}{2}\overline{\varpi}_{00}.
$$ 
Note that in order to keep the associativity  
\begin{equation}\label{inorderto}
(\varpi_{00}{*}\frac{1}{i\h}u{\ctt}v)
{*}\overline{\varpi}_{00}=
\varpi_{00}{*}
(\frac{1}{i\h}u{\ctt}v{*}\overline{\varpi}_{00}), 
\end{equation}
we have to define 
$$
\frac{1}{2}\varpi_{00}{*}\overline{\varpi}_{00}=
{-}\frac{1}{2}\varpi_{00}{*}\overline{\varpi}_{00}=0.
$$

\subsection{Failure of binary operations}

\par\bigskip\par\bigskip
\noindent
%
\unitlength 0.1in
\begin{picture}(23.3000,9.1600)(2.5000,-15.9600)
%
\special{pn 8}%
\special{pa 380 794}%
\special{pa 408 770}%
\special{pa 436 748}%
\special{pa 464 724}%
\special{pa 492 702}%
\special{pa 520 680}%
\special{pa 548 660}%
\special{pa 576 638}%
\special{pa 604 620}%
\special{pa 632 602}%
\special{pa 660 584}%
\special{pa 688 570}%
\special{pa 716 556}%
\special{pa 744 544}%
\special{pa 774 534}%
\special{pa 802 524}%
\special{pa 830 518}%
\special{pa 858 514}%
\special{pa 886 514}%
\special{pa 914 514}%
\special{pa 942 518}%
\special{pa 970 524}%
\special{pa 998 532}%
\special{pa 1026 542}%
\special{pa 1054 554}%
\special{pa 1082 568}%
\special{pa 1110 584}%
\special{pa 1138 600}%
\special{pa 1166 618}%
\special{pa 1194 638}%
\special{pa 1222 658}%
\special{pa 1250 678}%
\special{pa 1278 700}%
\special{pa 1306 722}%
\special{pa 1332 746}%
\special{pa 1360 768}%
\special{pa 1388 792}%
\special{pa 1416 814}%
\special{pa 1442 838}%
\special{pa 1470 860}%
\special{pa 1498 882}%
\special{pa 1524 904}%
\special{pa 1552 926}%
\special{pa 1580 946}%
\special{pa 1606 966}%
\special{pa 1634 984}%
\special{pa 1660 1002}%
\special{pa 1688 1018}%
\special{pa 1716 1034}%
\special{pa 1742 1048}%
\special{pa 1770 1060}%
\special{pa 1798 1070}%
\special{pa 1826 1080}%
\special{pa 1854 1086}%
\special{pa 1882 1092}%
\special{pa 1910 1096}%
\special{pa 1938 1096}%
\special{pa 1966 1096}%
\special{pa 1996 1094}%
\special{pa 2024 1090}%
\special{pa 2052 1084}%
\special{pa 2082 1076}%
\special{pa 2110 1066}%
\special{pa 2140 1056}%
\special{pa 2170 1044}%
\special{pa 2198 1030}%
\special{pa 2228 1016}%
\special{pa 2258 1002}%
\special{pa 2288 986}%
\special{pa 2316 968}%
\special{pa 2346 950}%
\special{pa 2376 932}%
\special{pa 2406 912}%
\special{pa 2436 892}%
\special{pa 2466 872}%
\special{pa 2496 852}%
\special{pa 2526 832}%
\special{pa 2556 810}%
\special{pa 2580 794}%
\special{sp}%
%
\special{pn 8}%
\special{pa 390 794}%
\special{pa 418 818}%
\special{pa 446 842}%
\special{pa 474 866}%
\special{pa 502 890}%
\special{pa 528 912}%
\special{pa 556 934}%
\special{pa 584 956}%
\special{pa 612 976}%
\special{pa 640 994}%
\special{pa 668 1012}%
\special{pa 694 1030}%
\special{pa 722 1044}%
\special{pa 750 1058}%
\special{pa 778 1068}%
\special{pa 806 1078}%
\special{pa 834 1086}%
\special{pa 862 1090}%
\special{pa 890 1094}%
\special{pa 918 1094}%
\special{pa 946 1090}%
\special{pa 972 1086}%
\special{pa 1000 1078}%
\special{pa 1028 1070}%
\special{pa 1056 1058}%
\special{pa 1084 1044}%
\special{pa 1112 1030}%
\special{pa 1140 1014}%
\special{pa 1168 996}%
\special{pa 1196 978}%
\special{pa 1224 956}%
\special{pa 1252 936}%
\special{pa 1280 914}%
\special{pa 1306 892}%
\special{pa 1334 870}%
\special{pa 1362 846}%
\special{pa 1390 822}%
\special{pa 1418 800}%
\special{pa 1444 776}%
\special{pa 1472 754}%
\special{pa 1498 730}%
\special{pa 1526 708}%
\special{pa 1554 688}%
\special{pa 1580 666}%
\special{pa 1608 646}%
\special{pa 1634 628}%
\special{pa 1662 610}%
\special{pa 1690 592}%
\special{pa 1716 576}%
\special{pa 1744 562}%
\special{pa 1772 550}%
\special{pa 1800 538}%
\special{pa 1828 530}%
\special{pa 1856 522}%
\special{pa 1884 516}%
\special{pa 1912 512}%
\special{pa 1940 510}%
\special{pa 1968 512}%
\special{pa 1996 514}%
\special{pa 2026 518}%
\special{pa 2054 524}%
\special{pa 2084 530}%
\special{pa 2112 540}%
\special{pa 2142 550}%
\special{pa 2170 562}%
\special{pa 2200 576}%
\special{pa 2230 590}%
\special{pa 2260 604}%
\special{pa 2288 620}%
\special{pa 2318 638}%
\special{pa 2348 656}%
\special{pa 2378 674}%
\special{pa 2408 694}%
\special{pa 2438 712}%
\special{pa 2468 732}%
\special{pa 2498 754}%
\special{pa 2528 774}%
\special{pa 2558 794}%
\special{pa 2570 804}%
\special{sp}%
%
\special{pn 8}%
\special{ar 1970 800 62 62  0.0000000 6.2831853}%
%
\special{pn 8}%
\special{sh 1}%
\special{ar 390 790 10 10 0  6.28318530717959E+0000}%
\special{sh 1}%
\special{ar 1410 810 10 10 0  6.28318530717959E+0000}%
\special{sh 1}%
\special{ar 1410 810 10 10 0  6.28318530717959E+0000}%
%
\special{pn 8}%
\special{sh 1}%
\special{ar 2560 790 10 10 0  6.28318530717959E+0000}%
\special{sh 1}%
\special{ar 2560 790 10 10 0  6.28318530717959E+0000}%
\special{sh 1}%
\special{ar 2560 790 10 10 0  6.28318530717959E+0000}%
%
\special{pn 13}%
\special{pa 1970 790}%
\special{pa 2180 240}%
\special{fp}%
\put(7.5000,-4.6000){\makebox(0,0)[lb]{$a$}}%
\put(7.8000,-12.4000){\makebox(0,0)[lb]{$b$}}%
\put(17.3000,-5.4000){\makebox(0,0)[lb]{$b$}}%
\put(18.2000,-12.1000){\makebox(0,0)[lb]{$a$}}%
\put(2.5000,-7.8000){\makebox(0,0)[lb]{$1$}}%
\put(3.3000,-8.3000){\makebox(0,0)[lb]{$\bullet$}}%
\put(3.6000,-8.0000){\makebox(0,0)[lb]{\vector(2,1){3}}}
\put(24.3000,-10.4000){\makebox(0,0)[lb]{$1$}}%
\put(25.1000,-7.1000){\makebox(0,0)[lb]{$-1$}}%
\put(21.7000,-3.5000){\makebox(0,0)[lb]{\footnotesize{slit}}}%
\put(5.7000,-13.5000){\makebox(0,0)[lb]
{\tiny{Only the sign differs at the crossed point }}}%

\end{picture}%
\hfill
\parbox[b]{.6\linewidth}{In the previous section  we saw that   
$$
({\e}_0)^2{=}1,\quad ({\e}^*)^2{=}-1, \quad ({\e}')^2{=}-1.
$$
In spite of this, we have $({\e}^*)^{-1}{=}{\e}'$. This is because 
when $t$ is replaced by $-t$, the lines of singular points are 
switched upside down, and this is proved by 
Cauchy's integration theorem. 
At a first glance it looks to contradict the exponential law.}  

Indeed, this is the reason why the polar element 
and $*$-exponential functions of quadratic forms are  viewed as  
double valued elements in generic ordered expressions (cf.\cite{OMMY4}).
Intertwiners are defined only as 2-to-2 mappings on such spaces. 
However, computations such as $\sqrt{a}\sqrt{b}{=}\sqrt{ab}$ arrows us 
to treat these double valued elements safely. As a result, we have 
some extraordinary ``blurred groups''  such as a double covering 
group of $S\!L(2,{\mathbb C})$. (Cf. \cite{OMMY5}.)  An extraordinary 
phenomenon mentioned above is then a {\bf natural conclusion} when we treat 
$*$-exponential functions of quadratic forms as strictly single 
valued elements.  

Thus, we have to conclude that these elements such as ${\e}_0,\,
{\e}^*, \, {\e}'$ and $e_1, e_2, e_3$ cannot be used as 
members of binary operations. We have to accept that there 
are many such elements in the extended Weyl algebra because of 
double-branched singular points. 
However, if one treat every element together with a path avoiding 
singular points from the origin $t=0$ where we assign $1$ always, then  
we obtain the value of $*$-exponential functions such as 
${:}e_*^{[0\sim t]\frac{1}{i\h}
\langle{\pmb u}g, {\pmb u}g\rangle_*}{:}_{_{K}}$ univalent way. 
By this way, one may give some groupoid structure for the space 
of $*$-exponential functions with paths of quadratic forms. But this
is too complicated to treat the objects safely, as we have to use two sheets and  
slits setting between singular points.

\subsubsection{Path-connecting products by restricting  paths}

One way to treat these safely is to restrict the paths 
although every element is given together with a path avoiding 
singular points from the origin $t=0$. 

We consider here in various products of elements  
$a(x){=}e_*^{x\frac{1}{i\h}u^2{+}v^2}$, 
$b(y){=}e_*^{iy\frac{1}{i\h}u{\ctt}v}$, 
$c(z){=}e_*^{iz\frac{1}{i\h}u^2{-}v^2}$ for $x,y,z{\in}{\mathbb R}$
such as 
$a(x_1)c(z_1)b(y_1)a(x_2)c(z_2)a(x_2)\cdots$.

To treat these we take the 3-dimensional lattice $L$ such that  
$$
cL{=}\{(x,y,z); \,\,\text{two of them are integers and the other is a real number} \}
$$
where $c$ is a scaling unit.
We assume there is no singular point of $a(x){*}b(y){*}c(z)$ on $cL$.

\medskip
The set $\{(x,y,z); \text{{\bf{only one of them is an integer, others are real}}}\}$
will be called the wall. We assume further that the singular set of $a(x){*}b(y){*}c(z)$
intersect the wall transversally. 

As the singular set forms a subset of complex codimension one,
singular set on the wall must be discrete set. It is hard to fix this
set, but by  observations in \cite{OMMY4} we see that if a box in the
wall contain a singular point, then travelling around the edge of the
box makes the sign change.

Now project $cL$ to ${\mathbb R}^2$ so that each cube projects down as
hexagon with 6 triangles inside. 
We assume for simplicity that every triangle contains at most one
singular point. 

In the next section, we set a virtual, but somewhat realistic
distribution of singular points,  and we investigate how 
path-connecting products are defined.

\subsection{Virtual experiment for binary operations} 

Here put several relations to these products, which might violate the
associativity. First, we consider $\pm$-sign at every
vertex. Denote in the figure of the next page
\parbox{.10\linewidth}{
\setlength{\unitlength}{1mm}
\begin{picture}(10,10)
\put(5,5){\color{red}{\circle*{2}}} 
\thicklines
\put(5,5){\vector(1,0){5}}
\put(5,5){\vector(0,1){5}}
\put(5,5){\vector(-1,-1){5}}
\put(11,5){$a$}
\put(5,10){$b$}
\put(0,2){$c$}
\end{picture}
}
, where $(a, b, c)$ are replaced sometimes by $(u, d, s)$.  
Consider paths starting at the (red) origin to some other lattice
point along any of three lines directed $a^{\pm 1}, b^{\pm 1}, c^{\pm 1}$ 
at each lattice point. We denote these by   
\bigskip
\parbox{.08\linewidth}{
\setlength{\unitlength}{.2mm}
\begin{picture}(15,38)
\put(0,0){\color{red}{\circle*{5}}} 
\thicklines
\put(0,0){\vector(1,0){20}}
\put(20,0){\vector(0,1){20}}
\put(20,20){\vector(1,1){20}}
\end{picture}
}
${=}c^{-1}{*}b{*}a{*}1$, 
\qquad 
\parbox{.08\linewidth}{
\setlength{\unitlength}{.2mm}
\begin{picture}(15,38)
\put(10,15){\color{red}{\circle*{5}}} 
\thicklines
\put(10,15){\vector(-1,-1){20}}
\put(-10,-5){\vector(1,0){20}}
\put(10,-5){\vector(1,0){20}}
\put(30,-5){\vector(0,1){20}}
\put(30,15){\vector(0,1){20}}
\put(30,35){\vector(-1,0){20}}
\end{picture}
}
$=a^{-1}{*}b{*}b{*}a{*}a{*}c{*}1$. \,\,\,(Be careful about the order.)

\bigskip
Paths are considered by parallel translations when we want to connect
two paths. Define the product for instance by    
$$
(b{*}c^{-1}{*}a{*}c{*}1){*}(c^{-1}{*}b{*}a{*}1){=}
b{*}c^{-1}{*}a{*}b{*}a{*}1
$$ 
where we set $c{*}c^{-1}{=}c^{-1}{*}c{=}1$. These are called a free
path connecting product.
Apparently these form a group and $(a^{\pm 1}, b^{\pm 1}, c^{\pm 1})$
is the generator. We denote this group by $\Gamma$. 
In what follows considering the $\pm$ sign for each element, we treat   
${\mathbb Z}_2{\otimes}\Gamma$.


Relations are set on each small triangle. If a triangle 
$\alpha\beta\gamma$ contains $\bullet$, or {\tiny{${\bullet}$}} then we set
$\alpha\beta\gamma{=}-1$ and if not we set $\alpha\beta\gamma{=}1$.
If bullets are distributed at random, then one can not fix the 
sign for almost all paths. Our main concern is to what extent we can control the $\pm$ sign, if 
certain periodic conditions are imposed.

%
%
%

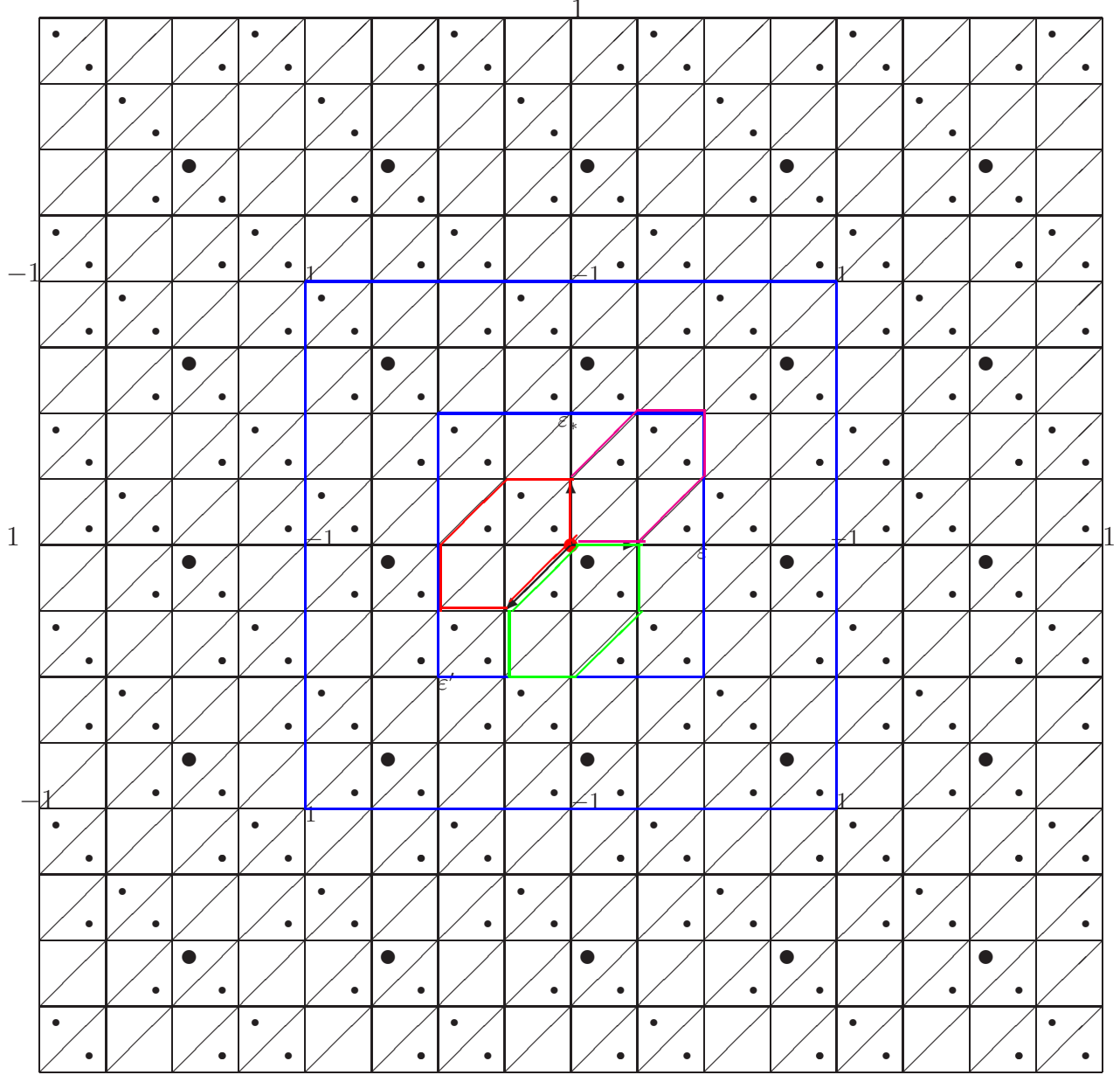
\begin{center}
\setlength{\unitlength}{1mm}
\begin{picture}(160,160)
\multiput(0,0)(0,10){17}{\line(1,0){160}}
\multiput(0,0)(10,0){17}{\line(0,1){160}}
\thinlines
\put(0,0){\line(1,1){160}}
\put(0,10){\line(1,1){150}}
\put(0,20){\line(1,1){140}}
\put(0,30){\line(1,1){130}}
\put(0,40){\line(1,1){120}}
\put(0,50){\line(1,1){110}}
\put(0,60){\line(1,1){100}}
\put(0,70){\line(1,1){90}}
\put(0,80){\line(1,1){80}}
\put(0,90){\line(1,1){70}}
\put(0,100){\line(1,1){60}}
\put(0,110){\line(1,1){50}}
\put(0,120){\line(1,1){40}}
\put(0,130){\line(1,1){30}}
\put(0,140){\line(1,1){20}}
\put(0,150){\line(1,1){10}}
\put(10,0){\line(1,1){150}}
\put(20,0){\line(1,1){140}}
\put(30,0){\line(1,1){130}}
\put(40,0){\line(1,1){120}}
\put(50,0){\line(1,1){110}}
\put(60,0){\line(1,1){100}}
\put(70,0){\line(1,1){90}}
\put(80,0){\line(1,1){80}}
\put(90,0){\line(1,1){70}}
\put(100,0){\line(1,1){60}}
\put(110,0){\line(1,1){50}}
\put(120,0){\line(1,1){40}}
\put(130,0){\line(1,1){30}}
\put(140,0){\line(1,1){20}}
\put(150,0){\line(1,1){10}}
\put(80,80){\color{red}{\circle*{2}}}
\put(80,160){$1$}
\put(160,80){$1$}
\put(-5,80){$1$}
\put(-5,120){$-1$}
\put(-3,40){$-1$}
\multiput(0,0)(0,30){6}{
{\multiput(7.5,2.5)(30,0){6}{\circle*{1}}}}
\multiput(0,0)(0,30){6}{
{\multiput(2.5,7.5)(30,0){6}{\circle*{1}}}}

\multiput(0,20)(0,30){4}{
{\multiput(7.5,2.5)(30,0){6}{\circle*{1}}}}
\multiput(20,0)(0,30){6}{
{\multiput(7.5,2.5)(30,0){5}{\circle*{1}}}}
\multiput(10,10)(0,30){5}{
{\multiput(7.5,2.5)(30,0){5}{\circle*{1}}}}

\multiput(10,20)(0,30){5}{
{\multiput(2.5,7.5)(30,0){5}{\circle*{1}}}}

\multiput(10,20)(0,30){5}{
{\multiput(7.5,2.5)(30,0){5}{\circle*{1}}}}

\multiput(20,10)(0,30){5}{
{\multiput(2.5,7.5)(30,0){5}{\circle*{2}}}}

\multiput(20,0)(0,30){5}{
{\multiput(7.5,2.5)(30,0){5}{\circle*{1}}}}
\multiput(20,10)(0,30){5}{
{\multiput(7.5,2.5)(30,0){5}{\circle*{1}}}}


\thicklines
\put(80,80){\vector(1,0){10}}
\put(80,80){\vector(0,1){10}}
\put(80,80){\vector(-1,-1){10}}
\put(99,78){\footnotesize{$\e$}}
\put(78,98){\footnotesize{$\e_*$}}
\put(60,58){\footnotesize{$\e'$}}
\put(119,80){\footnotesize{$-1$}}
\put(80,120){\footnotesize{$-1$}}
\put(40,38){\footnotesize{$1$}}
\put(40,120){\footnotesize{$1$}}
\put(120,120){\footnotesize{$1$}}
\put(40,80){\footnotesize{$-1$}}
\put(80,40){\footnotesize{$-1$}}
\put(120,40){\footnotesize{$1$}}
\color{blue}{
\put(60,60){\line(1,0){40}}
\put(60,60){\line(0,1){40}}
\put(60,100){\line(1,0){40}}
\put(100,60){\line(0,1){40}}
\put(40,40){\line(1,0){80}}
\put(40,40){\line(0,1){80}}
\put(40,120){\line(1,0){80}}
\put(120,40){\line(0,1){80}}}
\color{red}{
\put(59,70.5){\line(1,0){10}}
\put(59,70){\line(0,1){10}}
\put(59,80){\line(1,1){10}}
\put(69,90){\line(1,0){10}}
\put(69.5,71.5){\line(1,1){10}}
\put(78.5,80){\line(0,1){10}}
}
\color{green}{
\put(68,60){\line(1,0){10}}
\put(78,80){\line(1,0){10}}
\put(68,60){\line(0,1){10}}
\put(78,60){\line(1,1){10}}
\put(87.5,70){\line(0,1){10}}
\put(68.5,70){\line(1,1){10}}
}
\color{magenta}{
\put(86,80.5){\line(1,1){10}}
\put(96,90.5){\line(0,1){10}}
\put(76,90.5){\line(1,1){10}}
\put(86,100.5){\line(1,0){10}}
\put(77,80.5){\line(1,0){10}}
}
\end{picture}
\end{center}

\medskip
\noindent
$(1)$: First of all, the case there is no singular point $\bullet$, nor
{\tiny{${\bullet}$}}. 
Set as follows:
$$
a^{2}{*}1{=}1,\quad b^{2}{*}1{=}1,\quad c^{2}{*}1{=}1,\quad
a{*}b{*}c{*}1{=}1.
$$
Then, $1{=}c^{-1}{*}b^{-1}{*}a^{-1}{*}1{=}c{*}b{*}a{*}1$, and
$c=b{*}a{=}c^{-1}$. Hence, $a{*}b{=}c$. Thus we obtain the Klein's 4 group $K_4$. 
Replacing $(a, b, c)$ by $(-a, -b, -c)$ makes the changing relation 
$a{*}b{*}c{*}1{=}1$ to $a{*}b{*}c{*}1{=}-1$. This is the case where 
every triangle contains a singular point $\bullet$, 
but the generated group is isomorphic. The 
group ring $RK_4$ of $K_4$ over $\mathbb C$ is 
by identifying ${\mathbb R}\oplus{\mathbb R}ia$ with ${\mathbb C}$ 
$$
{\mathbb C}{\otimes}({\mathbb R}\oplus{\mathbb R}ia){\oplus}
{\mathbb C}{\otimes}\big(b{*}({\mathbb R}\oplus{\mathbb R}ia)\big){=}
{\mathbb C}\oplus {\mathbb C}{*}b,\quad (z{+}wb){*}(z'{+}w'b){=}zz'{+}ww'{+}(zw'{+}wz'){*}b.
$$
Hence, the group of all invertible elements is 
${\mathbb C}_*\times{\mathbb C}_*{*}b$. Its compact part is
$U(1)\times U(1)$.  
$RK_4$ has a natural nondegenerate bilinear form 
$\langle x,y\rangle$ as the coefficient of the identity 
element of $xy$. This satisfies 
$$
\langle x,y{*}z\rangle{=}\langle x{*}y,z\rangle, \quad
\langle a,a\rangle{=}\langle b,b\rangle{=}\langle a{*}b,a{*}b\rangle{=}1.
$$

\medskip
\noindent
$(2)$: Next, the case every triangle contains a singular point. We set then
$$
a^{2}{*}1{=}-1,\quad b^{2}{*}1{=}-1,\quad c^{2}{*}1{=}-1,\quad
a{*}b{*}c{*}1{=}-1.
$$
Then, $1{=}{-}c^{-1}{*}b^{-1}{*}a^{-1}{*}1{=}c{*}b{*}a{*}1$, and 
$b{*}a{*}1{=}-c{*}1{=}c^{-1}{*}1$. Hence $a{*}b{*}1{=}c{=}{-}b{*}a{*}1$.
Similarly,
$b{*}c{*}1{=}a{*}1$, $c{*}a{*}1{=}b{*}1$ are obtained easily. 
Thus we obtain the quaternion group. 

Replacing $(a, b, c)$ by $(-a, -b, -c)$ makes the changing relation 
$a{*}b{*}c{*}1{=}-1$ to $a{*}b{*}c{*}1{=}1$.  
The cases $(1)$ and $(2)$ are already appeared in \cite{OMMY5}. 

The group ring over ${\mathbb R}$ (resp. $\mathbb C$) of the quaternion group is the
quaternion field ${\mathbb H}$ (resp. $\mathbb C\otimes{\mathbb H}$). 
The compact part of all invertible elements is the group 
$SU(2)$ (resp. $U(2)$).

\bigskip
\noindent
$(3)$: The cases $(1)$ and $(2)$ are already appeared in
\cite{OMMY5}. On the other hand, we found in \cite{OMMY5} 
there is a class of expression parameters such that square roots of polar elements  
have strange properties that 
$$
a^{4}{*}1{=}-1,\quad b^{4}{*}1{=}-1,\quad c^{4}{*}1{=}1,\quad
a{*}b{*}c{*}1{=}1,\quad c{*}a{*}b{*}1{=}-1.  
$$
(See also \cite{ommy9}.)  To treat this case, we now consider the
consider the paths shown in Magical Lattice with singular points in
the previous page.  

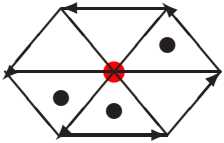
\setlength{\unitlength}{.7mm}
\parbox{.15\linewidth}{
\begin{picture}(15,30)(-7,-5)
\put(15,15){\color{red}{\circle*{4}}} 
\thicklines
\put(15,15){\line(1,0){20}}
\put(15,15){\line(-1,0){20}}
\put(15,15){\line(5,-6){10}}
\put(15,15){\line(5,6){10}}
\put(15,15){\line(-5,6){10}}
\put(15,15){\line(-5,-6){10}}
\put(7,5){\vector(-1,-1){3}}
\put(5,3){\vector(1,0){20}}
\put(25,27){\vector(-1,0){20}}
\put(25,3){\line(5,6){10}}
\put(32,12){\vector(1,1){3}}
\put(35,15){\line(-5,6){10}}
\put(27,25){\vector(-1,1){3}}
\put(5,27){\line(-5,-6){10}}
\put(-3,16.5){\vector(-1,-1){3}}
\put(-5.5,15){\line(5,-6){10}}
\put(25,20){\circle*{3}}
\put(15,7.5){\circle*{3}}
\put(5,10){\circle*{3}}
\end{picture}
}
\hfill 
\parbox[c]{.7\linewidth}{ 
These read at each triangle, the product of the three edges makes $1$ when 
there is no singular point ($\bullet$) inside, and it makes $-1$ when  
there is a singular point ($\bullet$) inside.
as a matter of cause, such relations destroys the possibility of binary operations.
The the relation requested for the path connected product is 
$c{*}b{*}a{*}1{=}-1$ at the right lower triangle, but 
there are right lower triangles with no $\bullet$ inside.
Same holds for $b{*}a{*}c{*}1{=}1$ at the left lower triangle.
}

\medskip 
Such a trouble is caused by the discordance between the periodicity 
of the tiling by triangles and that of bullets.
Any way, it is impossible to use $a, b, c$ as fundamental  
elements of binary operations. However, it may be possible 
to make binary operations between pairs of $a, b, c$. 
Note that once binary operations  
are established  by path connecting products on a 
family of suitably combined elements, then the associativity  
holds automatically.

\medskip
First setting ${\e}{=}a^2,$ ${\e^*}{=}b^2$,  ${\e'}{=}c^2$, we
investigate the system generated by ${\e},{\e^*},{\e'}$. 
(See inside the large blue box.) 

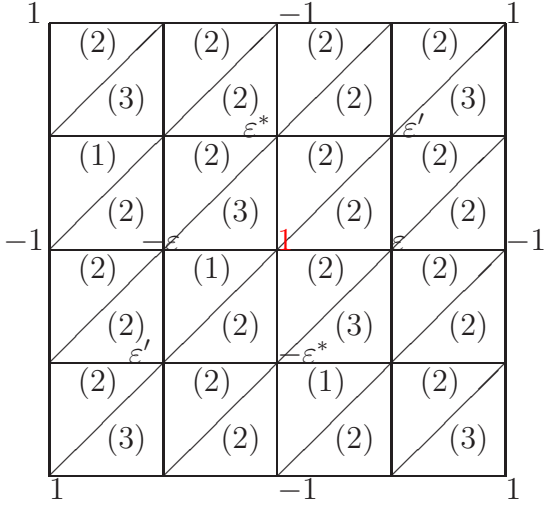
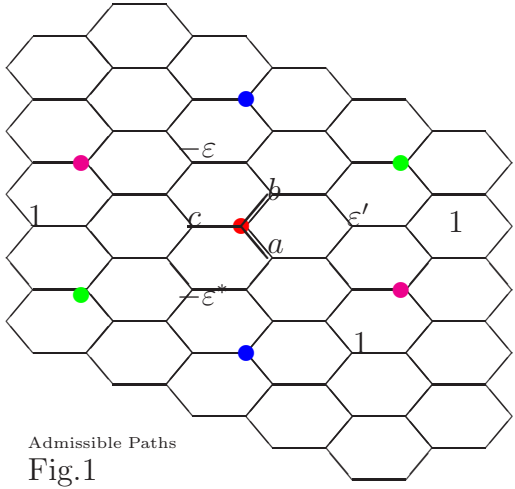
\setlength{\unitlength}{1.5mm}
\begin{picture}(50,50)(0,-5)
\multiput(0,0)(0,10){5}{\line(1,0){40}}
\multiput(0,0)(10,0){5}{\line(0,1){40}}
\put(0,0){\line(1,1){40}}
\put(0,10){\line(1,1){30}}
\put(0,20){\line(1,1){20}}
\put(0,30){\line(1,1){10}}
\put(10,0){\line(1,1){30}}
\put(20,0){\line(1,1){20}}
\put(30,0){\line(1,1){10}}
\put(0,-2){$1$}
\put(20,20){\color{red}{$1$}}
\put(-4,20){$-1$}
\put(-2,40){$1$}
\put(20,-2){$-1$}
\put(40,-2){$1$} 
\put(20,40){$-1$}
\put(40,40){$1$}
\put(40,20){$-1$} 
\put(7,10){${\e'}$}
\put(31,30){${\e'}$}
\put(8,20){$-{\e}$} 
\put(30,20){${\e}$}
\put(20,10){$-{\e^*}$}
\put(17,30){${\e^*}$} 
\multiput(2.5,37.5)(10,-10){4}{$(2)$}
\multiput(5,32.5)(10,-10){4}{$(3)$}
\multiput(12.5,37.5)(10,-10){3}{$(2)$}
\multiput(15,32.5)(10,-10){3}{$(2)$}
\multiput(2.5,27.5)(10,-10){3}{$(1)$}
\multiput(5,22.5)(10,-10){3}{$(2)$}
\multiput(2.5,17.5)(10,-10){2}{$(2)$}
\multiput(5,12.5)(10,-10){2}{$(2)$}
\multiput(22.5,37.5)(10,-10){2}{$(2)$}
\multiput(25,32.5)(10,-10){2}{$(2)$}
\put(2.5,7.5){$(2)$}
\put(32.5,37.5){$(2)$}
\put(5,2.5){$(3)$}
\put(35,32.5){$(3)$}
\end{picture}
\hfill 
\parbox[b]{.55\linewidth}
{We see ${\e}^2{=}-1,\,\,{\e^*}^2{=}-1,\,\,{\e'}^2{=}1$. 
The number in each triangle is the 
cardinality of singular point sitting inside, 
and this pattern repeated periodically. 
Hence, we cannot obtain a consistent definition of path connecting 
products of these elements. For instance, counting the number of 
singular points in the path, we have to write   
$$
{\e}{*}\e^*{=}{\e}{*}{\e}^{*}, \quad
{\e}^*{*}{\e}{=}{\e}^{-1}{*}{\e^*}{=}-{\e}{*}{\e^*},
$$
$$ 
{\e}{*}{\e}'{=}{\e}{*}{\e}', \quad {\e}'{*}{\e}{=}{\e}^{-1}{*}{\e}'
{=}{-}{\e}{*}{\e}'.
$$
Hence in ${\mathfrak K}_{im}$ expression, these polar elements cannot 
be a member of binary operation, 
while in ${\mathfrak K}_{re}$ expressions, the polar element 
is a single element such that $\e^2{=}1$.
Recall that ${\e}{=}a^2,$ ${\e^*}{=}b^2$, ${\e'}{=}c^2$. (Cf.\cite{OMMY5})} 

\noindent
\newsavebox{\firstpattern}
\setlength{\unitlength}{0.7mm}
\begin{picture}(100,100)(0,-10)
\savebox{\firstpattern}{
\begin{picture}(20,18)
\put(5,0){\line(1,0){10}}
\put(15,0){\line(5,6){5}}
\put(20,6){\line(-5,6){5}}
\put(15,12){\line(-1,0){10}}
\put(5,0){\line(-5,6){5}}
\put(0,6){\line(5,6){5}}
\end{picture}}
\multiput(0,30)(15,-6){6}{\usebox{\firstpattern}}
\multiput(0,42)(15,-6){6}{\usebox{\firstpattern}}
\multiput(0,54)(15,-6){6}{\usebox{\firstpattern}}
\multiput(0,66)(15,-6){6}{\usebox{\firstpattern}}
\multiput(0,18)(15,-6){5}{\usebox{\firstpattern}}
\multiput(15,72)(15,-6){5}{\usebox{\firstpattern}}
\put(46,42){{\color{red}{\circle*{3}}}}
\thicklines
\put(46,42){\line(-1,0){10}}
\put(46,42){\line(5,6){5}}
\put(46,42){\line(5,-6){5}}
\put(36,42){$c$}
\put(51,47){$b$}
\put(51,37){$a$}
\put(6,42){$1$}
\put(67,18){$1$}
\put(85,41){$1$}
\put(34,27){$-{\e}^*$}
\put(34,55){$-{\e}$}
\put(66,42){${\e}'$}
\put(6,0){\tiny{Admissible Paths}}
\put(6,-6){Fig.1}
\put(76,30){{\color{magenta}{\circle*{3}}}}
\put(76,54){{\color{green}{\circle*{3}}}}
\put(47,66){{\color{blue}{\circle*{3}}}}
\put(47,18){{\color{blue}{\circle*{3}}}}
\put(16,29){{\color{green}{\circle*{3}}}}
\put(16,54){{\color{magenta}{\circle*{3}}}}
\end{picture}
\hfill
\parbox[b]{.55\linewidth}{
\medskip
Note that the number of singular points in every small triangle in
Magical lattice is one or zero, and such a difference destroys the
binary operations. However this means only that we cannot use 
$a, b, c$ for fundamental elements of binary operations, just as
$\{{\e}, {\e}^*,  {\e}'\}$. 
On the other hand, note that every hexagon in the
Magical lattice contains three singular points.  Fig.1 is a 
tilings containing the origin at the common edges of three tiles.
Now we permit to use 
only invertible paths sitting on the edges of hexagons on a tiling.
Single paths $a,\, b,\, c$ are not involved, for these are not 
invertible within the admissible paths.

Now we see all closed paths around a hexagon make $-1$, hence 
we have relations as follows:}
$$
b^{-1}{*}c{*}a^{-1}{*}b{*}c^{-1}{*}a{*}1{=}-1, \quad 
c^{-1}{*}a{*}b^{-1}{*}c{*}a^{-1}{*}b{*}1{=}-1,\quad 
a^{-1}{*}b{*}c^{-1}{*}a{*}b^{-1}{*}c{*}1{=}-1.
$$
It does not depend on the orientation of closed paths, but such relations are forbidden: 
$$
b{*}a{*}1{=}{-}c^{-1}{*}1,\,\, c{*}b{*}1{=}{-}a^{-1}{*}1,\,\, a{*}c{*}1{=}{-}b^{-1}{*}1,\,\,  
a{*}b{*}1{=}c^{-1}{*}1,\,\, b{*}c{*}1{=}a^{-1}{*}1,\,\,c{*}a{*}1{=}b^{-1}{*}1. 
$$
Consider now compositions of two generators not included in the above:
For simplicity and by minding the similarity to the quark model, we denote these as follows:
$$
\bar{s}u{=}c^{-1}{*}a{*}1, \quad \bar{u}d{=}a^{-1}{*}b{*}1, \quad \bar{d}s{=}b^{-1}{*}c{*}1,
$$
$$
\bar{u}s{=}a^{-1}{*}c{*}1, \quad \bar{d}u{=}b^{-1}{*}a{*}1,\quad \bar{s}d{=}c^{-1}{*}b{*}1.  
$$
The second row are inverses of the first row, and 
$$
(\bar{d}s){*}(\bar{s}u){=}\bar{d}u, \quad (\bar{u}s){*}(\bar{s}d){=}\bar{u}d. 
$$
As the tiling and singularities are distributed periodically, 
these form a group by the path connecting product. Here, the origin {\color{red}{$\bullet$}}
has $\pm$ sign. We fix $+${\color{red}{$\bullet$}} is the
identity element of the group.

We denote this group by $\Gamma_{uds}$. Note that polar elements ${\e}, {\e^*}, {\e'}$ are not 
on the admissible paths. By tracing in Magical lattice, it is easy to see 
$$
(\bar{s}u)_*^4{=}{-1}{=}(\bar{u}s)_*^4, \quad
(\bar{u}d)_*^4{=}1{=}(\bar{d}u)_*^4,\quad  
(\bar{d}s)_*^4{=}{-1}{=}(\bar{s}d)_*^4,\quad 
(\bar{s}u)_*{*}(\bar{u}d)_*{*}(\bar{d}s)_*{=}1.
$$
Moreover as every hexagon contains three singular points, we have
another relation   
$$
(\bar{d}s)_*{*}(\bar{u}d)_*{*}(\bar{s}u)_*{=}-1,       
$$
and these show that they anti-commute each other: i.e.
$$
(\bar{s}u){*}(\bar{u}d){=}-(\bar{u}d){*}(\bar{s}u),\quad 
(\bar{u}d){*}(\bar{s}d){=}-(\bar{s}d){*}(\bar{u}d),\quad 
(\bar{u}s){*}(\bar{d}s){=}-(\bar{d}s){*}(\bar{u}s), e.t.c.
$$ 
It follows that the squares   
$(\bar{s}u)_*^2,\,\,(\bar{u}d)_*^2,\,\,(\bar{d}s)_*^2$ commute with
all other elements. Hence these generate the center 
${\mathcal Z}_2{\cong}{\mathbb Z}_4{\times}{\mathbb Z}_2$:
\begin{equation}\label{Z4Z2}
{\mathcal Z}_2{=}\{\pm 1,\,\,\pm(\bar{s}u)_*^2,\,\,
\pm(\bar{u}d)_*^2,\,\,\pm (\bar{s}u)_*^2{*}(\bar{u}d)_*^2\}
\quad({\text{6 colored points in Fig.1}}).
\end{equation}

Let $\pi:\Gamma_{uds}\to \Gamma_{uds}/{\mathcal Z}_2$ be the
projection onto the factor group. We have then 
$$
\begin{aligned}
&(\pi(\bar{s}u))^2{=}(\pi(\bar{u}d))^2{=}(\pi(\bar{d}s))^2{=}1,\\ 
&
\pi(\bar{s}u){*}\pi(\bar{u}d){=}-\pi(\bar{u}d){*}\pi(\bar{s}u),\quad 
\pi(\bar{u}d){*}\pi(\bar{d}s){=}-\pi(\bar{d}s){*}\pi(\bar{u}d),\quad 
\pi(\bar{s}u){*}\pi(\bar{d}s){=}-\pi(\bar{d}s){*}\pi(\bar{s}u). 
\end{aligned}
$$
Hence we have 
\begin{prop}\label{QTN}
${\mathcal Z}_2$ is the center of $\Gamma_{uds}$, and 
$\pi(i\bar{s}u),\,\,\pi(i\bar{u}d),\,\,\pi(i\bar{d}s)$ generates the
quaternion group. Its group ring over $\mathbb C$ is $\mathbb
C{\otimes}{\mathbb H}$. 
\end{prop}

It is convenient to represent these as follows by using Pauli
matrices: Set $\alpha{=}e^{\frac{\pi i}{4}}$, 
$\beta{=}e^{\frac{3\pi i}{4}}$ and  
\begin{equation}\label{udsuds}
\bar{s}u{=}
\begin{bmatrix}
0&i\alpha& 0\\
i\alpha&0&0 \\
0&0&\alpha
\end{bmatrix},\quad 
\bar{d}s{=}
\begin{bmatrix}
i\alpha& 0&0\\
0&-i\alpha&0 \\
0&0&\beta
\end{bmatrix},\quad
\bar{u}d{=}
\begin{bmatrix}
0& -1&0\\
1&0&0 \\
0&0&-1
\end{bmatrix}.
\end{equation}
Note that ${\mathcal Z}_2$ is not the polar elements. It is
interesting that we obtain a kind of double cover of the quaternion
group. Recall that $\{1, {\e}, {\e}^{*}, {\e}'\}$ does not form a group.

\bigskip
\noindent
{\bf Other 5 isomorphic groups}.  Beside $\Gamma_{uds}$, there are 5
groups which is isomorphic to  $\Gamma_{uds}$:
$$
u\Gamma_{uds}\bar{u},\,\, d\Gamma_{uds}\bar{d},\,\, s\Gamma_{uds}\bar{s},\,\, 
\bar{u}^2\Gamma_{uds}u^2,\,\, \bar{d}^2\Gamma_{uds}d^2.
$$
These 6 groups all together seem to be deeply related to the octernions.


\begin{thebibliography}{OM}






\bibitem{BF}{F.Bayen,\,M,Flato,\,C.Fronsdal,\,A.Lichnerowicz,\,D.Sternheimer,\,}
{\it Deformation theory and quantization I, II}, Ann. Phys. 111, (1977),
61-151.  

\bibitem{B}
F. Berezin, {\it General concept of quantization,} Comm. Math. Phys.  
8 (1975), 153-174. 

%



\bibitem{GF}{I.M.Gel'fand,\,D.B.Fuchs,\,\,}
{\it Cohomologies of Lie algebra of tangential vector fields of a smooth
  manifold I, II}, Functional Anal. Appl. 3. 4, 
(1969), (1970), 194-210, 257-282. 



\bibitem{He}{L. H{\"o}rmander}
\newblock{\it Fourier Integral operators, I}, 
\newblock{Acta Math. 127 (1971), 79-183.}


\bibitem{He2}{L. H{\"o}rmander},
\newblock{\it The Weyl calculus of pseudo-differential operators}, 
\newblock{Commun. Pure. Appl. Math. 32 (1979) 359-443}

\bibitem{Hi}{N. Hitchin},
\newblock{\it Lectures on special Lagrangian submanifolds}, 
\newblock{arXiv:math.DG/9907034vl 6Jul, 1999.
\newblock{AMS/IP Stud Adv. Math. 23 (2001) 151-182}}



\bibitem{Ko}{M. Kontsevitch} {Deformation quantization of Poisson
manifolds, I}, {qalg/9709040}
\newblock{Lett. Math. Phys. 66 (2003) 157-216}.

%

\bibitem{om3}{H. Omori},\,\,{\sc Infinite dimensional Lie groups}, AMS
  Translation Monograph 158, 1997.   

%
%
%





\bibitem{OMMY7}
{H.Omori, Y.Maeda, N.Miyazaki and A.Yoshioka :}
\newblock{\em Strange phenomena related to ordering problems in quantizations},  
  \newblock{Jour. Lie Theory, Vol 13, No 2, (2003)}, 481-510. 


\bibitem{OMMY3}
{H.Omori, Y.Maeda, N.Miyazaki and A.Yoshioka :}
\newblock{\it Deformation of expressions for elements of algebras  (I)}, 
 -(Jacobi's theta functions and $*$-exponential functions)-
{arXiv:1104.2109} 
 

\bibitem{OMMY4}
{H.Omori, Y.Maeda, N.Miyazaki and A.Yoshioka :}
\newblock{\it Deformation of expressions for elements of algebras  (II)}, 
-(Weyl algebra of $2m$-generators)-
{arXiv:1105.1218} 


\bibitem{OMMY5}
{H.Omori, Y.Maeda, N.Miyazaki and A.Yoshioka :}
\newblock{\it Deformation of expressions for elements of algebras  (III)}, 
-Generic product formula for ${*}$-exponentials of quadratic forms-
{arXiv:1107.2474}


\bibitem{ommy6}
{H.Omori, Y.Maeda, N.Miyazaki and A.Yoshioka :}
\newblock{\it Deformation of expressions for elements of algebras  (IV)}
-Matrix elements and related integrals-
{arXiv:1109.0082} 


\bibitem{ommy7}
{H.Omori, Y.Maeda, N.Miyazaki and A.Yoshioka :}
\newblock{\it Deformation of expressions for elements of algebras (V)}
-Diagonal matrix calculus and $*$-special functions-
{arXiv:1111.1806} 


\bibitem{ommy8}
{H.Omori, Y.Maeda, N.Miyazaki and A.Yoshioka :}
\newblock{\it Deformation of expressions for elements of algebras (VI)}
-Vacuum representation of Heisenberg algebra-,{arXiv:1204.5566}



\bibitem{ommy9}
{H.Omori, Y.Maeda, N.Miyazaki and A.Yoshioka :}
\newblock{\it Deformation of expressions for elements of algebras (VII)}
-Vacuum/Psudo-vacuum Representation-,{arXiv:1210.3426}


\bibitem{PSG}
{A.Pressley, G. Segal}
\newblock{\sc Loop Groups}, \newblock{Oxford Sci. Publ.}, 1986.   
 

 
 



\end{thebibliography}
\end{document}